\documentclass[journal]{IEEEtran}
\IEEEoverridecommandlockouts
\usepackage{cite}
\usepackage{amsmath,amssymb,amsfonts,amsthm}
\usepackage{algorithmic}
\usepackage{graphicx}
\usepackage{textcomp}
\usepackage{comment}
\usepackage{url}
\usepackage{authblk}
\usepackage{array}
\usepackage{multirow}
\usepackage{subfigure}
\usepackage[ruled,linesnumbered]{algorithm2e}

\usepackage[table]{xcolor}
\usepackage{booktabs}
\newtheorem{theorem}{Theorem}
\newtheorem{lemma}{Lemma}

\newcommand{\tabincell}[2]{\begin{tabular}{@{}#1@{}}#2\end{tabular}}

\def\BibTeX{{\rm B\kern-.05em{\sc i\kern-.025em b}\kern-.08em
    T\kern-.1667em\lower.7ex\hbox{E}\kern-.125emX}}
\usepackage{balance}

\begin{document}

\title{Pioneering Deterministic Scheduling and Network Structure Optimization for Time-Critical Computing Tasks in Industrial IoT}
\author{Yujiao Hu, \textit{IEEE Member}, Yining Zhu, Huayu Zhang, Yan Pan, Qingmin Jia, Renchao Xie, \textit{IEEE Senior Member}, Gang Yang, F. Richard Yu, \textit{IEEE Fellow}

\thanks{Yujiao Hu, Huayu Zhang, Qingmin Jia and Renchao Xie are with Future Network Research Center, Purple Mountain Laboratories, Nanjing 211111, China (email: huyujiao@pmlabs.com.cn, zhanghuayu@pmlabs.com.cn, jiaqingmin@pmlabs.com.cn, Renchao\_xie@bupt.edu.cn).
Yining Zhu and Gang Yang are with the School of Computer Science, Northwestern Polytechnical University, Xi'an 710029, China (email: yiningzhu@nwpu.edu.cn, yang.gang@gmail.com).
Yan Pan is with Science and Technology on Information Systems Engineering Laboratory, National University of Defense Technology, China (email:panyan@nudt.edu.cn)
Renchao Xie is also with the State Key Laboratory of networking and Switching Technology, Beijing University of Posts and Telecommunications, Beijing 100876, China.
F. Richard Yu is with the Department of Systems and Computer Engineering, Carleton University, Ottawa, Canada (e-mail: richard.yu@carleton.ca).  

This work was supported in part by the National Nature Science Foundation of China under Grants 62171046, 61876151 and 62032018. 
}
}

\maketitle

\begin{abstract}	
The Industrial Internet of Things (IIoT) has become a critical technology to accelerate the process of digital and intelligent transformation of industries. As the cooperative relationship between smart devices in IIoT becomes more complex, getting deterministic responses of IIoT periodic time-critical computing tasks becomes a crucial and nontrivial problem. However, few current works in cloud/edge/fog computing focus on this problem. This paper is a pioneer to explore the deterministic scheduling and network structural optimization problems for IIoT periodic time-critical computing tasks. We first formulate the two problems and derive theorems to help quickly identify computation and network resource sharing conflicts. Based on this, we propose a deterministic scheduling algorithm, \textit{IIoTBroker}, which realizes deterministic response for each IIoT task by optimizing the fine-grained computation and network resources allocations, and a network optimization algorithm, \textit{IIoTDeployer}, providing a cost-effective structural upgrade solution for existing IIoT networks. Our methods are illustrated to be cost-friendly, scalable, and deterministic response guaranteed with low computation cost from our simulation results.

\end{abstract}

\begin{IEEEkeywords}
Industrial Internet of Things, periodic time-critical computing tasks, deterministic scheduling, network structure optimization, computation and network resource sharing 
\end{IEEEkeywords}

\section{Introduction}
The Industrial Internet of Things (IIoT) links all types of industrial equipment through networks and can further support data collection, exchange and analysis \cite{qiu2020edge}. IIoT has become a critical technology to help industries reduce cost and enhance productivity, as well as enable digital and intelligent transformation. For example, the manufacturing industry introduces IIoT technologies to connect all smart devices in the production workshop, such as flexible robotic arms, automated guided vehicles (AGV), industrial cameras, in order to enhance the intelligence and automation of production lines. As the industrial digital transformation process accelerates, it can be foreseen that the devices scale of IIoT will become much larger. 
Along with the application of IIoT into industries, such as manufacturing\cite{gupta2022industrial,tao2017iihub}, transport\cite{pretorius2020industrial} and energy \cite{mao2021energy}, many periodic time-critical computing tasks are generated by networked devices, such as sensor data analysis\cite{ghahramani2020ai,dai2020big} and motion planning for robots\cite{zacharia2020agv,lu2021collision}. Timely completion of the periodic computing tasks is crucial for ensuring the correct and efficient operations of IIoT.
Moreover, as the cooperative relationships between smart devices in IIoT become more complex, there is an increasing expectation for the response delays of certain time-critical computing tasks to be deterministic.

Cloud computing, edge computing and fog computing are primary approaches to provide computation resources for the time-critical computing tasks in the development of IIoT. Some works have proposed computation offloading approaches to empower IIoT. For example, \cite{chen2019energy} develops an fog computing oriented energy-optimal dynamic computation offloading scheme to minimize energy consumption with consideration of a desired energy overhead and delay. \cite{chen2019cooperative} and \cite{hong2019multi} break the normal assumption that all IIoT devices can connect to edge servers or cloud data centers directly, and build their game-theoretic formulations in which the IIoT devices make computation-offloading strategies individually. Some learning-based approaches \cite{ren2020deep,rahman2020deep} and SDN-based methods \cite{tang2022sdn,naeem2020sdn} are explored as well. 

However, in the end-edge-cloud cooperative architecture, no matter it is a single hop connection or a multi-hop connection between IIoT devices and remote computing servers, it relies on best-effort networks to forward data. The transmission time in the best effort network is unstable and unpredictable. This leads to the entire response delays of some tasks exceed their deadlines, even though the task latency may be guaranteed as much as possible at the computation level. In addition, because of the dynamic load balancing mechanism and high priority preemption mechanism at the computation level, the waiting time, calculation time and execution server of IIoT tasks are different in every period. This leads to uncertainty and unpredictability in the computing time of IIoT tasks. 

Motivated by the challenges mentioned above, we start to explore a local scheme to provide deterministic and predictable response delay guarantee for IIoT periodic computing tasks. One simple way is configuring a dedicated server for each task and building communication networks to achieve near-zero transmission latency between every pair of dedicated servers and source IIoT devices of tasks. However, the deployment cost of such method is not only too expensive for the enterprises, but also a waste of computing resources. Therefore, it is crucial to explore a task scheduling approach that can both ensure deterministic responses for IIoT tasks and allow to share local servers and networks. Furthermore, to reduce the deployment cost of computation and network resources while meeting task requirements, it is essential to study IIoT network structure optimization methods, so that local servers, IIoT devices and network routers can be organized in an efficient architecture. 

Prior to this work, the deterministic scheduling problem is studied in time sensitive networks (TSN). The scheduling targets, however, are network traffics instead of periodic time-critical computing tasks. 
TSN is a set of IEEE 802.1 standards that defines mechanisms to provide deterministic services through IEEE 802 networks \cite{finn2018introduction,messenger2018time,farkas2018time,bello2019perspective}. Deterministic services include guaranteed packet transport with bounded latency, low packet delay variation, and low packet loss. Some algorithms have been proposed to make deterministic scheduling schemes for network traffics \cite{craciunas2017overview,atallah2019routing,ashjaei2021time}. The core idea is to divide each network link into time slots, and then allocate transmission time slots for each traffic through specific algorithms. The algorithms usually fully utilize the limited buffer ($<1MB$) of switches and routers, temporarily caching the traffic packets during forwarding, to avoid traffic conflicts and increase bandwidth utilization. 
Compared to deterministic scheduling for TSN traffics that only works at the network level, the deterministic scheduling of IIoT periodic time-critical computing tasks is more complex, since it requires to simultaneously plan at both the network level and computation level. Even the network-level scheduling of IIoT computing tasks differs from TSN traffic scheduling. Because the data packets of IIoT computing tasks are usually more than 1MB, which indicates that caching the data packets on routers/switchers during forwarding as the way done in TSN traffic scheduling is not feasible. 

To our best knowledge, our work is the first to explore deterministic scheduling and network structure optimization issues for periodic and time-critical computing tasks in IIoT. 
In this paper, we first theoretically analyze the conditions of resource sharing conflicts at the computation level and network level. Then we derive theorems and lemmas to assist deterministic scheduling and network structural optimization algorithms design. 
Following that, a resource-efficiency heuristic scheduling approach \textit{IIoTBroker} is proposed to compute a global resource allocation scheme, which guarantees deterministic response delay for each IIoT computing task. The scheme consists of two parts: the server and corresponding computation slots that are allocated to each IIoT task, as well as the bidirectional routing paths planned for each IIoT task (from IIoT devices to servers and from servers to IIoT devices) and the transmission slots occupied for forwarding data packets of the IIoT task on every routing link. 
After that, a demand-aware IIoT network structure optimization approach \textit{IIoTDeployer} is proposed, 
which offers an improvement suggestion based on current IIoT network to industrial enterprises, so that the required computation and network resources by IIoT tasks can be provided in a cost-friendly way. 
Finally, a set of simulations are presented to illustrate the efficiency of proposed approaches. The results demonstrate that IIoTBroker can provide higher-quality scheduling decisions to ensure deterministic response for IIoT tasks than baselines. The computation efficiency of IIoTBroker is higher than others as well, since IIoTBroker consumes shorter runtime. IIoTDeployer provides more cost-friendly IIoT network structure optimization solutions than baselines and can be compatible with scenarios where the number of IIoT tasks and the cost settings for adding servers and links are different. 

Our main contributions are summarized as follows.
\begin{itemize}
    \item We identify crucial but underexplored challenges in the intelligent IIoT, which are the deterministic scheduling and network structure optimization for IIoT periodic time-critical computing tasks. 
    \item We derive theorems and lemmas that can quickly identify computation and network resource sharing conflicts. Based on that, we propose a deterministic scheduling algorithm, namely \textit{IIoTBroker}, to allocate computing time slots and transmission time slots for each IIoT computing task to ensure its deterministic response. 
    \item We propose an algorithm, \textit{IIoTDeployer}, to generate cost-friendly IIoT network structure optimization scheme which minimizes upgrade cost while ensuring the deterministic responses for all IIoT tasks.
    \item We carry out abundant simulations to demonstrate that IIoTBroker can adapt to any problem scales and make cost-friendly effective scheduling decisions to ensure deterministic response for all IIoT tasks. IIoTDeployer can provide cost-friendly IIoT network structure upgrade solutions for scenarios with different problem scales and cost settings. 
\end{itemize}

This paper is structured as follows. We discuss related work in Section~\ref{section:related-work} and introduce  the preliminaries in Section~\ref{section:perliminaries}. In Section~\ref{section:theoretical-modeling}, we give our theoretical models. In Section~\ref{section:IIoTBroker} and Section~\ref{section:IIoTDeployer}, \textit{IIoTBroker} and \textit{IIoTDeployer} are proposed and explained in detail. We evaluate our algorithms in Section~\ref{section:experiments} and conclude our paper in Section~\ref{section:conclusion}. 

\section{Related Work}\label{section:related-work}
\subsection{Cloud/Edge/Fog-based IIoT Computation Offloading}
To overcome the computation resource constraints of IIoT devices, cloud/edge/fog-based IIoT computation offloading has been widely researched. 
Chen et al. \cite{chen2019cooperative} develop a game-theory-based distributed computation offloading solution to minimize the economic cost of IIoT devices for blockchain-empowered IIoT, and develops a distributed algorithm to help IIoT devices quickly reach the Nash equilibrium.  
Hong et al. \cite{hong2019multi} formulates multi-hop cooperative communication problems based on game theory, and proposes two QoS-aware distributed algorithms to make the game terminate in an Nash equilibrium through a finite improvement step. 
Ren et al. \cite{ren2020deep} builds a learning-based model for each IIoT device to make it identify its fog server based on network and device states, then builds a greedy algorithm to make fog server determine which IIoT tasks should be further offloaded to the cloud. 
Rahman et al. \cite{rahman2020deep} designs a deep reinforcement learning based controller to intelligently decides whether to deal with the IIoT task locally or offload it to a fog/cloud server, as well as the optimal amount of computation and power resources that should be allocated.
Tang et al. \cite{tang2022sdn} introduces software-defined network technology to help determine the most suitable offloading path and computational server, so that the response latency of IIoT tasks can be minimized. 
Naeem et al. \cite{naeem2020sdn} formulates the quaility-of-service (QoS) enabled routing optimization problem as a max-flow min-cost problem, and proposes a concurrent planning mechanism for calculating optimal forwarding paths considering the QoS requirements of each flow. 
Peng et al. \cite{peng2022intelligent} proposes new models for scientific workflow and concurrent workflow, and further designs an end-edge-cloud collaborative computation offloading method in accordance with NSGA-III to realize multi-object optimization, including energy consumption and time constrain of tasks, resource utilization and load balance of edge servers. 

\subsection{Deterministic Scheduling of TSN traffics}
Given TSN topology and traffic specifications, TSN traffic scheduling is to compute routing policies of traffics and control parameters in TSN device, these parameters mainly include forwarding table, gate control list (GCL), filtering policy and so on. Constraint programming based on integer linear programming (ILP) formulations is the most basic computing method. For example, Steiner et al. \cite{steiner2010evaluation} pioneered to formulate the time-triggered frame scheduling problem in multi-hop networks based satisfiability modulo theories, which is a classic combinatorial problem. Based on the formulation, the offsets of each frame along each hop on its routing path can be computed. Then the offsets are transferred into the GCL of TSN switches. After that, many TSN traffic scheduling researches follow this idea. 
Zhou et al. \cite{zhou2022time} proposed an integer linear formulation based method for scheduling time-triggered traffic with preemption supports, which can improve the schedulability compared to non-preemptive scheduling. 
Atallah et al. \cite{atallah2019routing} and Gavrilut et al. \cite{gavrilut2017fault} presented an ILP-based method for routing and scheduling time-triggered traffic with multi-path supports, which can enhance the fault tolerance and load balancing of TSN. However, for large-scale networks or traffics, Constraint programming is a NP-hard computing complexity problem, which suffers from scalability problems. 

To make mass TSN traffic scheduling feasible, heuristic or meta-heuristic algorithms have been proposed. Gavriluct et al. \cite{gavriluct2018scheduling} proposed a greedy randomized adaptive search procedure (GRASP) for synthesizing GCLs for both time-triggered and audio-video bridging traffics. 
Nayak et al. \cite{nayak2017incremental} proposed an integration method between TSN networks and software-defined networking (SDN), utilizing the global view and flexible control of SDN to optimize scheduling of flows in a TSN network.
Pahlevan et al. \cite{pahlevan2018genetic} presented a genetic algorithm that combined the routing and scheduling constraints and generated static global schedules, improving the schedulability, time-triggered transmission efficiency and resource utilization.
Prados et al. \cite{prados2020learnet} introduced reinforcement learning technique to dynamically adjust bandwidth allocation and priority assignment for asynchronous deterministic networks, with consideration of network congestion and status changes. 
 
\begin{figure}
    \centering
    \includegraphics[scale=0.25]{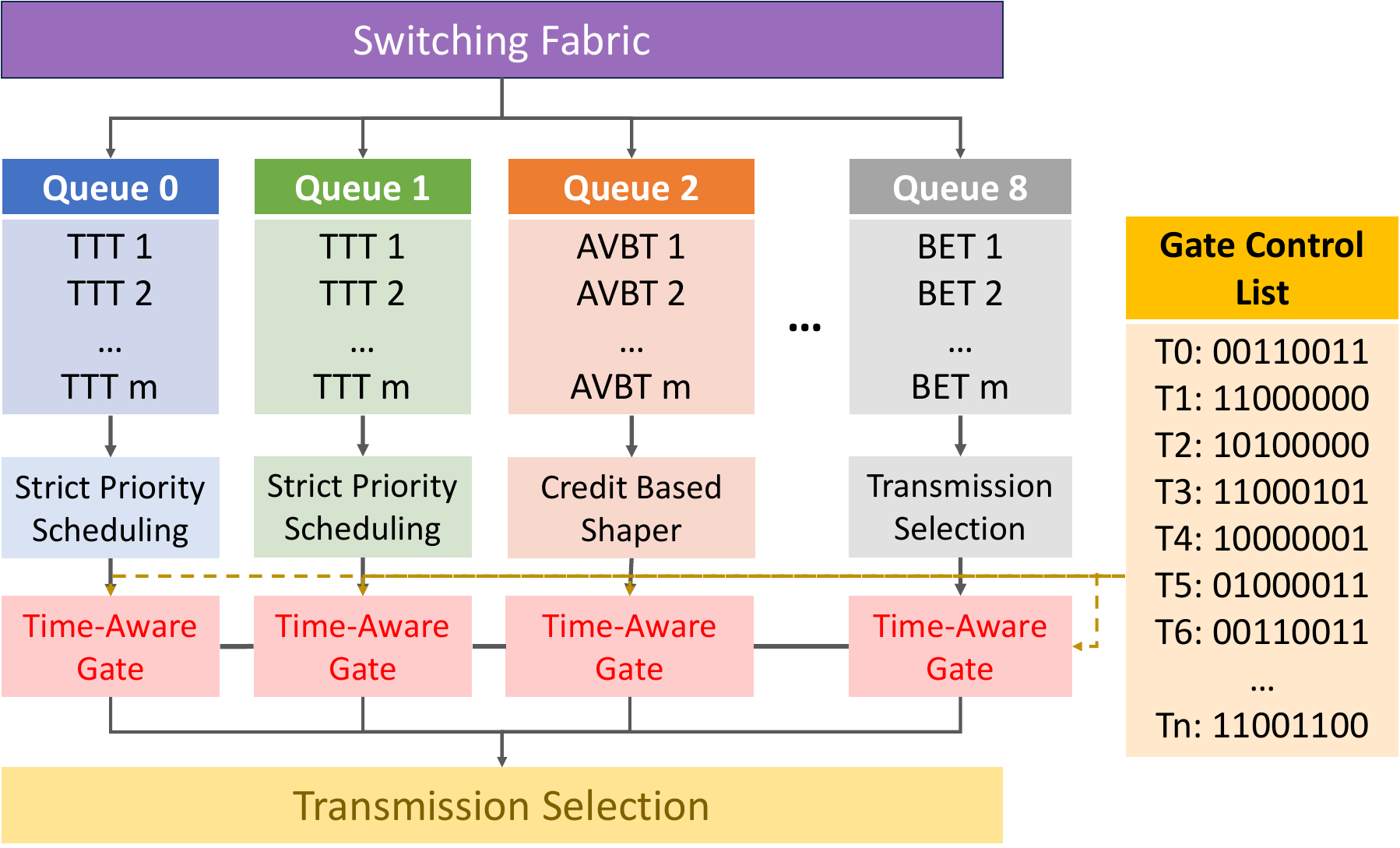}
    \caption{Illustration of how the IEEE 802.1Qbv TAS works on egress hardware. TSN traffic can be classified into three types: time-triggered traffic (TTT), audio video bridging traffic (AVBT), and best effort traffic (BET). With TAS, each queue is controlled by the gate control list that determines which queue is open. 1 indicates that the queue is open, while 0 indicates the queue is closed. }
    \label{fig:fig2}
\end{figure}
\section{Preliminaries}\label{section:perliminaries}
\subsection{TSN and IEEE 802.1Qbv}
The Time-Sensitive Networking (TSN), which guarantees packet transmission with bounded latency, low packet delay variation, and low packet loss, is considered as an enhancement to Ethernet and a promising technology to help accelerate the digitalization of industries.
IEEE 802.1 TSN task group has put forward standards in perspectives of clock synchronization, frame preemption, network management, and scheduled traffic enhancement. Especially, IEEE 802.1Qbv defines a time-aware shaper (TAS) for scheduling TSN traffics, which is closely related to our work. TAS manages the forwarding time of queues in each output port of a switch by a gate control list (GCL), as illustrated in Fig.~\ref{fig:fig2}. With TAS, the traffic scheduling problem becomes equivalent to allocating transmission time slots for each traffic flow over the hyper period\footnote{Hyper period refers to the least common multiple (LCM) of the periods of multiple periodic TSN traffics.}. 

Inspired by the equivalent transformation of the deterministic scheduling problem of TSN traffics, we convert the deterministic scheduling problem of IIoT periodic time-critical computing tasks to a fine-grained time slot allocation problem on computation and network resources in IIoT. 


\subsection{Computing Resource Sharing}
Computers have indeed supported the sharing of computing resources among multiple tasks, whether those tasks are executed sequentially or in parallel. Moreover, computers have also been designed with interrupt mechanisms to allow high-priority tasks to quickly acquire resources. However, the interrupt mechanism may increase the computation time of tasks due to process switching overhead, and disrupt the deterministic and controlled system status. 

Therefore, this paper assumes that the computing process of IIoT tasks can never be interrupted. Furthermore, to prevent multiple tasks from competing for computing resources, which leads to an unpredictable and uncontrolled increase in the computation time for each task, we assume that servers in IIoT can only process computing tasks in sequence. Then the computing-level planning for IIoT tasks can be modeled as an on-demand allocation of computation slots across all servers. 

\subsection{The Lifecycle of IIoT Computing Tasks} 

\begin{figure}
    \centering
    \includegraphics[scale=0.26]{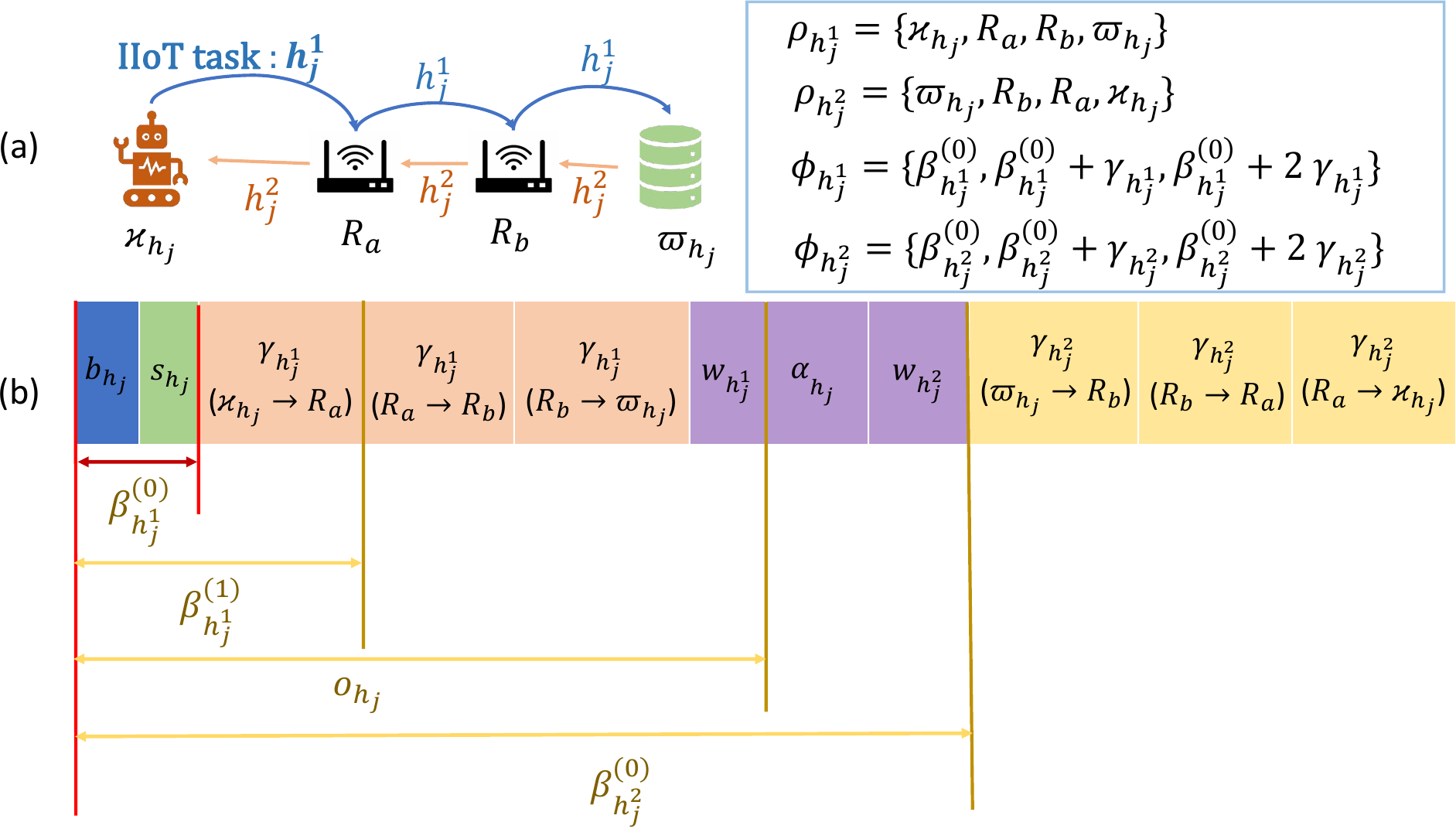}
    \caption{An example of an IIoT periodic time-critical computing task's lifecycle in one period. 
    (a) visualizes the important life stages of the IIoT task $h_j$. 
    (b) describes the lifecycles of $h_j$ from a temporal perspective. }
    \label{fig:iiot-life-cycle}
\end{figure}


We present an instance in Fig.~\ref{fig:iiot-life-cycle} to illustrate the lifecycle of an IIoT periodic time-critical computing task in its one period.

Fig.~\ref{fig:iiot-life-cycle}(a) visualizes the important life stages of the IIoT task $h_j$.  An IIoT computing task $h_j$ is generated by an IIoT device $\varkappa_{h_j}$, meanwhile, the IIoT server $\varpi_{h_j}$ is designated to realize data analysis for $h_j$. Then $h_j^1$ containing data awaiting analysis is transmitted to $\varpi_{h_j}$ through the routing path $\rho_{h_j^1}$. $R_a$ and $R_b$ are IIoT routers. Subsequently, $\varpi_{h_j}$ launches specific application programs to analyze data of $h_j$, and then packets the computing results into $h_j^2$. Finally, $\varpi_{h_j}$ returns $h_j^2$ back through the routing path $\rho_{h_j^2}$. The lifecycle of $h_j$ concludes after $\varkappa_{h_j}$ receives the result packet $h_j^2$. 

Fig.~\ref{fig:iiot-life-cycle}(b) describes of the entire lifecycle of $h_j$ from a temporal perspective. 
The notations are summarized in Table~\ref{tab:math}. With clock synchronization, all IIoT tasks have a same initial  time, which is set to 0. Relatively to it, at $b_{h_j}$, $h_j$ is generated by the device $\varkappa_{h_j}$. After waiting for $s_{h_j}$, $h_j^1$ departure from $\varkappa_{h_j}$. It goes through $\rho_{h_j^1}$ and takes $3\gamma_{h_j^1}$ to arrive the designated server $\varpi_{h_j}$. Because the pre-allocation of time slots on $\varpi_{h_j}$, $h_j$ has to wait for $w_{h_j^1}$ before being finished. The computing process costs $\alpha_{h_j}$. The computing results will be transmitted through $\rho_{h_j^2}$ after waiting for $w_{h_j^2}$. The transmission of $h_j^2$ consumes $3\gamma_{h_j^2}$ slots.

\begin{table}[t]
    \centering
    \caption{List of Main Notations. }
    \renewcommand{\arraystretch}{1.2}
    \label{tab:math}
    \begin{tabular}{c|l}
         \toprule
         Notation & \tabincell{c}{Explanation}  \\ 
         \midrule
         $B_g$ & \tabincell{l}{The bandwidth of the IIoT network.} \\ 
         $h_j$ & An IIoT periodic time-critical computing task. \\
         $\varkappa_{h_j}$ & The source IIoT device of $h_j$. \\
         $\varpi_{h_j}$ & \tabincell{l}{The designated server for $h_j$. } \\
         $h_j^1$, $h_j^2$ & \tabincell{l}{Data packets related to $h_j$. $h_j^1$ contains data awaiting \\ analysis, $h_j^2$ contains results.} \\
         $b_x$ & \tabincell{l}{The generation time of $x$, where $x \in \{h_j^1, h_j^2\}$}. \\
         $P_x$ & \tabincell{l}{The period of $x$}. \\
         $D_x$ & \tabincell{l}{The deadline of $x$.} \\
         $B_x$ & \tabincell{l}{The data size of $x$.} \\
         $s_x$ & \tabincell{l}{The waiting duration of $x$ to be transmitted at $\varkappa_{x}$.} \\
         $w_{h_j^1}$ & \tabincell{l}{The waiting duration to be executed for $h_j^1$.}\\
         $w_{h_j^2}$ & \tabincell{l}{The waiting duration to be transmitted back for $h_j^2$.} \\
         $\gamma_{x}$ & \tabincell{l}{The one-hop transmission duration of $x$. $\gamma_{x} = B_x/B_g$}. \\
         $\alpha_{x}$ & \tabincell{l}{The computing duration of $x$.} \\
         $\rho_{x}$ & \tabincell{l}{The routing path of $x$.} \\
         $\beta_x^{(n)}$ & \tabincell{l}{The departure time of $x$ at the $n^{th}$ waypoint along $\rho_{x}$.} \\
         $\phi_{x}$ & \tabincell{l}{The departure time set of $x$ along $\rho_{x}$. }\\
         $o_{h_j}$ & \tabincell{l}{The starting computing time of $h_j$ on $\varpi_{h_j}$.} \\ 
         \bottomrule
    \end{tabular}
\end{table}

\section{Theoretical Modeling} \label{section:theoretical-modeling}
The purpose of this paper is to answer two questions. 
\textit{1) How to schedule existing IIoT computation and network resources to ensure deterministic responses for IIoT periodic time-critical computing tasks? } 
\textit{2) How to optimize the structure of existing IIoT network based on the personalized digital upgrading requirements of industrial enterprises?}

To answer the first question, we must solve the fine-grained resource time slots allocation problem (RTAP), which considers realizing computation-level scheduling and network-level scheduling. At the computation level, RTAP considers assigning an IIoT server for each IIoT task and arrange its execution time slots on the server. At the network level, RTAP considers planning a bidirectional routing paths and arrange transmission time slots to forward data of the task between the task source and the server. 
RTAP is a NP-Hard combinatorial optimization problem, since the RTAP can be modeled as a bin-packing problem, which is proved to be NP-Hard \cite{steiner2010evaluation}. The computing complexity of RTAP grows significantly with the number of IIoT tasks and IIoT servers. 
Moreover, while solving the RTAP, there are some strict constraints that must be satisfied, including that (a) IIoT tasks be completed before their deadlines; (b) IIoT tasks that share the same server cannot occupy the same time slots; (c) IIoT tasks cannot be processed until all data are received by the server; (d) IIoT tasks that share the same routing link cannot occupy the same time slots; (e) the constraints (a-d) must be satisfied in every period. 

To answer the second question, we must address the \textit{IIoT network structure optimization problem (NSOP)}. The NSOP is a combinatorial optimization problem that considers which links and which servers should be added to the current IIoT network in order to support deterministic responses of all IIoT tasks. A feasible improved network of NSOP indicates that there is at least one feasible solution for RTAP on the network. Therefore, RTAP is a subproblem of the NSOP, and NSOP has a higher computation complexity than RTAP.  

In the following subsections, we will first formulate RTAP and NSOP, and then theoretically analyze the IIoT resource sharing conflicts. 

\subsection{RTAP Formulation} \label{section:RTAP-formulations}
An IIoT network $G$ is defined as $G=(\mathcal{Y}, \mathcal{R}, \mathcal{C}, \mathcal{E})$. $\mathcal{Y}=\{Y_k\}_{k=1}^K$ refers to the set of $K$ IIoT devices that will generate IIoT periodic time-critical computing tasks. $\mathcal{R}=\{R_t\}_{t=1}^T$ denotes a set of $T$ IIoT routers that will forward data packets. $\mathcal{C}=\{C_q\}_{q=1}^Q$ refers to a set of $Q$ IIoT servers that will realize IIoT tasks. $\mathcal{E}$ denotes the set of connections between routers, servers and devices. 
An IIoT periodic time-critical computing task is denoted as $h_j$, and $h_j$ has its attribute set $A_{h_j} = \{b_{h_j}, P_{h_j}, D_{h_j}, B_{h_j^1}, B_{h_j^2}, \alpha_{h_j}\}$. The set that includes $N$ IIoT periodic time-critical computing tasks is denoted $\mathcal{H}$ and $\mathcal{H}=\{h_j\}_{j=1}^N$. The corresponding attribute set is denoted as $\mathcal{A}$ and $\mathcal{A}=\{A_{h_j}\}_{h_j \in \mathcal{H}}$. 

Given $G$, $\mathcal{H}$ and $\mathcal{A}$, the original objective of RTAP is to compute a scheduling solution $\pi$, so that based on $\pi$, all IIoT tasks can receive deterministic responses before their deadlines in every period. Here $\pi = \{\varpi_{h_j}, o_{h_j}, \rho_{h_j^1}, \phi_{h_j^1}, \rho_{h_j^2}, \phi_{h_j^2}\}_{h_j \in \mathcal{H}}$. 
Considering that the more IIoT servers used, the more energy consumed, and the higher the maintenance and upgrade costs, we add a high-level optimization objective for RTAP. That is, we require $\pi$ not only to ensures the deterministic response of IIoT tasks, but also to minimize the number of used servers. 
The formulation of RTAP is presented as \eqref{eq:objective}-\eqref{eq:ICC-MCC}. 
$\zeta(\pi)$ counts the number of servers that are allocated to IIoT tasks based on $\pi$. 
$L(\rho_{h_j^1})$ and $L(\rho_{h_j^2})$ count the number of hops in $\rho_{h_j^1}$ and $\rho_{h_j^2}$, respectively. 
The definitions of other notations can be found in Table~\ref{tab:math}. 
Note that the generation and execution behaviours of IIoT tasks are repeated in each period. Thus, while formulating the problem and getting a feasible solution, we only focus on their first period. 


\begin{align}
    & \min_{\pi}~~ \zeta(\pi)  \label{eq:objective} \\
    s.t. \nonumber  \\
    & b_{h_j} \leq \beta_{h_j^1}^{(0)}~,~~~~\forall h_j \in \mathcal{H} \label{eq:c1}\\ 
    & \beta_{h_j^1}^{(0)} + L(\rho_{h_j^1}) \cdot \gamma_{h_j^1} \leq o_{h_j}~,~~~~\forall h_j \in \mathcal{H} \label{eq:c2}\\
    & o_{h_j} + \alpha_{h_j} \leq \beta_{h_j^2}^{(0)}~,~~~~\forall h_j \in \mathcal{H} \label{eq:c3}\\
    & \beta_{h_j^2}^{(0)} +  L(\rho_{h_j^2}) \cdot \gamma_{h_j^2} < D_{h_j}~,~~~~\forall h_j \in \mathcal{H} \label{eq:c4}\\
    & \beta_{h_j^1}^v = \beta_{h_j^1}^u + \gamma_{h_j^1},~~ \forall (u\rightarrow v) \in \rho_{h_j^1}~,~~~~\forall h_j \in \mathcal{H} \label{eq:c5}\\
    & \beta_{h_j^2}^v = \beta_{h_j^2}^u + \gamma_{h_j^2},~~ \forall (u\rightarrow v) \in \rho_{h_j^2}~,~~~~\forall h_j \in \mathcal{H} \label{eq:c6}\\
    & no~resource~sharing~conflicts 
    \label{eq:ICC-MCC}
\end{align}

A feasible $\pi$ must satisfy constrains \eqref{eq:c1}-\eqref{eq:ICC-MCC}. 
Constraints \eqref{eq:c1}-\eqref{eq:c5} guarantee the temporal relationship of $h_j$ among the two transmission phases and the computing phase, and ensure the task to be completed before its deadline. 
To be specific, in \eqref{eq:c1}-\eqref{eq:c5}, $\beta_{h_j^1}^{(0)}$ refers to the first waypoint along the routing path $\rho_{h_j^1}$, that is, the source IIoT device of $h_j$. $\beta_{h_j^2}^{(0)}$ refers to the first waypoint along the routing path $\rho_{h_j^2}$, that is, the designated IIoT server of $h_j$. 
\eqref{eq:c1} denotes the departure time of $h_j$ from its source IIoT device cannot earlier than the time it is generated. 
\eqref{eq:c2} denotes that the time when IIoT server $\varpi_{h_j}$ starts calculating $h_j$ must be later than the time when $h_j^1$ arrives $\varpi_{h_j}$. 
\eqref{eq:c3} shows that the departure time of $h_j^2$ from $\varpi_{h_j}$ must be later than the time when data analysis of $h_j$ is completed. 
\eqref{eq:c4} represents the time $h_j^2$ arrives the source IIoT device of $h_j$ must earlier than its deadline. 

Constraints \eqref{eq:c5}-\eqref{eq:c6} maintain the temporal relationship of $h_j^1$ and $h_j^2$ during routing, where $v$ is the closest waypoint following $u$ in the routing path $\rho_{h_j^1}$ and $\rho_{h_j^2}$. Since data packets of $h_j$ cannot be cached in the routers, the equation relation in \eqref{eq:c5}-\eqref{eq:c6} must hold for a feasible solution. 

Constraint \eqref{eq:ICC-MCC} requires $\pi$ to avoid computation and network resource sharing conflicts for any two tasks, which will be explained in section~\ref{section:sharing_conflicts}. The specific example in Fig.~\ref{fig:iiot-life-cycle} can help clarify the temporary relationships in the formulation. 

While $\pi$ being applied in the practical IIoT applications, at the computation level, the opening and closing of time slots can be controlled by introducing a runtime monitor with timing function; at the network level, the allocation of time slots on each link can be supported by IEEE 802.1Qbv protocols.

\subsection{NSOP Formulation} \label{section:NSOP-formulation}
When the capacity of current IIoT network cannot support all IIoT periodic time-critical computing tasks, it becomes necessary to upgrade the IIoT network by adding communication links and/or servers. Assume that $G_0$ is the current IIoT network, and $\mathcal{G}=\{G_1, G_2, ..., G_k\}$ is a hypothesis set. Each hypothesis is a candidate IIoT network structure to support all IIoT tasks. With consideration of upgrading costs, the problem can be formulated as \eqref{eq:nsop_objective}-\eqref{eq:nsop_st}, where $c_l$ and $c_s$ are the cost of adding a link and a server, respectively, and $\eta(G_i)$ and $\kappa(G_i)$ compute the number of added links and servers of $G_i$ compared with $G_0$. \eqref{eq:nsop_objective} aims to select $G_i$, so that the upgrading cost from $G_0$ to $G_i$ can be minimized. Constrain \eqref{eq:nsop_st} represents that $g_i$ must be capable to respond to all IIoT tasks. 
\begin{align}
    & \min_{G_i \in \mathcal{G}}~~c_l \cdot \eta(G_i) + c_s \cdot \kappa(G_i) \label{eq:nsop_objective} \\
    s.t. ~~ &A~feasible~RTAP~solutin~\pi~exists~on~G_i \label{eq:nsop_st}
\end{align}

\subsection{Theoretical Analysis on Resource Sharing Conflicts} \label{section:sharing_conflicts}
In the following, we will conduct a theoretical analysis to some special cases where IIoT tasks fail to share computation and network resources. 
\begin{figure}
    \centering
    \includegraphics[scale=0.25]{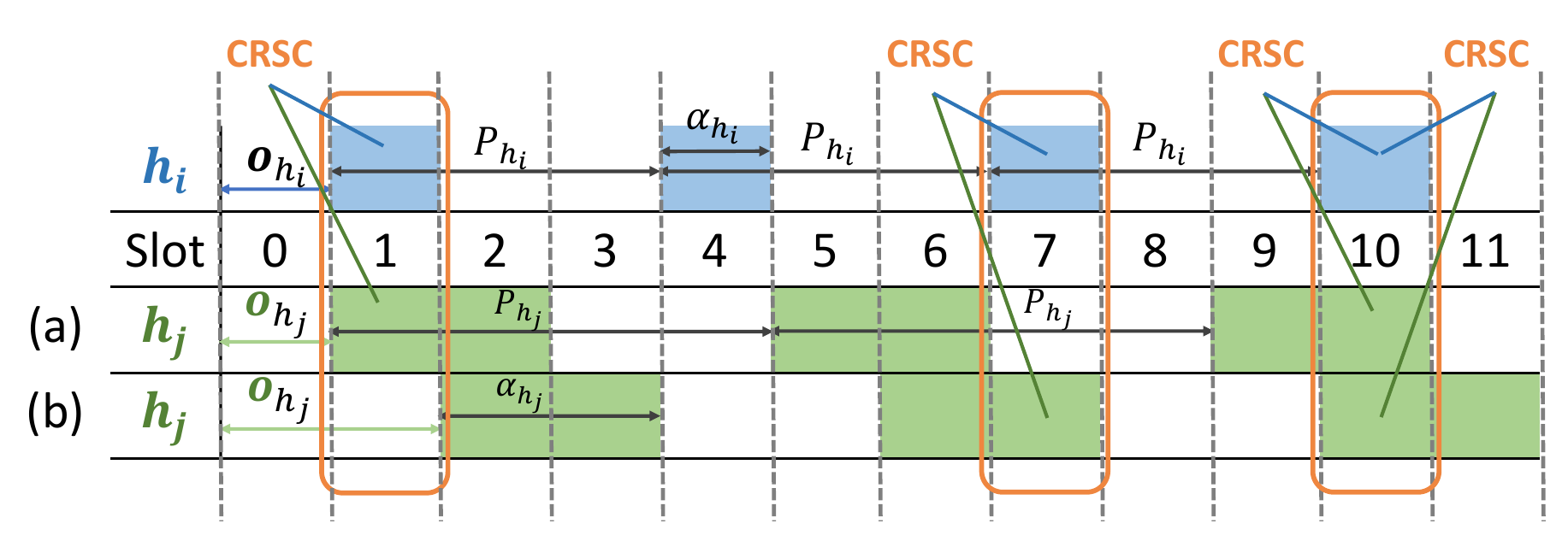}
    \caption{An example of computing resource sharing conflicts of task $h_i$ and $h_j$. Blue and green blocks refer to the occupied time slots by $h_i$ and $h_j$, respectively.}
    \label{fig:sharing-conflicts}
\end{figure}

\subsubsection{Analysis on sharing computing resources}
Two IIoT tasks $h_i$ and $h_j$ being allocated to the same server may suffer from computation resource sharing conflicts (CRSC), which refer to the execution duration of $h_i$ and $h_j$ are overlapped. In Fig.~\ref{fig:sharing-conflicts}, we present some cases that two IIoT tasks happen sharing conflicts. In addition, two conditions can be used to make sure CRSC occur. (a) There exist integers $m$ and $n$ making \eqref{eq:1} true, which indicates that $h_j$ start to be processed after $h_i$ but before $h_i$ being completed. (b) There exist integers $m$ and $n$ making \eqref{eq:2} true, which indicates that $h_i$ start to be processed after $h_j$ but before $h_j$ being completed.


\begin{equation}
    o_{h_i} + mP_{h_i} \leq o_{h_j} + nP_{h_j} < o_{h_i} + mP_{h_i} + \alpha_{h_i}
    \label{eq:1}
\end{equation}
\begin{equation}
    o_{h_j} + nP_{h_j} \leq o_{h_i} + mP_{h_i} < o_{h_j} + nP_{h_j} + \alpha_{h_j}
    \label{eq:2}
\end{equation}

Based on \eqref{eq:1}-\eqref{eq:2}, we further theoretically analyze some special conditions under which CSRC happen.  

\begin{theorem}\label{theorem:crsc1}
    If there are integers $m$ and $n$ such that $A \cap B \neq \emptyset$, where $A = [nP_{h_j}, (n+1)P_{h_j})$ and $B = [o_{h_i} + mP_{h_i}, o_{h_i} + \alpha_{h_i} + mP_{h_i})$, make $V = [ \max(o_{h_i} + mP_{h_i}-nP_{h_j}, 0), \min(P_{h_j},  o_{h_i} + \alpha_{h_i} + mP_{h_i}-nP_{h_j}))$. If $o_{h_j} \in V$, CSRC will happen. 
\end{theorem}

\begin{proof}
    Assume $h_i$ has already been allocated to a server and $h_j$ wants to share the same server with $h_i$. If $o_{h_j} \in V$, there must exist $(m,n)$ making $o_{h_i} + mP_{h_i} \leq o_{h_j} + nP_{h_j} < o_{h_i} + \alpha_{h_i} + mP_{h_i}$, which indicates $h_j$ will commence its operation during $h_i$ being processed in the $n_{th}$ period of $h_j$, thus CRSC occur.
\end{proof}

Using Theorem~\ref{theorem:crsc1}, we can quickly obtain infeasible search space of $o_{h_j}$, so that the search space of $o_{h_j}$ can be reduced. 

\begin{theorem}\label{theorem:crsc2}
    For any IIoT periodic computing tasks $h_i$ and $h_j$ sharing a same server, let their hyper period $hp = LCM(P_{h_i}, P_{h_j})$, $a = hp/P_{h_i}$, $b=hp/P_{h_j}$,  $A=\{(o_{h_i} + mP_{h_i}, o_{h_i} + \alpha_{h_i} + mP_{h_i})\}_{0\leq m < a}$, $B = \{(o_{h_j} + nP_{h_j}, o_{h_j} + \alpha_{h_j} + nP_{h_j})\}_{0\leq n < b}$, if $\exists(x_i, y_i) \in A$ and $\exists(x_j, y_j) \in B$ such that $(x_j - y_i) \cdot (y_j - x_i) < 0$, then CRSC occurs. 
\end{theorem}

\begin{figure}[h!]
    \centering
    \includegraphics[scale=0.25]{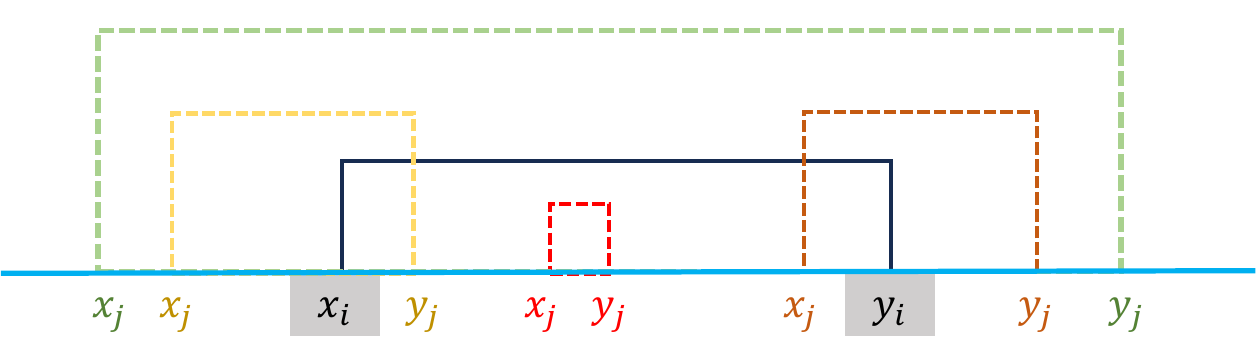}
    \caption{Situation enumeration when $h_i$ and $h_j$ cannot share computing resource.}
    \label{fig:fig5}
\end{figure}

\begin{proof}
   Fig.~\ref{fig:fig5} provides a comprehensive overview of the possible conflict situations between $h_i$ and $h_j$. $(x_j - y_i) < 0$ indicates that $h_j$ starts running before $h_i$ is completed, and $(y_j - x_i) > 0$ signifies that $h_j$ is realized after $x_i$ starts running. With $(x_j - y_i) < 0 \& (y_j - x_i) < 0$, CRSC is guaranteed. When $(x_j - y_i) > 0$, since $y_i > x_i$ and $y_j > x_j$, $(y_j - x_i) > 0$, no conflicts occur for this combination $(x_i, y_i, x_j, y_j)$. 
    Note that, the range from $x_i$ to $y_i$ is defined as including $x_i$ but excluding $y_i$. In other words, the interval is left-closed and right-open. As a results, the situation $(x_j - y_i) \cdot (y_j - x_i) = 0$ is not considered in Fig.~\ref{fig:fig5}. However, if $y_i$ was included in the range as well, then when $(x_j - y_i) \cdot (y_j - x_i) = 0$, conflicts will occur.  
\end{proof}

Using Theorem~\ref{theorem:crsc2}, we can quickly predict whether computation resource sharing conflicts will happen between $h_i$ and $h_j$ with $o_{h_i}$ and $o_{h_j}$. 

\subsubsection{Analysis on sharing network Links} \label{section:net-conflict-analysis}
When two IIoT periodic tasks $h_i$ and $h_j$ share a network link ($u \rightarrow v$) to transmit data, they may also encounter resource sharing conflicts with conditions similar to those observed while sharing computation resources. Therefore, we establish several theorems as follows. 

\begin{theorem}\label{theorem:net-rsc1}
    If there exist integers $m$ and $n$ such that $[nP_{h_j}, (n+1)P_{h_j}) \cap [\beta_{h_i}^u + mP_{h_i}, \beta_{h_i}^u + \gamma_{h_i} + mP_{h_i}) \neq \emptyset $, make $V = [ \max(\beta_{h_i}^u + mP_{h_i} - nP_{h_j}, 0), \min(P_{h_j},  \beta_{h_i}^u + \gamma_{h_i} + mP_{h_i} - nP_{h_j}))$. If $\beta_{h_j}^u \in V$, network resource sharing conflicts will occur.
\end{theorem}

\begin{theorem}\label{theorem:net-rsc2}
    Let hyper period $hp = LCM(P_{h_i}, P_{h_j})$, $a = hp/P_{h_i}$, $b=hp/P_{h_j}$,  $A=\{(\beta_{h_i}^u + mP_{h_i}, \beta_{h_i}^u + \gamma_{h_i} + mP_{h_i})\}_{0\leq m < a}$, $B = \{(\beta_{h_j}^u + nP_{h_j}, \beta_{h_j}^u + \gamma_{h_j} + nP_{h_j})\}_{0\leq n < b}$. If $\exists(k_x, z_x) \in A$ and $\exists(k_y, z_y) \in B$ such that $(k_y- z_x) \cdot (z_y - k_x) < 0$, network resource sharing conflicts occur. 
\end{theorem}

Since $h_j$ cannot be cached at routers, its offsets along the routing path are determined at its source. It is more effective to directly control the search space of $\beta_{h_j}^{(0)}$, rather than exploring the search space for each hop individually. However, the search space of $\beta_{h_j}^{(0)}$ is constrained by the slot allocation situation along the subsequent hops. Based on the discussion above, we derive Lemma~\ref{lemma:1} to help quickly obtain the infeasible solution space of $\beta_{h_j}^{(0)}$. 

\begin{lemma}\label{lemma:1}
    Let $u \rightarrow v$ is the $J^{th}$ hop in the routing path $\rho_{h_j}$, and consider a set of flows $A=\{x,y,z...\}$ that shares this link. For $\forall a \in A$, if there exist integers $m$ and $n$ such that $[nP_{h_j}, (n+1)P_{h_j}) \cap [\beta_a^u + mP_a, \beta_a^u + \gamma_a + mP_a) \neq \emptyset $, make $V_a = [K, L)$ where $K = \max(\beta_a^u + mP_a - nP_{h_j} - (J-1)\gamma_{h_j}, 0)$ and $ L = \max(\min(P_{h_j},  (\beta_a^u + \gamma_a + mP_a - nP_{h_j})-(J-1)\gamma_{h_j}, 0))$. $V = \cup_{a \in A} V_a$, 
If $\beta_{h_j}^{(0)} \in V$, network resource sharing conflicts will occur.
\end{lemma}

\begin{proof}
    According to the Theorem~\ref{theorem:net-rsc2},  for $\forall a \in A$, let $V_a^u = [ \max(\beta_a^u + mP_a - nP_{h_j}, 0), \min(P_{h_j},  \beta_a^u + \gamma_a + mP_a - nP_{h_j}))$. If $\beta_{h_j}^u \in V_a^u$, network resource sharing conflicts occurs. Considering that $h_j$ has traversed $J-1$ hops from task source $\varkappa_{h_j}$ to $u$, the transmission time $T$ is given by $T = (J-1)\gamma_{h_j}$. Thus $\beta_{h_j}^{\varkappa_{h_j}} = \beta_{h_j}^u - T$. The range $V_a$ is essentially the translated version of $V_a^u$ shifted by $T$ slots to the left. Therefore, if $\beta_{h_j}^u \in V_a^u$ leads to conflicts, then $\beta_{h_j}^{(0)} \in V_a$ will result in conflicts as well. Similarly, $\beta_{h_j}^{(0)} \in V-V_a$ will result in conflicts. 
\end{proof}

\section{IIoTBroker} \label{section:IIoTBroker}
Based on the theoretical analysis, we propose a heuristic algorithm, \textit{IIoTBroker}, to address RTAP. IIoTBroker can be divided into computation-level scheduling and network-level scheduling. We will first introduce the assumptions and notations of IIoTBroker, and then describe scheduling details of IIoTBroker.

\subsection{Assumptions and Notation Description}
IIoTBroker makes an assumption that each IIoT device can only generate one IIoT periodic time-critical computing task, in order to prevent network link sharing conflicts at the first hop.  If an IIoT device has the demands to generate $k(k>1)$ tasks, it can be realized by replicating the IIoT device $k$ times. $\varkappa_{h_j}$ denotes the task source of $h_j$. Each $h_j$ has an attribute set $A_{h_j}$, $A_{h_j}=\{b_{h_j}, P_{h_j}, D_{h_j}, B_{h_j^1}, B_{h_j^2}, \alpha_{h_j}\}$. Because we only consider homogeneous IIoT servers in this paper, the $\alpha_{h_j}$ keeps consistent on different IIoT servers. 

Except that, IIoTBroker extends the IIoT network $G$ defined in Section~\ref{section:RTAP-formulations} to an operational graph that not only possess a device set $\mathcal{Y}=\{Y_k\}_{k=1}^K$, a router set $\mathcal{R}=\{R_t\}_{t=1}^T$, a server set $\mathcal{C}=\{C_q\}_{q=1}^Q$ and a connection set $\mathcal{E}$, but also maintain records of fine-grained time slot allocations for each network link and server.

\subsection{Computation-level Scheduling of IIoTBroker}

The computation-level scheduling of IIoTBroker is presented in Algorithm~\ref{alg:computation-level-IIoTBroker}. Given the operational graph $G$ and a specific IIoT computing task $h_j$, Algorithm~\ref{alg:computation-level-IIoTBroker} will generate a set of feasible computation-level scheduling solutions for $h_j$, denoted as $\Pi_{h_j}^c$. We use the superscript $c$ to indicate that $\Pi_{h_j}^c$ is a computation-level decision set. The reason we generate multiple solutions instead of one is that the computation-level solutions will be used to guide network-level planning. Sometimes, for some computation-level solutions, their corresponding network-level planning fails due to network link sharing conflicts. In this case, multiple computation-level solutions can ensure more opportunities to realize fine-grained network-level planning. 

\begin{algorithm}
\caption{Computation-level Scheduling of IIoTBroker}
\label{alg:computation-level-IIoTBroker}
\KwIn{$G$; $h_j$; $A_{h_j}$}
\KwOut{$\Pi_{h_j}^c$}
\ForEach{$C_q \in \mathcal{C}$}{\label{alg:cls-p1}
Evaluate the transmission distance $\xi_{h_j}^{q}$ between $\varkappa_{h_j}$ and $C_q$ based on $G$   \\ \label{alg:cls-p2}
Obtain lower/upper bounds of $o_{h_j}^{q}$ based on $\xi_{h_j}^{q}$, which is $\Delta_{h_j}^{q}=[\delta_{min}, \delta_{max})$ \\ \label{alg:cls-p3}
Obtain infeasible solutions of  $o_{h_j}^q$, denoted as $\Theta_{h_j}^{q}$, based on Theorem~\ref{theorem:crsc1}\\ \label{alg:cls-p4}
Obtain feasible solution space $\Omega_{h_j}^{q}$ by $\Delta_{h_j}^{q} - \Theta_{h_j}^{q}$ \\ \label{alg:cls-p5}
\ForEach{$\delta \in \Omega_{h_j}^{q}$}{ \label{alg:cls-p6}
        \If{Time slots $[\delta, \delta+\alpha_{h_j})$ is free on $C_q$}{ \label{alg:cls-p7}
            append $(C_q, \delta)$ to $\Pi_{h_j}^c$ \label{alg:cls-p8}
        } \label{alg:cls-p9}
    }\label{alg:cls-p10}
}\label{alg:cls-p11}
\end{algorithm}

The details of computation-level scheduling of IIoTBroker are presented as follows.

Line~\ref{alg:cls-p2} predicts the transmission distance between the task source $\varkappa_{h_j}$ and the IIoT server $C_q$, which is used to constrain value range of $o_{h_j}^{q}$. $o_{h_j}^{q}$ refers to the starting computing time of $h_j$ on $C_q$. In the experiments, we use \textit{Dijkstra algorithm} to get a shortest distance, because the shortest distance can ensure the shortest transmission time at the ideal situation. Based on this, we can obtain the lower bound of $o_{h_j}^{q}$. However, it's worth noting that, theoretically, other path planning algorithms can be used to predict the distance as well. 

Line~\ref{alg:cls-p3} obtains the lower/upper bounds of $o_{h_j}^{q}$ as \eqref{eq:cls-p3-1}-\eqref{eq:cls-p3-2}. $\delta_{min}$ computed by \eqref{eq:cls-p3-1} represents the earliest time that $h_j$ arrives at $C_q$. $\delta_{max}$ is derived as \eqref{eq:cls-p3-2}. It reserves the computing time and result transmission time of $h_j$ before its deadline. 
\begin{align}
    & \delta_{min} = b_{h_j} + \xi_{h_j}^{q} \cdot \gamma_{h_j^1} \label{eq:cls-p3-1} \\
    & \delta_{max} = D_{h_j} - \xi_{h_j}^{q} \cdot \gamma_{h_j^2} - \alpha_{h_j} \label{eq:cls-p3-2}
\end{align}

Line~\ref{alg:cls-p4} obtains infeasible solutions of $o_{h_j}^{q}$ based on Theorem~\ref{theorem:crsc1} illustrated in Section~\ref{section:sharing_conflicts}. The detailed processes are presented in Algorithm~\ref{alg:cls-infeasible}.  $LCM(x,y)$ refers to computing the least common multiple of $x$ and $y$. 

Line~\ref{alg:cls-p5} computes feasible solution space $\Omega_{h_j}^{q}$ by $\Delta_{h_j}^{q} - \Theta_{h_j}^{q}$. 

Lines~\ref{alg:cls-p6}-\ref{alg:cls-p10} further judge whether $\delta \in \Omega_{h_j}^{q}$ is feasible. Time slots $[\delta, \delta+\alpha_{h_j})$ is free on $C_q$ indicating that the continuous free execution time is long enough to support $h_j$. 

If $\Pi_{h_j}^c$ is empty, it indicates that $h_j$ cannot be allocated to any IIoT server. Otherwise, it indicates that computation-level scheduling for $h_j$ succeeds. 


\begin{algorithm}
\caption{Computing Infeasible Space of $\Theta_{h_j}^{q}$}
\label{alg:cls-infeasible}
\ForEach{$h_i$ that has been allocated to $C_q$}{   
    $\tau = LCM(P_{h_i}, P_{h_j})$ \\
    $\varepsilon = \tau / P_{h_i}$ \\
    \ForEach{$\varrho$ in range($\varepsilon$)}{
        $x = \varrho * P_{h_i} + o_{h_i}$, $y = x + \alpha_{h_i}$ \\
        $k = x / P_{h_j}$, $z = y / p_{h_j}$ \\
        \If{$\lceil k \rceil - \lceil z \rceil == 0 $}{
            $a = x \% P_{h_j}$, $b = y \% P_{h_j}$ \\
            Add $range(a:b)$~to~$\Theta_{h_j}^{q}$ \\
        }
        \If{$\lceil z \rceil - \lceil k \rceil > 0 $}{
            $a = x \% P_{h_j}$, $b = y \% P_{h_j}$ \\
            Add $range(a:P_{h_j})$~to~$\Theta_{h_j}^{q}$ \\
            Add $range(0:b)$~to~$\Theta_{h_j}^{q}$ \\
        }
    }
}
\end{algorithm}

\subsection{Network-level Scheduling of IIoTBroker}
The network-level scheduling of IIoTBroker is presented in Algorithm~\ref{alg:network-level-IIoTBroker}. Before running the algorithm, we need to make some preparations.
First, we compute the computation cost and predict the response delay for each solution $\pi_{h_j}^c$ in the set  $\Pi_{h_j}^c$. 
The computation cost of $\pi_{h_j}^c$ is computed as follows: if $h_j$ shares a server with other tasks, then cost is 0; if $h_j$ uses a new free server, then cost is 1. 
The response delay of $\pi_{h_j}^c$ is evaluated as $o_{h_j} + \alpha_{h_j} + \xi_{h_j}^{*} \cdot \gamma_{h_j^2}$, where $\xi_{h_j}^{*}$ refers to the transmission distance between the scheduled server $\varpi_{h_j}$ as $\pi_{h_j}^c$ and the source IIoT device $\varkappa_{h_j}$. 
Then we sort items in $\Pi_{h_j}^c$ as the order of computation cost first and response delay second. After that, items in $\Pi_{h_j}^c$ are enumerated as the sorted order. By doing this, the solutions with low cost and response latency will get priority to be selected. 
Based on $\pi_{h_j}^c$, task $h_j$ has two stages, forwarding task $h_j^1$ and returning task $h_j^2$. Let $b_{h_j^1}$ and $b_{h_j^2}$ denote the generation time of $h_j^1$ and $h_j^2$. Then, $b_{h_j^1} = b_{h_j}$, and the deadline of $h_j^1$, denoted as $D_{h_j^1}$, is $o_{h_j}$. The generation time of $h_j^2$, denoted as $b_{h_j^2}$, equals to  $o_{h_j} + \alpha_{h_j}$, and the deadline $D_{h_j^2} = D_{h_j}$. 

After that, Algorithm~\ref{alg:network-level-IIoTBroker} can give a feasible network-level plan. Because the operations of searching departure time $\beta_{h_j^1}^{(0)}$ and $\beta_{h_j^2}^{(0)}$ for $h_j^1$ and $h_j^2$ are same, we introduce $r_j$ to denote any one of them. 

\begin{algorithm}
\caption{Network-level Scheduling of IIoTBroker}
\label{alg:network-level-IIoTBroker}
\KwIn{$G$; $r_j$ ($r_j \in \{h_j^1, h_j^2\}$); $\pi_{h_j}^c$}
\KwOut{$\beta_{r_j}^{(0)}$}
Plan routing paths on $G$ for $r_j$, $\rightarrow$ $\Lambda_{r_j}$ \\ \label{alg:nls-p1}
\ForEach{routing path $\lambda \in \Lambda_{r_j}$}{\label{alg:nls-p2}
    Obtain lower/upper bounds of $\beta_{r_j}^{(0)}$ based on $\lambda$ and $\pi_{h_j}^c$, $\rightarrow$ $\Delta_{r_j}^{\lambda}=[\delta_{min}, \delta_{max})$ \\ \label{alg:nls-p3}
    Obtain infeasible solutions of  $\beta_{r_j}^{(0)}$ based on Lemma~\ref{lemma:1},  $\rightarrow \Theta_{r_j}^{\lambda}$ \\ \label{alg:nls-p4}
    Obtain feasible solution space $\Omega_{r_j}^{\lambda}$ by $\Delta_{r_j}^{\lambda} - \Theta_{r_j}^{\lambda}$ \\ \label{alg:nls-p5}
    \ForEach{$\delta \in \Omega_{r_j}^{\lambda}$}{ \label{alg:nls-p6}
        \If{Slots $[\delta, \delta+\gamma_{r_j})$ is free on ($\lambda^0 \rightarrow \lambda^1$)}{ \label{alg:nls-p7}
            $ \beta_{r_j}^{(0)} \leftarrow \delta $; $\rho_{r_j} \leftarrow \lambda$; return \label{alg:nls-p8}
        } \label{alg:nls-p9}
    } \label{alg:nls-p10}
} \label{alg:nls-p11}
\end{algorithm}

At the first round, we feed $h_j^1$ to  Algorithm~\ref{alg:network-level-IIoTBroker}. 
Line~\ref{alg:nls-p1} plans routing paths based on $G$ for $h_j^1$. 
In the experiments, we evaluate the allowed longest transmission distance $\eta$ by $(o_{h_j} - b_{h_j})/\gamma_{h_j^1}$, and use the depth-first search method with constrain $\eta$ to plan all feasible routing paths. Line~\ref{alg:nls-p2}-\ref{alg:nls-p11} enumerate path $\lambda$ as the order of distance ascending. Line~\ref{alg:nls-p3} obtains lower/upper bounds of $\beta_{r_j}^{(0)}$ as \eqref{eq:nls-p3-1}-\eqref{eq:nls-p3-2}. $L(\lambda)$ refers to the length of $\lambda$. 
\begin{align}
    & \delta_{min} = b_{r_j} \label{eq:nls-p3-1} \\
    & \delta_{max} = o_{r_j} - (L(\lambda) - 1) \cdot \gamma_{r_j}\label{eq:nls-p3-2}
\end{align}

Line~\ref{alg:nls-p4} obtains infeasible solutions of $\beta_{h_j^1}^{(0)}$ based on Lemma~\ref{lemma:1}. The operations are presented in Algorithm~\ref{alg:nls-infeasible}, where $\omega$ denotes the $u \rightarrow v$ is the $\omega^{th}$ hop in $\lambda$. 
Line~\ref{alg:nls-p5} obtains feasible space by $\Delta_{h_j^1}^{\lambda} - \Theta_{h_j^1}^{\lambda}$. 
Line~\ref{alg:nls-p6}-\ref{alg:nls-p10} further judge whether $\delta \in \Omega_{h_j^1}^{\lambda}$ is feasible. Free slots $[\delta, \delta+\gamma{h_j^1})$ on the first hop ($\lambda^0 \rightarrow \lambda^1$) of $\lambda$ indicates that $h_j^1$ can be transmitted through $\lambda$. Then $\rho_{h_j^1} = \lambda$ and $\beta_{h_j^1}^{(0)}=\delta$. 

If $\beta_{h_j^1}^{(0)}$ and $\rho_{h_j^1}$ are allocated successfully, we feed $h_j^2$ to Algorithm~\ref{alg:network-level-IIoTBroker} to search $\beta_{h_j^2}^{(0)}$ and $\rho_{h_j^2}$ . If $\beta_{h_j^2}^{(0)}$ and $\rho_{h_j^2}$  are assigned as well, the network-level scheduling is completed, a scheduling decision $\pi_{h_j}$ is obtained. $\pi_{h_j} = \{\varpi_{h_j}, o_{h_j}, \rho_{h_j^1}, \phi_{h_j^1}, \rho_{h_j^2}, \phi_{h_j^2}\}$, and $(\varpi_{h_j}, o_{h_j}) \leftarrow \pi_{h_j}^c$. 

However, if $\beta_{h_j^1}^{(0)}$ is not allocated based on $\pi_{h_j}^c$, then next $\pi_{h_j}^c$ is iterated. If network-level scheduling fails for all items in $\Pi_{h_j}^c$, then fine-grained scheduling for $h_j$ fails. 

\begin{algorithm}
\caption{Computing Infeasible Space of $\Theta_{r_j}^{\lambda}$}
\label{alg:nls-infeasible}
\KwIn{$G$; $r_j$; $\lambda$}
\KwOut{$\Theta_{r_j}^{\lambda}$}
\ForEach{$u \rightarrow v \in \lambda$}{
    \ForEach{$f_i$ that has been allocated to $u \rightarrow v$}{
        $\tau = LCM(P_{f_i}, p_{r_j})$ \\
        $\varepsilon = \tau / P_{f_i}$ \\
        \ForEach{$\varrho$ in range($\varepsilon$)}{
            $x = \varrho * P_{f_i} + \beta_{f_i}^u$, $y = x + \gamma_{f_i}$ \\
            $k = x / P_{r_j}$, $z = y / P_{r_j}$ \\
            $a = x \% P_{r_j} - \omega \cdot \gamma_{r_j}$ \\
            $b = y \% P_{r_j} - \omega \cdot \gamma_{r_j}$ \\
            \If{$\lceil k \rceil - \lceil z \rceil == 0 $}{
                Add $range(a:b)$~to~$\Theta_{r_j}^{\lambda}$ \\
            }
            \If{$\lceil z \rceil - \lceil k \rceil > 0 $}{
                Add $range(max(a, 0):P_{r_j} - \omega \cdot \gamma_{r_j})$~and~$range(0:max(b, 0))$~to~$\Theta_{r_j}^{\lambda}$ \\
            }
        }
    }
}
\end{algorithm}

\subsection{Updating Operational Graph $G$}
When $h_j$-oriented fine-grained scheduling is successful, the operation graph $G$ must be updated according to the scheduling decision $\pi_{h_j}$. Time slots $[o_{h_j}, o_{h_j} + \alpha_{h_j}) $ of IIoT server $\varpi_{h_j}$ in $G$ are reserved by $h_j$. For $\omega^{th}$ hop $(u \rightarrow v)$ in $\rho_{h_j^1}$, time slots $[\beta_{h_j^1}^{(0)} + \omega \cdot \gamma_{h_j^1}, \beta_{h_j^1}^{(0)} + (\omega + 1) \cdot \gamma_{h_j^1})$ on the link $(u \rightarrow v)$ are reserved for $h_j^1$. 
Similarly, for $\omega^{th}$ hop $(u \rightarrow v)$ in $\rho_{h_j^2}$, time slots $[\beta_{h_j^2}^{(0)} + \omega \cdot \gamma_{h_j^2}, \beta_{h_j^2}^{(0)} + (\omega + 1) \cdot \gamma_{h_j^2})$ on the link $(u \rightarrow v)$ are reserved for $h_j^2$. 

\section{IIoTDeployer}\label{section:IIoTDeployer}

According to the formulation of NSOP, we can find that the search space of $\mathcal{G}$ is very huge, since there are innumerable combinations of adding links and servers. It is intractable to enumerate all hypothesises and select the best one. Therefore, we decided to start with an \textit{ideal graph} and gradually remove some links and servers from the graph during the deterministic scheduling process over all IIoT tasks. 
After that, based on the cropped graph, we can obtain a new IIoT network structure. 
By comparing it with the original IIoT network structure, we can easily obtain an network upgrading plan. Based on this idea, we design \textit{IIoTDeployer} to produce a new IIoT network structure with minimized upgrading cost. 

We establish some requirements for a reasonable IIoT network structure. (a)~An IIoT device can connect to only one router. (b)~An IIoT server can only be connected to one router. (c)~One router can be connected to multiple routers. (d)~An IIoT device and an IIoT server cannot be directly connected. (e)~There is only one link between any two directly connected IIoT entities. These requirements are consistent with the most real IIoT environment. Some customized requirements that may violate (a)-(e) are allowed, but some operations must be slightly adjusted while using \textit{IIoTDeployer}. 

In addition, we divided network links into three categories: router-router links, router-server links and router-device links. The router-device links are deployed in advance, and cannot be changed. When we discuss adding/removing links into/from IIoT network, it refers to adding/removing router-router links and router-server links. The router-router links have a common deployment cost $c_l$, and usually $c_l \geq 1$ in our settings. The cost of adding a router-server link into IIoT network is independent of $c_l$, and is set to 1. 

\textit{IIoTDeployer} provides two IIoT network structure upgrading methods, link cost first upgrading (LCFU) and server cost first upgrading (SCFU). 
The idea of LCFU is to schedule all IIoT tasks by adding servers without modifying the original network topology. LCFU is much suitable for situations where adding links is more expensive than adding servers. However, LCFU-based scheduling may fail when the capacity of network links becomes bottleneck. Under the circumstances, SCFU can help realize successful scheduling.
The idea of SCFU is to prioritize increasing links so that IIoT tasks can reuse servers. If scheduling is unsuccessful, both links and servers can be added simultaneously. SCFU is more suitable for situations where deployment costs of servers are more expensive than links.
The details of LCFU and SCFU are presented as follows.

\subsection{Link Cost First Upgrading (LCFU)}

\begin{figure*}
    \centering
    \includegraphics[scale=0.31]{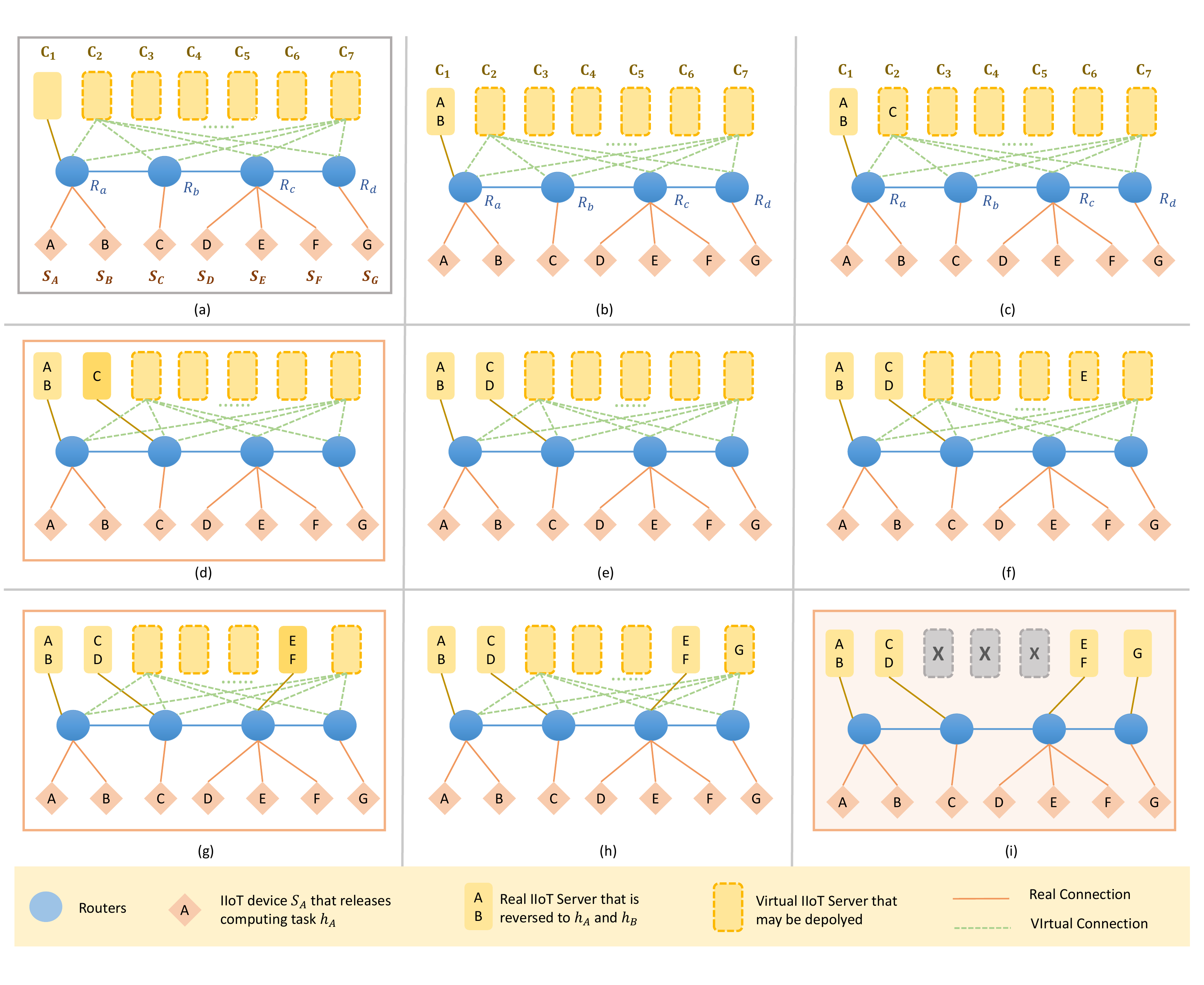}
    \caption{An illustration of the processes of LCFU. In this case, seven IIoT computing tasks are generated, and four routers and one server locate in the original IIoT network.  (a)~Build the \textit{ideal graph} $G$. (b)~IIoTBroker schedules $h_A$ and $h_B$ to $C_1$. (c)~IIoTBroker schedules $h_C$ to virtual server $C_1$.  (d)~The connection between $C_2$ and $R_b$ is retained, connections between $C_2$ and other routers are removed from $G$. $C_2$ becomes a real server.  (e)~IIoTBroker schedules $h_D$ to server $C_2$. (f)~IIoTBroker schedules $h_E$ to virtual server $C_6$. (g)~The connection between $C_6$ and $R_c$ is retained, and connections between $C_6$ and other routers are removed from $G$. $C_6$ becomes a real server. IIoTBroker schedules $h_F$ to server $C_6$. (h)~IIoTBroker schedules $h_G$ to the virtual server $C_7$. (i)~The connection between $C_7$ and $R_d$ is retained, connections between $C_7$ and other routers are removed from $G$. $C_6$ becomes a real server. Other unused servers $C_3$, $C_4$, $C_5$ and related virtual connections are removed from $G$.}
    \label{fig:link-cfu}
\end{figure*}

We present an instance to make the processes of LCFU clearer in Fig.~\ref{fig:link-cfu}.

The first step of LCFU is to build an \textit{ideal graph} based on the original network structure, and assign a usage cost for each (virtual or real) server and link. The \textit{ideal graph} is generated by two steps. In the first step, in order to guarantee sufficient computing resources, we introduce some virtual servers to make the total number of real servers and virtual servers to the same as number of IIoT tasks. In the second step,  we establish virtual connections between each virtual server and real router, since we cannot decide the positions of virtually added servers in advance. 
An illustration of the \textit{ideal graph} is shown in Fig.~\ref{fig:link-cfu}(a) where there are an original server and six virtual servers, as well as virtual connections among virtual servers and real routers. 
The usage costs of virtual servers, unused real server and used real server are set to $c_s$, 1 and 0 respectively. The usage costs of virtual connections and real connections are set to 1 and 0 respectively.  

After that, we use \textit{IIoTBroker} to provide multiple scheduling decisions for IIoT task $h_j$ on the \textit{ideal graph}. Cost of each decision is evaluated based on the usage costs of servers and connections. Moreover, the decision with lowest cost is always selected. Following that, based on the decision, usage costs of servers and connections are updated, as well as the \textit{ideal graph}. 
For example, in Fig.~\ref{fig:link-cfu}(b), IIoT tasks $h_A$ and $h_B$ are scheduled to the original server $C_1$, the \textit{ideal graph} remains unchanged, but the usage cost of $C_1$ changes from 1 to 0.
In Fig.~\ref{fig:link-cfu}(c), $h_C$ is assigned to the virtual server $C_2$, with routing paths $(S_C \rightarrow R_b \rightarrow C_2)$. Then in Fig.~\ref{fig:link-cfu}(d), the connection between $C_2$ and $R_b$ is retained, and connections between $C_2$ and other routers are removed. The usage cost of $C_2$ changes from $c_s$ to $0$. The usage cost of links $(R_b \rightarrow C_2)$ and $(C_2 \rightarrow R_b)$ change from 1 to 0.  

\begin{algorithm}
\caption{LCFU}
\label{alg:link-cfu}
\KwIn{Original IIoT network graph $G_0$; $\mathcal{H}$, $c_s$}
\KwOut{Cropped ideal graph $G$; $\{\pi_{h_j}^*\}_{h_j \in \mathcal{H}}$}
Build an \textit{ideal graph} $G$ based on $G_0$ \\
Build cost matrix $\Psi$ for servers and links \\
\ForEach{$h_j \in \mathcal{H}$}{
    IIoTBroker produces $\Pi_{h_j}$ on $G$ for $h_j$\\
    Evaluate cost for each $\pi_{h_j} \in \Pi_{h_j}$ based on $\Psi$ \\
    Select $\pi_{h_j}^*$ with lowest cost \\
    Remove links from $G$ based on $\pi_{h_j}^*$\\
    Update $\Psi$ based on $\pi_{h_j}^*$ \\
}
Remove unused virtual servers and links from $G$
\end{algorithm}

The scheduling-updating process iterates until all IIoT tasks are realized. At this time, we can obtain the final-version \textit{ideal graph}, which provides an IIoT network upgrading plan. For example, in Fig.~\ref{fig:link-cfu}(i), all IIoT tasks are scheduled, and three virtual servers and connections are retained in the \textit{ideal graph}. It indicates that the three virtual servers and connections should be added to the original IIoT network structure, in order to realize deterministic scheduling for all IIoT computing tasks of enterprises. 

The main processes of LCFU is presented in Algorithm~\ref{alg:link-cfu}. $G - G_0$ provides the upgrading scheme of current IIoT network. $ \{\pi_{h_j}^*\}_{h_j \in \mathcal{H}}$ provides scheduling decisions on the upgraded IIoT network to ensure deterministic response for all IIoT tasks.

\subsection{Server Cost First Upgrading (SCFU)}
The processes of SCFU are very similar with LCFU. The main difference lies in the construction of the \textit{ideal graph}. The initial ideal graph of SCFU adds virtual connections between any two routers on top of the initial ideal graph of LCFU. The purpose is to ensure both network and computing capacity on the ideal graph of SCFU are sufficient for IIoT tasks. The rules of setting and updating cost matrix $\Psi$ has some slight difference as well. Usage costs of virtual links between routers are $c_l$ $(c_l \geq 1)$. Usage costs of virtual links between virtual servers and routers are 1. Usage costs of real links are 0. Usage costs of virtual servers are $c_s$ $(c_s \geq 1)$. Usage costs of unused real servers and used real servers are 1 and 0 respectively. 
Based on the \textit{ideal graph} and cost matrix update rules of SCFU, the scheduling-updating processes follow that in LCFU. 

\begin{algorithm}
\caption{Combination of LCFU and SCFU}
\label{alg:combination-lcfu-scfu}
\KwIn{Original IIoT network graph $G_0$; $\mathcal{H}$, $c_s$, $c_l$}
\KwOut{$G$; $\{\pi_{h_j}^*\}_{h_j \in \mathcal{H}}$}
$G \leftarrow G_0$ \\
\ForEach{$h_j \in \mathcal{H}$}{ \label{alg:deployer-search-start}
    Build an \textit{ideal graph} $\pounds$ of LCFU based on $G$ \\ \label{alg:deployer-comb-lcfu1}
    Build cost matrix $\Psi$ of LCFU\\
    IIoTBroker produces $\Lambda_{h_j}$ on $\pounds$ for $h_j$\\
    Evaluate cost for each $\lambda_{h_j} \in \Lambda_{h_j}$ based on $\Psi$ \\ \label{alg:deployer-comb-lcfu2}
    Build an \textit{ideal graph} $\mho$ of SCFU based on $G$ \\ \label{alg:deployer-comb-scfu1}
    Build cost matrix $\Upsilon$ of SCFU \\
    IIoTBroker produces $\Theta_{h_j}$  on $\mho$ for $h_j$\\
    Evaluate cost for each $\xi_{h_j} \in \Theta_{h_j}$ based on $\Upsilon$ \\ \label{alg:deployer-comb-scfu2}
    Select $\pi_{h_j}^*$ with lowest cost from $\Lambda_{h_j} + \Theta_{h_j}$ \\ \label{alg:deployer-comb-select}
    Add $\pi_{h_j}^*$ required links and servers to $G$\\ \label{alg:deployer-comb-update}
} \label{alg:deployer-search-end}
\end{algorithm}

\subsection{Combination of LCFU and SCFU}
LCFU and SCFU can be combined to search a low-cost optimized IIoT network structure based on the current IIoT network structure $G$ and the upgrading cost $(c_s, c_l)$.  The processes are shown in Algorithm~\ref{alg:combination-lcfu-scfu}. At the beginning, $G$ is a copy of $G_0$. Then $G$ becomes the original IIoT network of every scheduling in Lines~\ref{alg:deployer-search-start}-\ref{alg:deployer-search-end}. 
Then for each $h_j$, LCFU and SCFU are used to search their own low-cost scheduling solution, shown in Line~\ref{alg:deployer-comb-lcfu1}-\ref{alg:deployer-comb-lcfu2} and Line~\ref{alg:deployer-comb-scfu1}-\ref{alg:deployer-comb-scfu2} respectively. Line~\ref{alg:deployer-comb-select} selects the decision with lowest cost. Line~\ref{alg:deployer-comb-update} adds corresponding servers and links to $G$ based on $\pi_{h_j}^*$. 

\section{Numerical Results} \label{section:experiments}
In this section, we conduct a variety of simulations to verify performance of our designed IIoTBroker and IIoTDeployer. 

\subsection{Simulation Settings}
\subsubsection{Common settings} We design simulations with different problem scales to validate whether the proposed algorithms can be directly generalized to any IIoT network. Here the problem scale refers to the number of IIoT periodic time-critical computing tasks. 

Given the number of IIoT periodic time-critical computing tasks in the simulation $N$, each IIoT periodic time-critical computing task $h_j$ is randomly generated as the following rules. $P_{h_j}$ is randomly sampled from $\{3000, 5000, 10000\}ms$. $D_{h_j} = P_{h_j}$. $b_{h_j}$ is uniformly sampled from range $[0, 100]$. $B_{h_j^1}$ is sampled from $\{1, 2, 5, 10\}MB$. $B_{h_j^2}=1MB$. $\alpha_{h_j}$ is randomly sampled from $\{500, 1000, 1500, 2000\}ms$. $\varkappa_{h_j}$ is uniformly sampled from $\{1, 2, ..., N\}$. The bandwidth of IIoT networks $B_g = 1GB$. 

\subsubsection{IIoTBroker-related settings} The number of IIoT rounter in IIoT networks is set to 10. Random 45 bidirectional connections are established between routers, so that the network topology is not too sparse to provide sufficient network transmission capacity, nor too dense,  making the routing path simple.
Each IIoT device is randomly connected to one router. To ensure the abundant computing capacity, the number of IIoT servers is same as the number of IIoT tasks. Each IIoT server randomly connects one router.   

\subsubsection{IIoTDeployer-related settings} 
\begin{figure}[h!]
    \centering
    \includegraphics[scale=0.15]{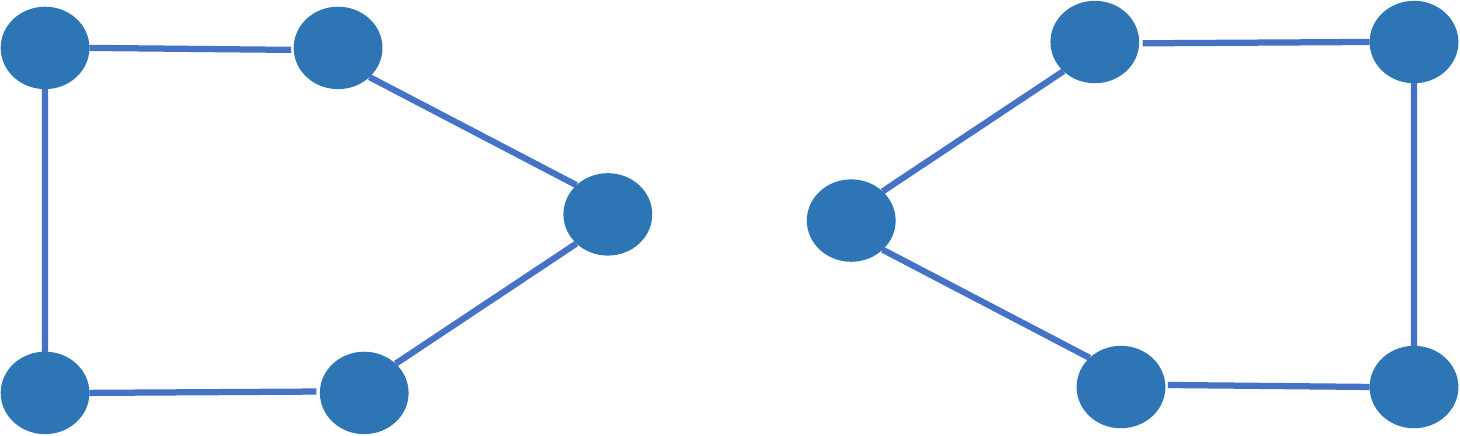}
    \caption{The original IIoT network topology of IIoTDeployer. A blue polka dot represents a router. }
    \label{fig:simulation-deployer-topo}
\end{figure}

The number of IIoT rounter in IIoT networks is set to 10. 
The original IIoT network topology is shown in Fig.~\ref{fig:simulation-deployer-topo}. As the number of IIoT tasks increases, the original network will be unable to satisfy the transmission requirements. Two isolated islands in the topology exacerbate this phenomenon. The number of IIoT servers in the original network $M$ is quiet small, and we set $M=\lceil N/5 \rceil$. Each IIoT device and each IIoT server randomly connect to one router.

\begin{table*}[h!]
	\caption{Evaluation Results of IIoTBroker on Different Problem Scales. (Best values are bold.)}
	\renewcommand{\arraystretch}{1.15}
	\begin{center}
		\setlength{\tabcolsep}{3pt}
		\begin{tabular}{c|ccc|ccc|ccc|ccc}
			\toprule
			\multicolumn{1}{c|}{} & \multicolumn{3}{c|}{Tnum=10} & \multicolumn{3}{c|}{Tnum=50} & \multicolumn{3}{c|}{Tnum=100} & \multicolumn{3}{c}{Tnum=200}\\
	Method & Snum $\downarrow$ & Utility $\uparrow$ & Delay(ms) $\downarrow$ &Snum $\downarrow$ & Utility $\uparrow$ & Delay(ms) $\downarrow$ & Snum $\downarrow$ & Utility $\uparrow$ & Delay(ms) $\downarrow$ & Snum $\downarrow$ & Utility $\uparrow$ & Delay(ms) $\downarrow$ \\ 
        \midrule
         IIoTBroker & \textbf{3.94}  & \textbf{67.53\%} & 2791.06  & \textbf{17.08}  & \textbf{77.23\%} & 3264.43  & \textbf{34.34}  & \textbf{78.22\%} & 3289.70  & \textbf{65.82}  & \textbf{80.28\%} & 3385.27   \\ 
        Nearest-FS & 6.00  & 44.37\% & 1888.73  & 22.90  & 57.45\% & 2492.23  & 40.70  & 65.95\% & 2760.04  & 73.54  & 71.85\% & 3028.53   \\ 
        Random-S & 7.36  & 35.54\% & 1717.47  & 35.88  & 36.61\% & 1809.27  & 69.16  & 38.81\% & 1854.87  & 130.84  & 40.37\% & 1964.02   \\ 
        Delay-FS & 9.92  & 26.27\% & \textbf{1297.07}  & 49.78  & 26.37\% & \textbf{1320.58}  & 99.60  & 26.93\% & \textbf{1311.90}  & 198.70  & 26.57\% & \textbf{1314.20}   \\ 
        DFNS-S & 9.92  & 26.27\% & \textbf{1297.07}  & 49.78  & 26.37\% & \textbf{1320.58}  & 99.60  & 26.93\% & \textbf{1311.90}  & 198.70  & 26.57\% & \textbf{1314.20}   \\
			\bottomrule
		\end{tabular}
		\label{table:IIoTBroker}
	\end{center}
\end{table*}

\begin{table*}
	\caption{Comparision Results of IIoTBroker and TSN(+) on Different Problem Scales\\  
	(\textit{Diff}: The value of TSN(+) minus the value of IIoTBroker on the same metric.)}
	\renewcommand{\arraystretch}{1.15}
	\begin{center}
		\setlength{\tabcolsep}{3pt}
		\begin{tabular}{c|cc|c|cc|c|cc|c|ccc}
			\toprule
			\multicolumn{1}{c|}{} & \multicolumn{3}{c|}{Tnum=10} & \multicolumn{3}{c|}{Tnum=50} & \multicolumn{3}{c|}{Tnum=100} & \multicolumn{3}{c}{Tnum=200}\\
	Metrics & IIoTBroker & TSN(+) & Diff & IIoTBroker &  TSN(+) & Diff & IIoTBroker &  TSN(+) & Diff & IIoTBroker &  TSN(+) & Diff \\ 
    \midrule
        Snum $\downarrow$ & \textbf{3.94}  & 4.04  & 0.10  & \textbf{17.08}  & 17.36  & 0.28  & \textbf{34.34}  & 34.70  & 0.36  & \textbf{65.82}  & 66.60  & 0.78  \\ 
        Utility $\uparrow$ & \textbf{67.53\%} & 65.80\% & -1.74\% & \textbf{77.23\%} & 75.98\% & -1.25\% & \textbf{78.22\%} & 77.36\% & -0.86\% & \textbf{80.28\%} & 79.33\% & -0.95\% \\
        Delay(ms) $\downarrow$ & 2791.06 & \textbf{2774.92}  & -16.13  & 3264.43  & \textbf{3254.76}  & -9.67  & \textbf{3289.70}  & 3311.29  & 21.59  & \textbf{3385.27}  & 3419.09  & 33.81  \\
        Runtime(s) $\downarrow$ &  \textbf{0.11}  & 0.31  & 0.20  & \textbf{0.84}  & 9.33  & 8.50  & \textbf{3.02}  & 35.44  & 32.42  & \textbf{11.20}  & 167.01  & 155.81  \\
			\bottomrule
		\end{tabular}
		\label{table:IIoTBroker-ablation}
	\end{center}
\end{table*}

\subsection{Performance Metrics}
\subsubsection{Common Metrics} Three metrics are used by both \textit{IIoTBroker} and \textit{IIoTDeployer}. 
\begin{itemize}
    \item \textit{Tnum}: The number of IIoT periodic time-critical computing tasks that are waiting for scheduling in the simulations. 
    \item \textit{Utility}: The average utilization rate of the scheduled servers. The higher the better.
    \item \textit{Delay(ms)}: The average response delay of all IIoT tasks, measured in millisecond. The lower the better.
    \item \textit{Runtime(s)}: The computing time of algorithms, measured in seconds. The lower the better.
\end{itemize}

\subsubsection{IIoTBroker-related Metrics} One metric is customized for \textit{IIoTBroker}. 
\begin{itemize}
    \item \textit{Snum}: The number of scheduled servers by \textit{IIoTBroker}. The lower the better. 
\end{itemize}

\subsubsection{IIoTDeployer-related Metrics} Three metrics are introduced to evaluate the performance of \textit{IIoTDeployer}.
\begin{itemize}
    \item \textit{ASnum}: The number of IIoT servers to be added. 
    \item \textit{ALnum}: The number of IIoT network links to be added. 
    \item \textit{TCost}: The total cost of upgrading IIoT network. 
\end{itemize}

\subsection{Baselines}
\subsubsection{Baselines of IIoTBroker}
\textit{IIoTBroker} is a pioneer to explore deterministic scheduling of IIoT periodic time-critical computing tasks, therefore, no previous related contributions can be directly used as baselines. Moreover, deterministic scheduling is a NP-Hard problem. It is intractable to get optimal solutions. Inspired by the construction idea of some classic heuristic algorithms, such as nearest-neighbor algorithm for traveling salesman problems, we construct some heuristic algorithms as baselines, including \textit{Delay-FS}, \textit{Nearest-FS}, \textit{DFNS-S} and \textit{Random-S}. 

The main difference between IIoTBroker and designed heuristic algorithms is the sorting principle of computation-level solutions obtained through Algorithm~\ref{alg:computation-level-IIoTBroker}, which determines the enumeration order of computation-level solutions to be fed into network-level scheduling processes. 
\begin{itemize}
    \item \textit{Delay-FS}: The computation-level solutions are sorted as the ascending order of response delay. In cases where solutions have the same response delays, they are further sorted in ascending order of computation cost.
    \item \textit{Nearest-FS}: The solutions are arranged in the ascending order of transmission distance between servers and the task source.  When transmission distances of multiple solutions are same, they are sorted as the ascending order of computation cost.
    \item \textit{DFNS-S}: The solution is sorted as the ascending order of response delay first and transmission distance second. 
    \item \textit{Random-S}: The solutions are randomly sorted. 
\end{itemize}

Besides that, we also extend a TSN traffics oriented deterministic scheduling approach as a baseline. To be specific, 
Lu et al. \cite{lu2022intelligent} derived some conditions to judge whether multiple TSN traffics can share a same link. After careful argumentation, we find that part of these conditions can be extended to deterministic scheduling of IIoT computing tasks.
However, Lu et al. have only realized one link sharing in \cite{lu2022intelligent}. When applying the approach to achieve complete planning, including computation-level and network-level scheduling, we found that the spatial complexity of the approach is quite high. Even the corresponding programs often encounter warnings of out of memory when running on a computer with $32GB$ memory. Therefore, we optimize the approach with our new derivations, i.e. Theorem~\ref{theorem:crsc2} and Theorem~\ref{theorem:net-rsc2}, so that the improved approaches can be used to solve deterministic scheduling of IIoT computing tasks. We call the improved approach \textit{TSN(+)}. Except for the sharing conflict avoidance operations, other operations of \textit{TSN(+)}, such as sorting and cost evaluation, are same with that of \textit{IIoTBroker}. 

\subsubsection{Baselines of IIoTDeployer}
IIoTDeployer is the first work that realizes IIoT network structure optimization based on the requirements of personalized IIoT periodic time-critical computing tasks in enterprises. To demonstrate the performance of IIoTDeployer, we further design three algorithms as baselines. 
\begin{itemize}
    \item \textit{NcostFS}: Compared with Algorithm~\ref{alg:combination-lcfu-scfu}, for each $h_j \in \mathcal{H}$, \textit{NcostFS} first schedules through LCFU. If LCFU scheduling fails, SCFU is adopted.
    \item \textit{ScostFS}: Compared with Algorithm~\ref{alg:combination-lcfu-scfu} which combines LCFU and SCFU to parallel schedule IIoT network and computing resources, \textit{ScostFS} only adopts SCFU to schedule resources.
    \item \textit{ONetFS}: \textit{ONetFS} first dispatches $h_j$ based on the current IIoT network. If dispatching fails, \textit{ONetFS} uses SCFU to add some servers and links into the IIoT network. 
\end{itemize}

\begin{figure}
	\centering
        \setcounter{subfigure}{0}
	\subfigure[]{
		\begin{minipage}{0.46\columnwidth}
			\centering
			\includegraphics[scale=.18]{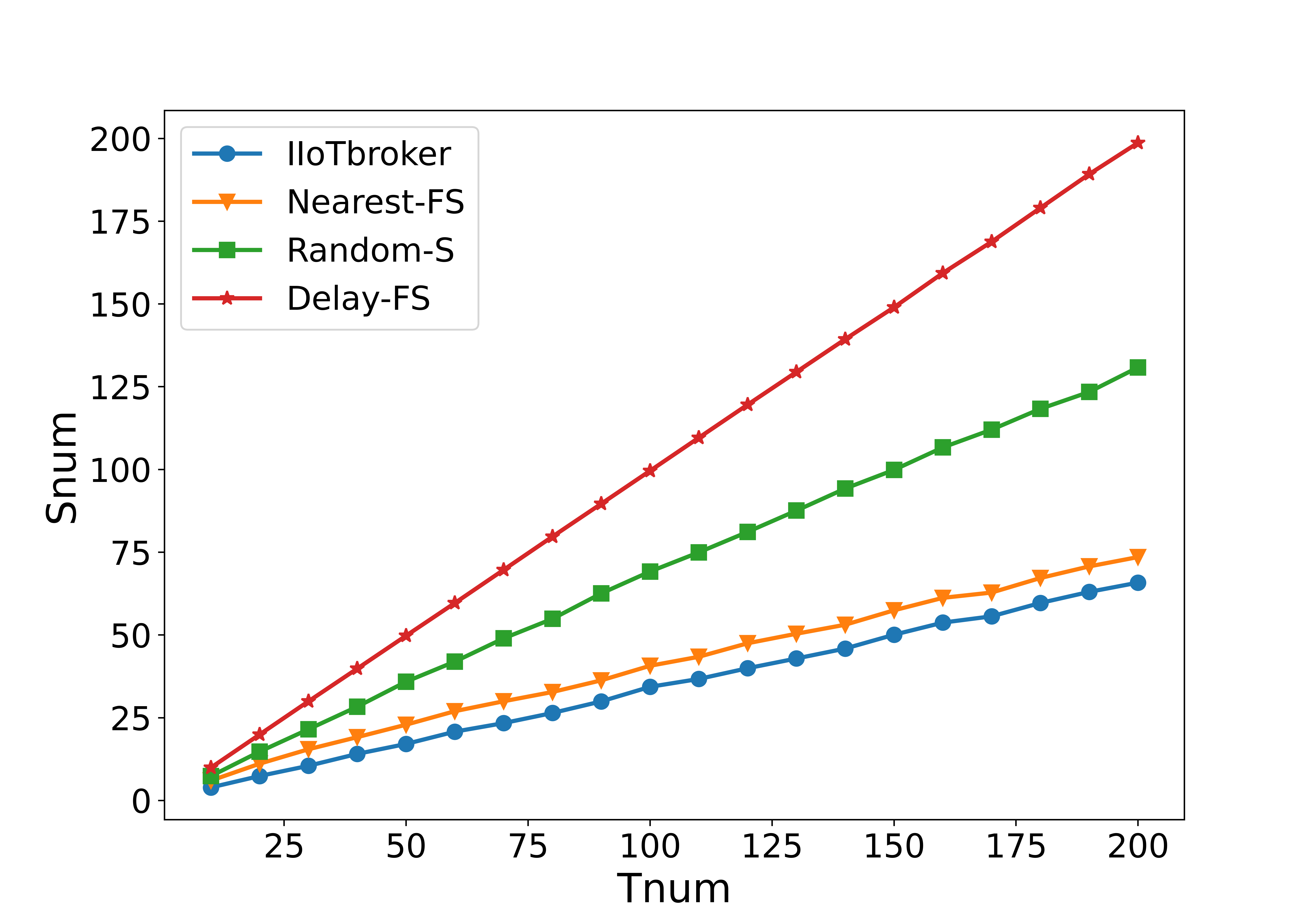}
		\end{minipage}
	}
        \subfigure[]{
		\begin{minipage}{0.46\columnwidth}
			\centering
			\includegraphics[scale=.18]{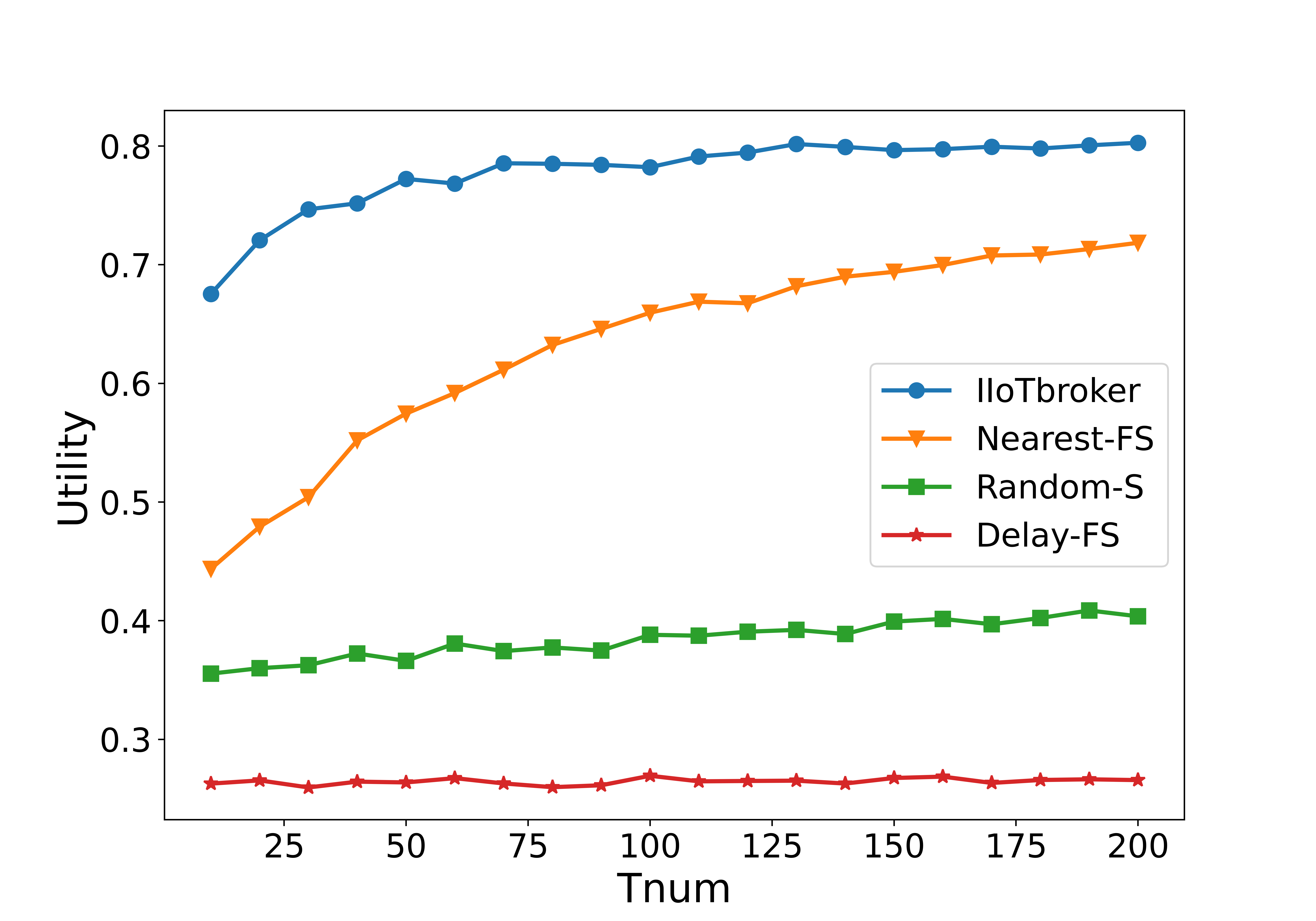}
		\end{minipage}
	}
        \subfigure[]{
		\begin{minipage}{0.46\columnwidth}
			\centering
			\includegraphics[scale=.18]{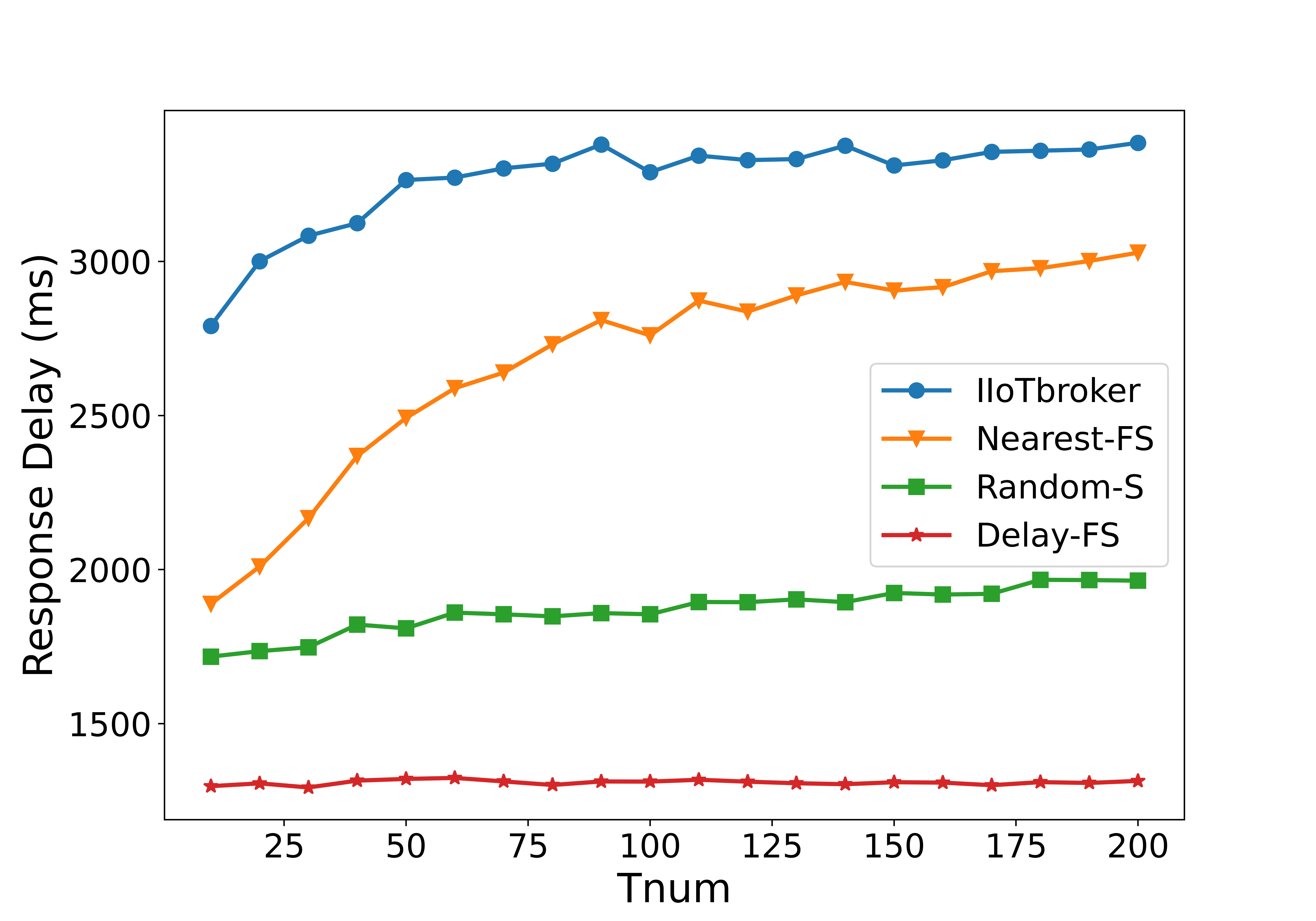}
		\end{minipage}
	}
        \subfigure[]{
		\begin{minipage}{0.46\columnwidth}
			\centering
			\includegraphics[scale=.18]{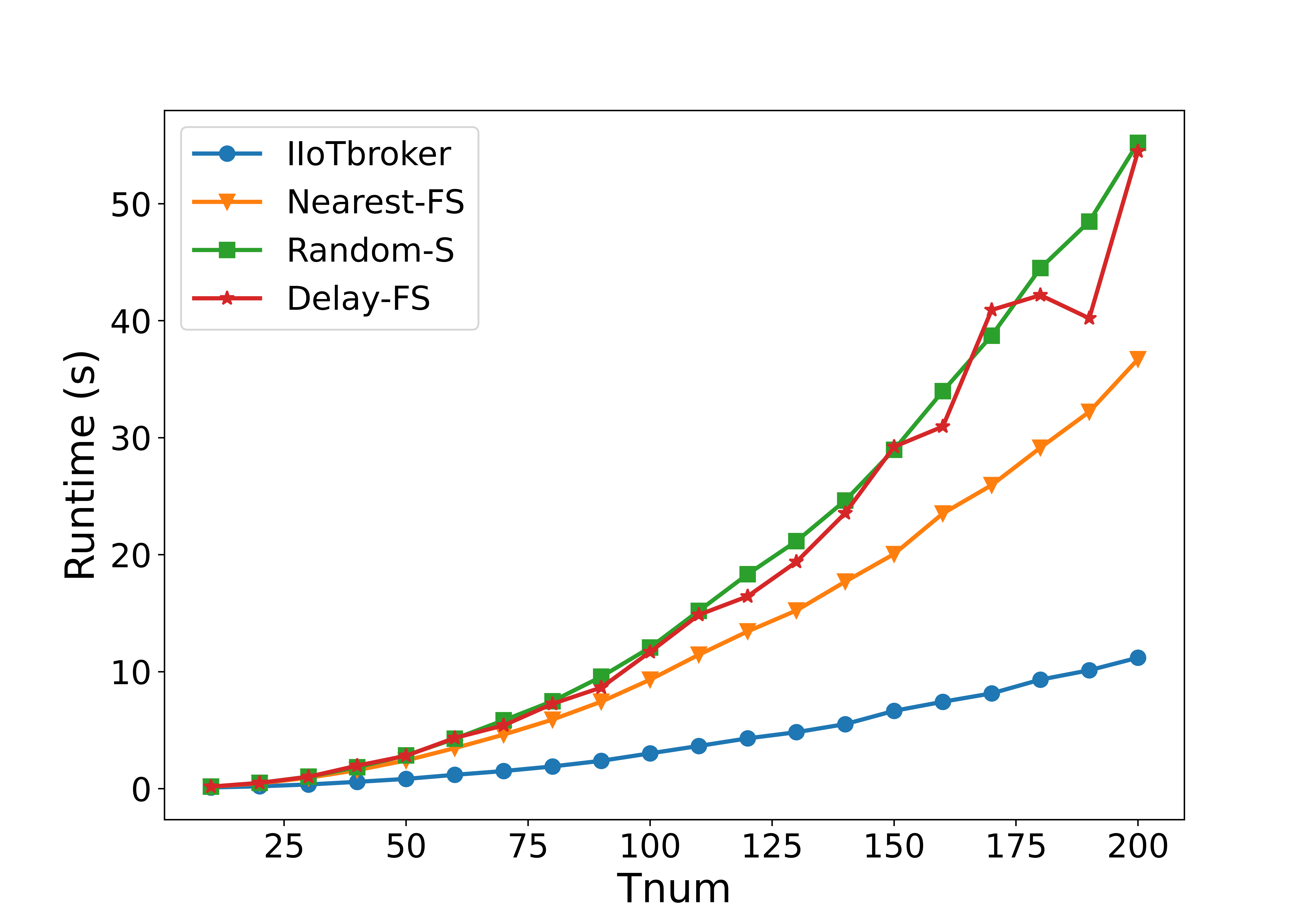}
		\end{minipage}
	}
	\caption{The relationship between the metric values of algorithms and the increase in the number of IIoT computing tasks.}
	\label{fig:IIoTBroker-varyTnum}
\end{figure}

\begin{figure}
	\centering
        \setcounter{subfigure}{0}
	\subfigure[]{
		\begin{minipage}{0.46\columnwidth}
			\centering
			\includegraphics[scale=.18]{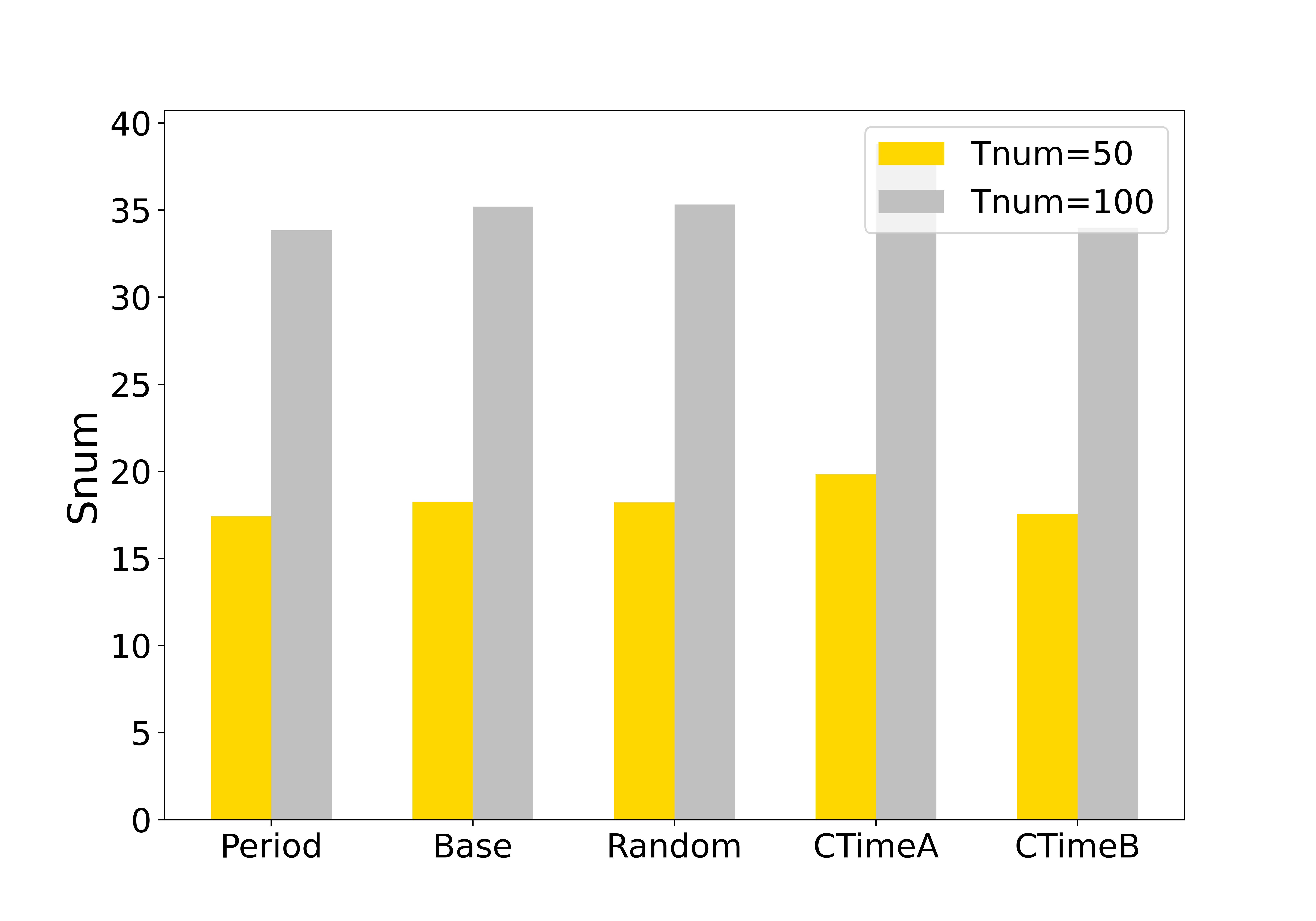}
		\end{minipage}
	}
        \subfigure[]{
		\begin{minipage}{0.46\columnwidth}
			\centering
			\includegraphics[scale=.18]{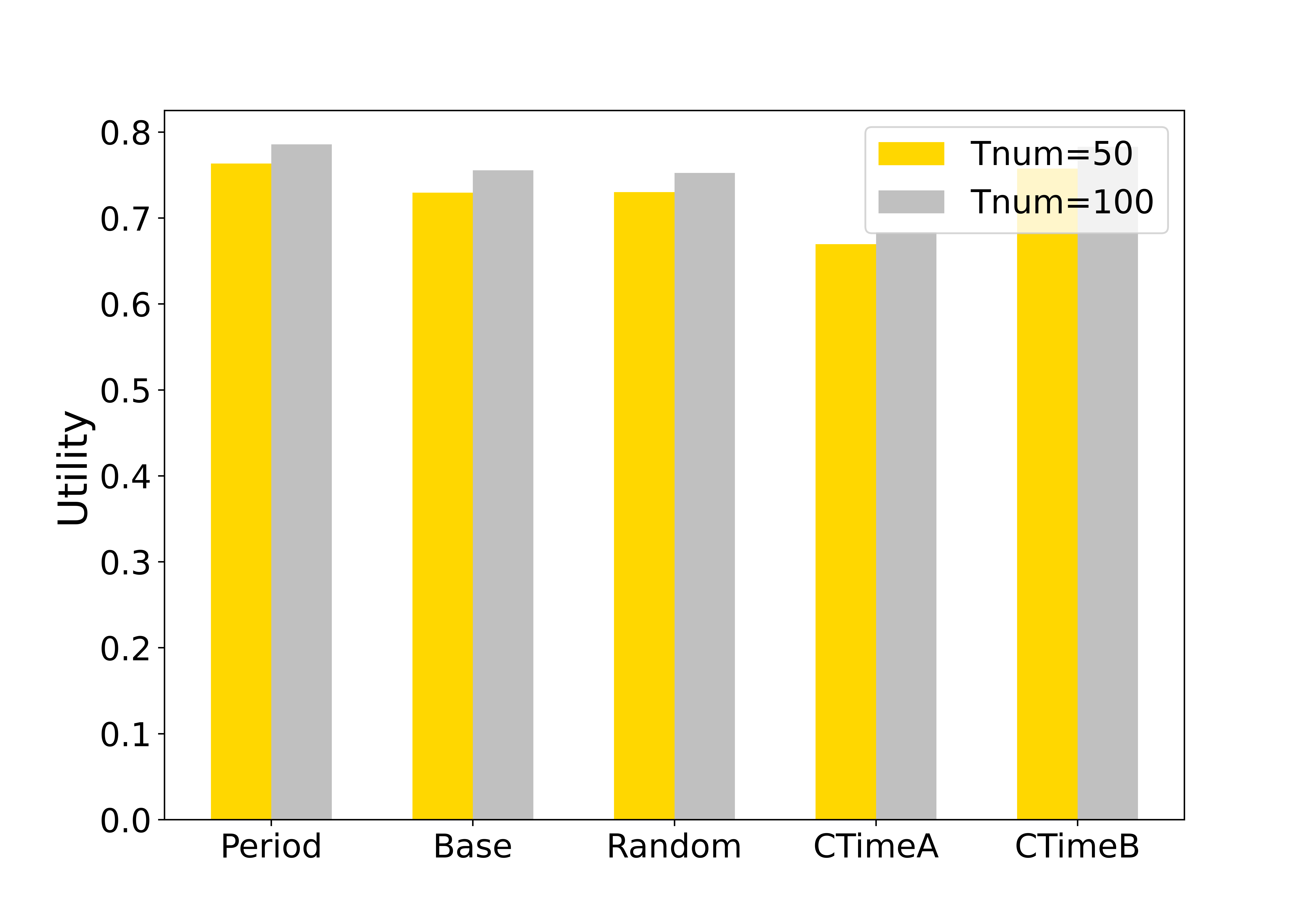}
		\end{minipage}
	}
        \subfigure[]{
		\begin{minipage}{0.46\columnwidth}
			\centering
			\includegraphics[scale=.18]{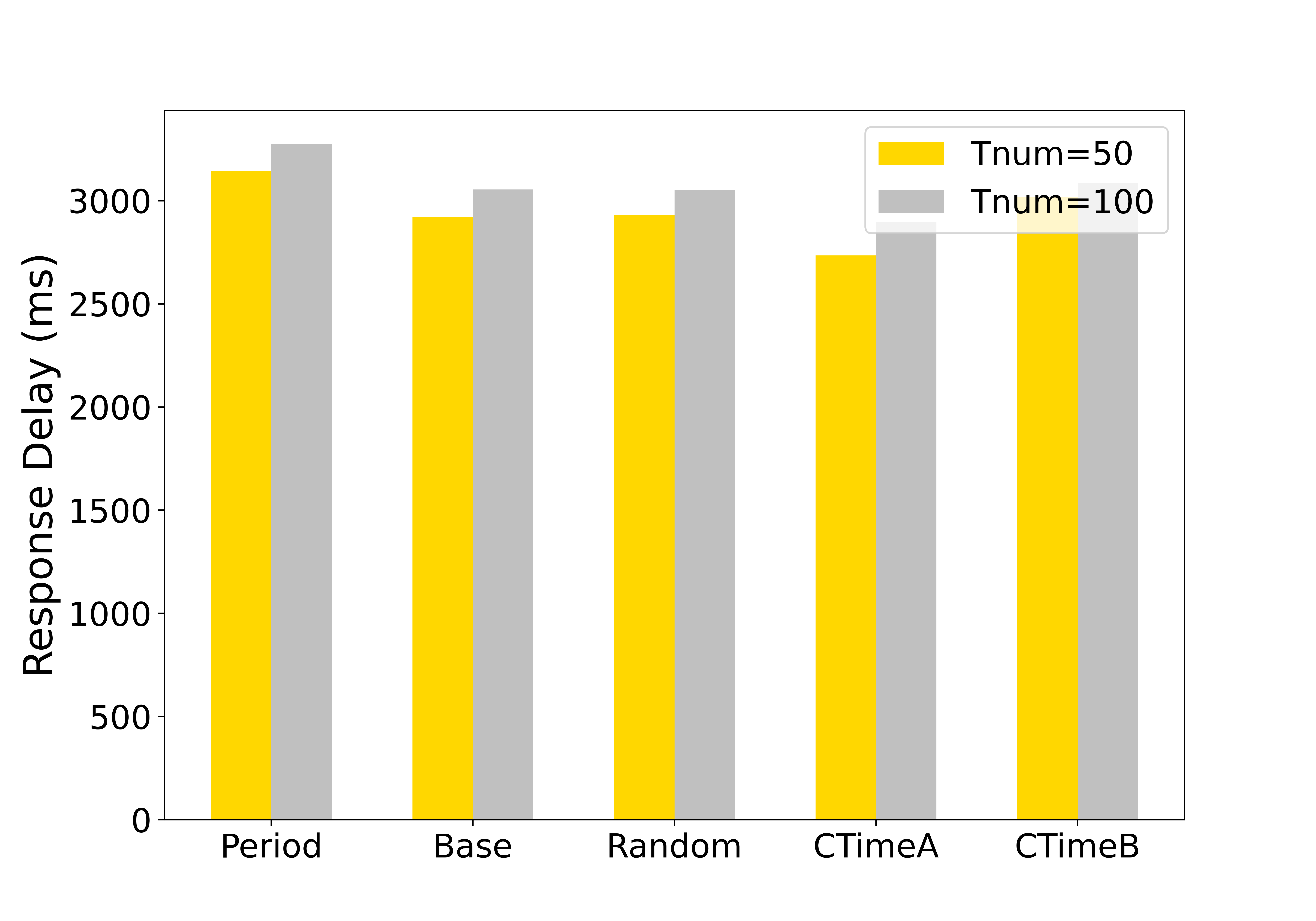}
		\end{minipage}
	}
        \subfigure[]{
		\begin{minipage}{0.46\columnwidth}
			\centering
			\includegraphics[scale=.18]{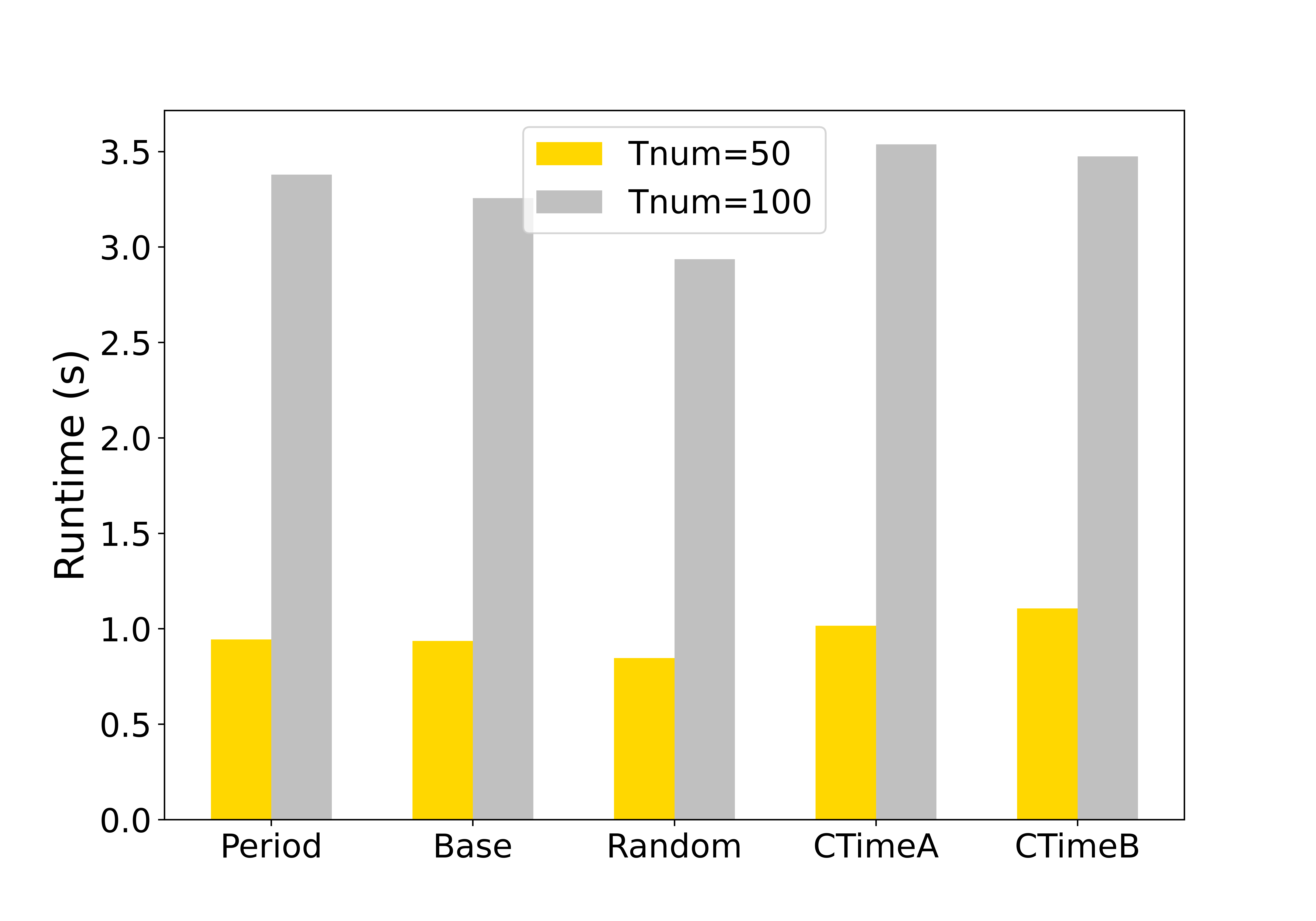}
		\end{minipage}
	}
	\caption{The metric values of IIoTBroker when IIoT computing tasks are sorted with different criteria. \textit{Period}: Sorting by period ascending. \textit{Base}: Sorting as generation time ascending. \textit{Random}: Sorting randomly. \textit{CTimeA}: Sorting in the ascending order of the computing time of IIoT tasks. \textit{CTimeB}: Sorting in the descending order of the computing time of IIoT tasks. }
	\label{fig:IIoTBroker-varysort}
\end{figure}

\begin{table*}[h!]
	\caption{Evaluation Results of IIoTdelpoyer on Problems with $Tnum \in \{10, 50, 100\}$ When $c_s=100 \& c_l=1$. (Best values are bold.)}
	\renewcommand{\arraystretch}{1.15}
	\begin{center}
		\setlength{\tabcolsep}{3pt}
		\begin{tabular}{c|cccc|cccc|cccc}
			\toprule
			\multicolumn{1}{c|}{} & \multicolumn{4}{c|}{Tnum=10} & \multicolumn{4}{c|}{Tnum=50} & \multicolumn{4}{c}{Tnum=100}\\
	 Metrics & IIoTBroker & NcostFS & ScostFS & ONetFS & IIoTBroker & NcostFS & ScostFS & ONetFS & IIoTBroker & NcostFS & ScostFS & ONetFS \\ 
  \midrule
        \textbf{Tcost} $\downarrow$ & \textbf{210.94} & 293.92  & 214.14  & 214.96 & \textbf{798.76}  & 903.76  & 803.60  & 820.70  & 1590.66  & 1642.64  & \textbf{1574.62}  & 1604.02  \\
    \midrule
        ASnum $\downarrow$ & \textbf{1.88} & 2.64  & \textbf{1.88}  & 1.92 & 6.90  & 7.92  & \textbf{6.84}  & 7.12  & 13.74  & 14.22  & \textbf{13.44}  & 13.88  \\
        ALnum $\downarrow$& 2.94 & \textbf{2.72}  & 6.14  & 2.96 & 8.76  & \textbf{8.16}  & 19.60  & 8.70  & 16.66  & \textbf{15.24}  & 30.62  & 16.02  \\
        Utility $\uparrow$ &  \textbf{66.03\%} & 55.04\% &  \textbf{66.03\%} & 65.25\% & 77.89\% & 73.43\% &  \textbf{78.15\%} & 76.88\% & 78.74\% & 77.57\% &  \textbf{79.45\%} & 78.36\% \\
        Delay(ms) $\downarrow$ & 2844.79 &  \textbf{2356.43}  & 2847.64  & 2835.42 & 3324.05  &  \textbf{3135.47}  & 3341.97  & 3302.90  & 3369.83  &  \textbf{3307.75}  & 3401.01  & 3353.84  \\
        Runtime(s) $\downarrow$ & 0.38  & 0.10  & 0.20  & 0.11  & 3.91  & 1.64  & 2.09  & 1.20  & 13.14  & 6.45  & 6.54  & 3.94 \\
			\bottomrule
		\end{tabular}
		\label{table:IIoTDeployer-cl=1}
	\end{center}
\end{table*}

\begin{table*}[h!]
	\caption{Evaluation Results of IIoTdelpoyer on Problems with $Tnum \in \{10, 50, 100\}$ When $c_s=100 \& c_l=1000$. (Best values are bold.)}
	\renewcommand{\arraystretch}{1.15}
	\begin{center}
		\setlength{\tabcolsep}{3pt}
		\begin{tabular}{c|cccc|cccc|cccc}
			\toprule
			\multicolumn{1}{c|}{} & \multicolumn{4}{c|}{Tnum=10} & \multicolumn{4}{c|}{Tnum=50} & \multicolumn{4}{c}{Tnum=100}\\
	  Metrics & IIoTBroker & NcostFS & ScostFS & ONetFS & IIoTBroker & NcostFS & ScostFS & ONetFS & IIoTBroker & NcostFS & ScostFS & ONetFS \\
   \midrule
        \textbf{Tcost} $\downarrow$& \textbf{279.78}  & \textbf{279.78}  & 3678.80  & 1101.80  & \textbf{985.80}  & \textbf{985.80}  & 12258.12  & 1897.10  & \textbf{2372.80}  & 2412.80  & 18574.62  & 3153.20  \\
        \midrule
        ASnum $\downarrow$ & 2.50  & 2.50  & \textbf{1.86}  & 1.98  & 7.60  & 7.60  & \textbf{6.84}  & 7.10  & 13.52  & 13.52  & \textbf{12.62}  & 13.20  \\ 
        ALnum $\downarrow$ & \textbf{2.58}  & \textbf{2.58}  & 5.64  & 2.88  & \textbf{7.90}  & \textbf{7.90}  & 18.38  & 8.18  & \textbf{14.40}  & 14.44  & 29.72  & 14.82  \\
        Utility $\uparrow$ & 55.36\% & 55.36\% &  \textbf{65.04\%} & 62.78\% & 74.36\% & 74.36\% &  \textbf{77.68\%} & 76.47\% & 78.00\% & 78.00\% &  \textbf{80.15\%} & 78.75\% \\ 
        Delay(ms) $\downarrow$ &  \textbf{2557.09}  &  \textbf{2557.27}  & 2900.17  & 2782.43  &  \textbf{3184.43}  & 3185.25  & 3381.47  & 3312.23  &  \textbf{3288.87}  & 3289.28  & 3382.18  & 3341.83  \\
        Runtime(s) $\downarrow$ & 0.39  & 0.12  & 0.23  & 0.13  & 3.87  & 1.74  & 2.31  & 1.24  & 13.58  & 7.20  & 7.10  & 4.34 \\
        \bottomrule
		\end{tabular}
		\label{table:IIoTDeployer-cl=1000}
	\end{center}
\end{table*}

\subsection{IIoTBroker-related Result Analysis}
We compare the performance of IIoTBroker with other four heuristic algorithms on problems of four different scales, and summarize the results in Table~\ref{table:IIoTBroker}. Compared with other algorithms, IIoTBroker can use fewest servers to support deterministic responses for all IIoT tasks, and realize highest average utilization of scheduled IIoT servers. Following IIoTBroker, Nearest-FS is the second algorithm to achieve fewer servers and higher utilization. Because when an IIoT server is the nearest for multiple IIoT devices, it earns some opportunities to be reused by multiple tasks. In this way, the number of scheduled servers is reduced, while the utilization is improved. Random-S can provide a not bad scheduling decision as well, but worse than IIoTBroker and Nearest-FS. Delay-FS and DFNS-S have similar performance. They cannot provide cost-friendly scheduling decisions, but can ensure lowest response delays through using as may resources as possible. As is shown in the simulation, Delay-FS and DFNS-S's scheduling tends to allocate only one IIoT task to an IIoT server in most cases. 

We also compare the performances of IIoTBroker to TSN(+), shown in Table~\ref{table:IIoTBroker-ablation}. From the perspectives of metrics \textit{Snum, Utility, Delay}, IIoTBroker still slightly outperform TSN(+). Furthermore, with consideration of \textit{Runtime}, we can easily find that IIoTBroker performs better in computational efficiency, even if we have already optimized TSN(+) in our derivations. Thus, IIoTBroker is not only an efficient algorithm, but also provides high-quality scheduling decisions. 

To demonstrate how the performance of IIoTBroker and other algorithms change with the problem scales, we evaluate the number of scheduled servers, utilization rate of the scheduled servers, average response delay, and computational time when the number of IIoT computing tasks $Tnum$ varies from 10 to 200 with steps 10, shown in Fig.~\ref{fig:IIoTBroker-varyTnum}. Fig.~\ref{fig:IIoTBroker-varyTnum}(a) shows that for all the four algorithms, the number of scheduled servers increases with $Tnum$. IIoTBroker, however, exhibits the slowest growth rate, indicating its excellent ability to save computation cost. Fig.~\ref{fig:IIoTBroker-varyTnum}(b) and Fig.~\ref{fig:IIoTBroker-varyTnum}(c) illustrate that IIoTBroker's utilization rate and average response delay are both the highest among the compared algorithms. This indicates that IIoTBroker sacrifices response time to ensure high utility and low cost. Opposite to IIoTBroker, Delay-FS provides timely response for IIoT tasks by allocating more IIoT servers. Fig.~\ref{fig:IIoTBroker-varyTnum}(d) shows that compared to baseline algorithms, IIoTBroker's computation time to get effective decisions increases much more slowly and steady with the number of IIoT computing tasks $Tnum$. 

\begin{figure*}[ht!]
	\centering
        \setcounter{subfigure}{0}
	\subfigure[$c_s=100, c_l=1$]{
		\begin{minipage}{0.6\columnwidth}
			\centering
			\includegraphics[scale=.23]{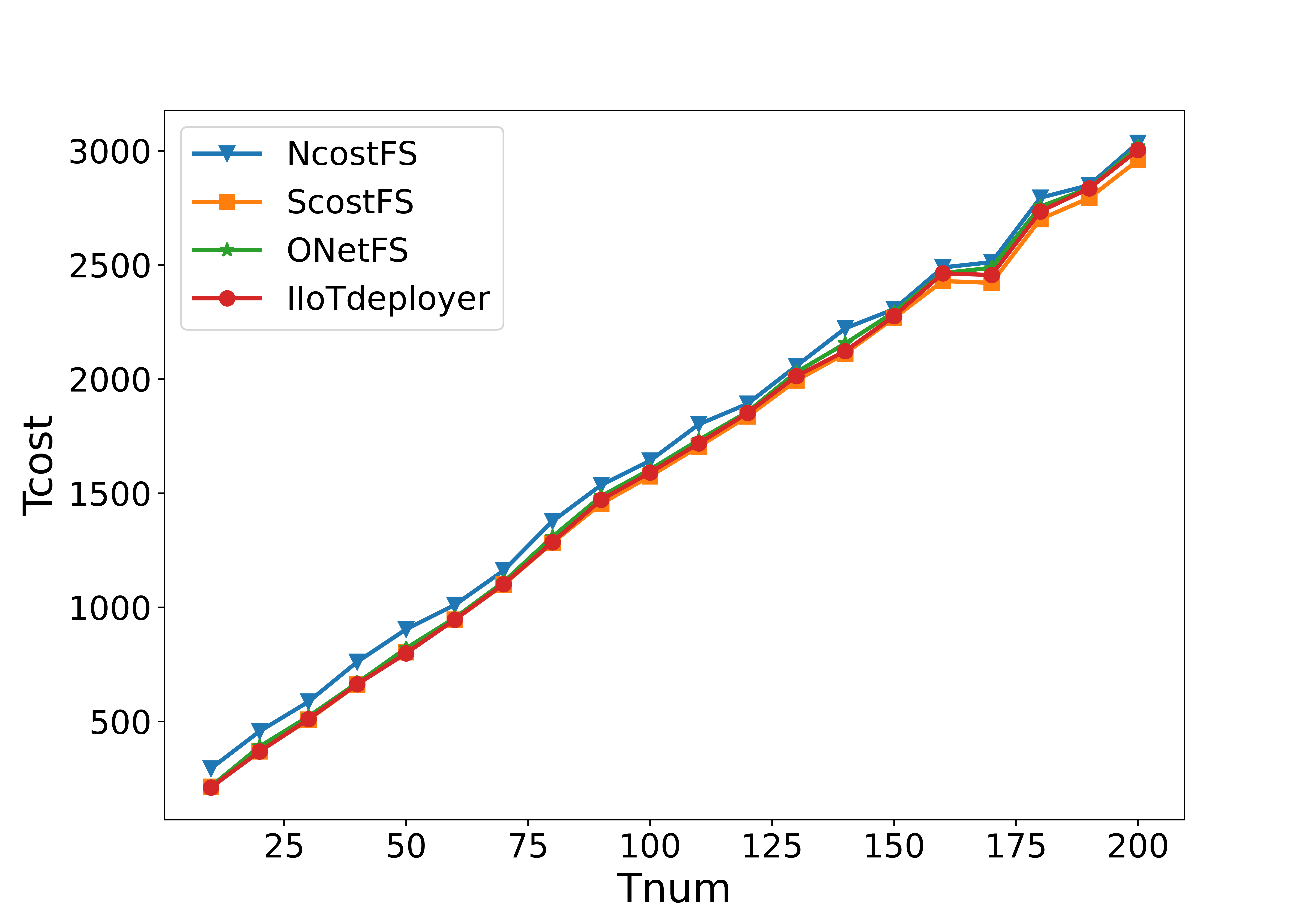}
		\end{minipage}
	}
        \subfigure[$c_s=100, c_l=1$]{
		\begin{minipage}{0.6\columnwidth}
			\centering
			\includegraphics[scale=.23]{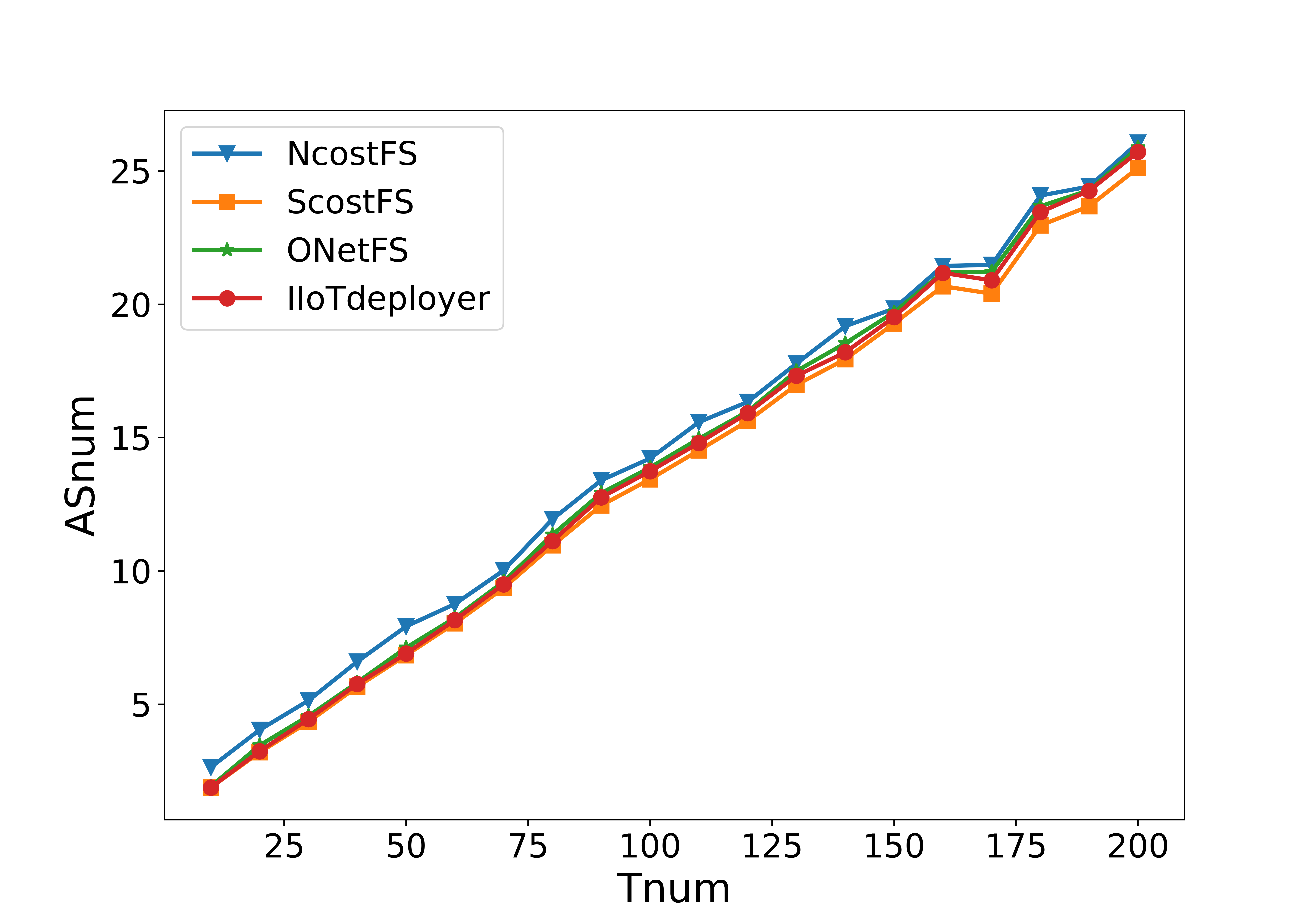}
		\end{minipage}
	}
        \subfigure[$c_s=100, c_l=1$]{
		\begin{minipage}{0.6\columnwidth}
			\centering
			\includegraphics[scale=.23]{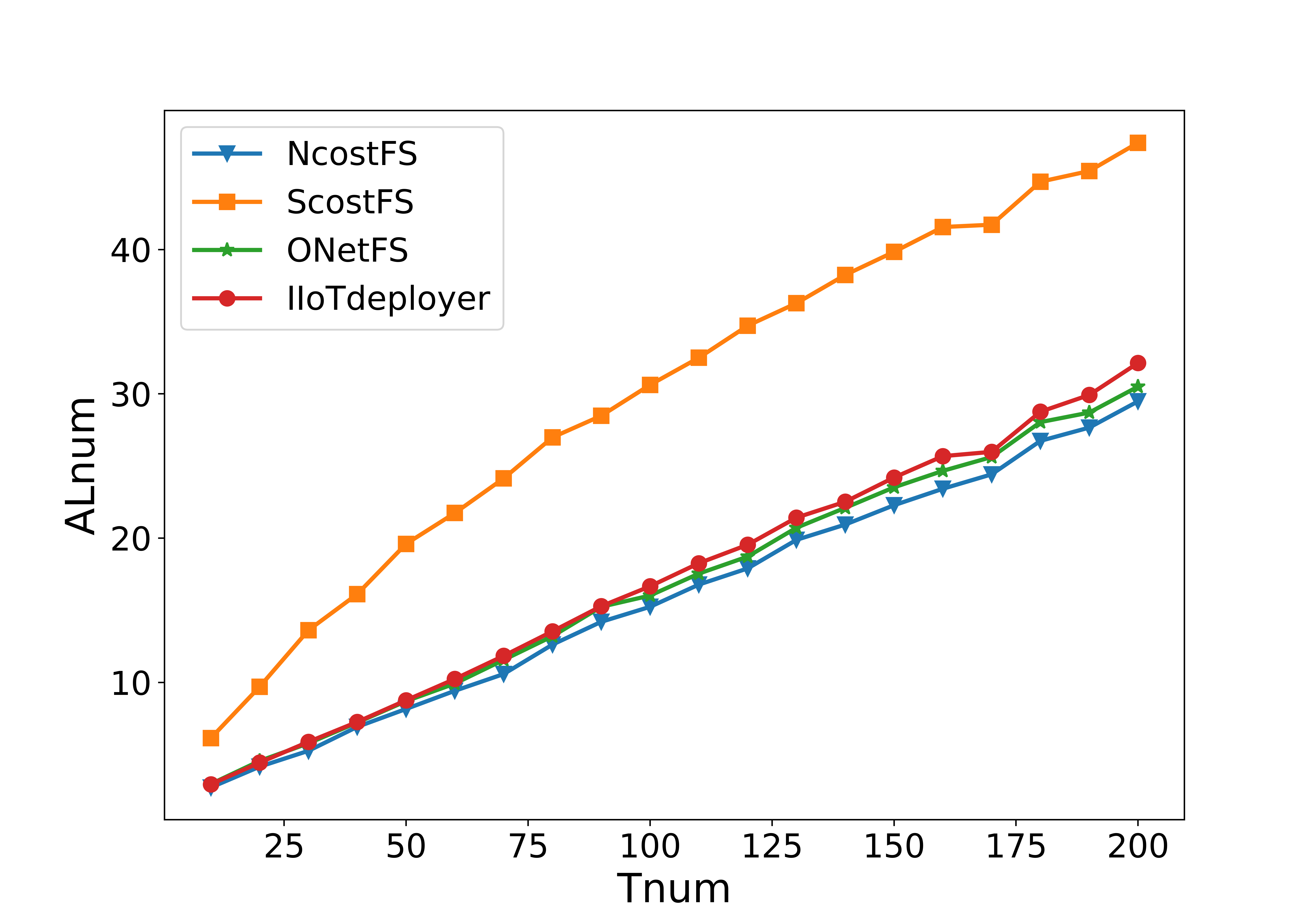}
		\end{minipage}
	}
    \subfigure[$c_s=100, c_l=1000$]{
		\begin{minipage}{0.6\columnwidth}
			\centering
			\includegraphics[scale=.23]{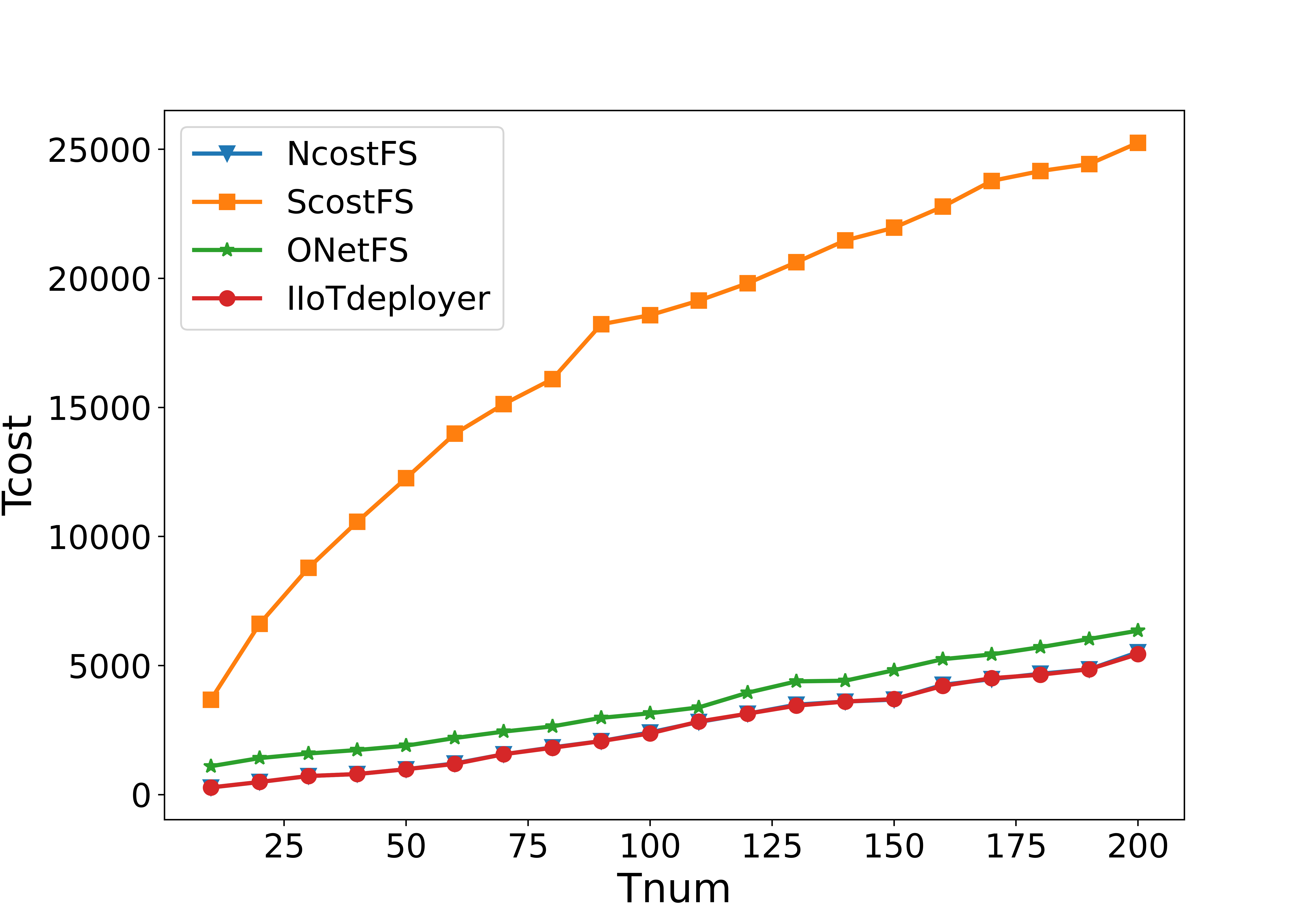}
		\end{minipage}
	}
        \subfigure[$c_s=100, c_l=1000$]{
		\begin{minipage}{0.6\columnwidth}
			\centering
			\includegraphics[scale=.23]{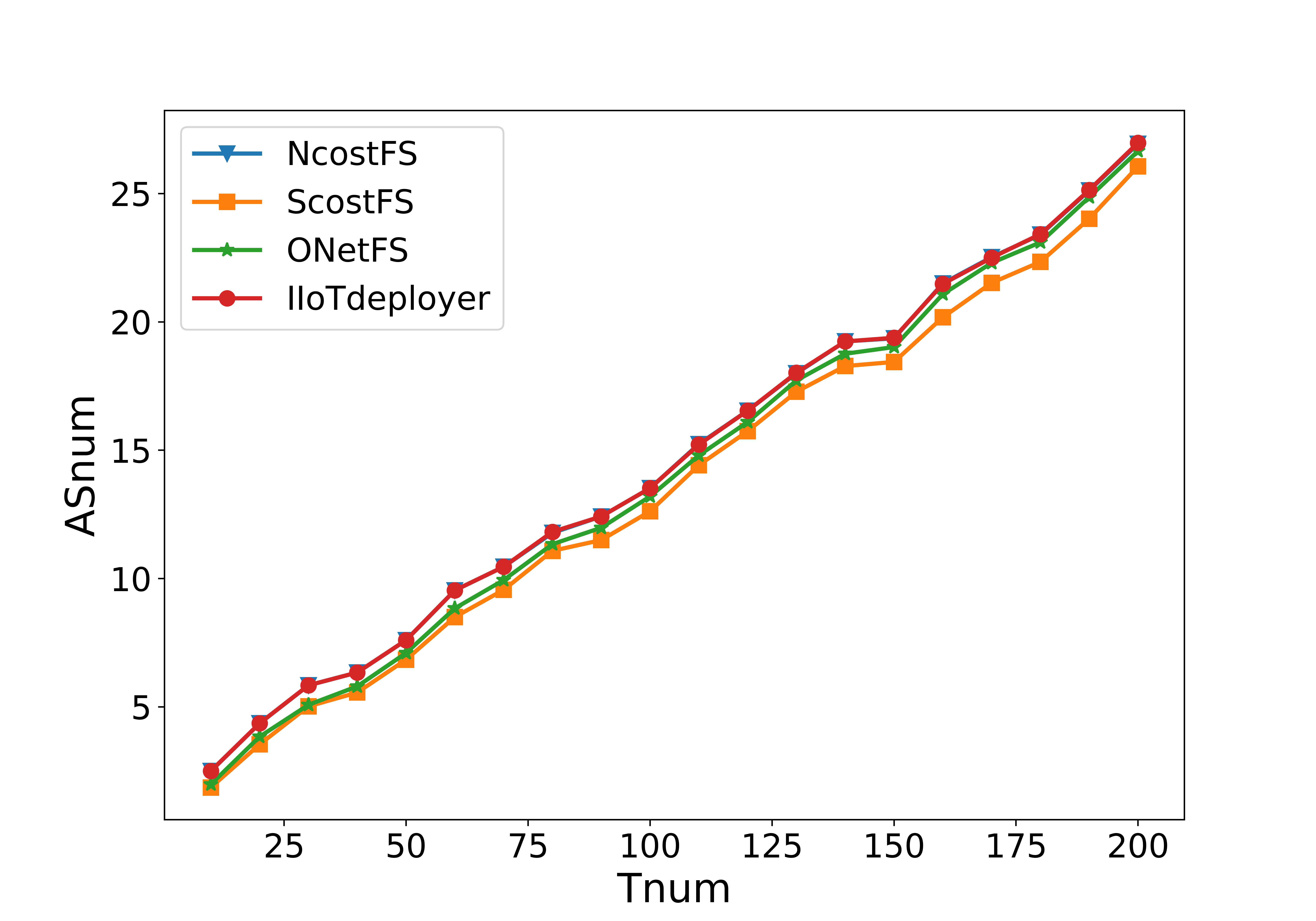}
		\end{minipage}
	}
        \subfigure[$c_s=100, c_l=1000$]{
		\begin{minipage}{0.6\columnwidth}
			\centering
			\includegraphics[scale=.23]{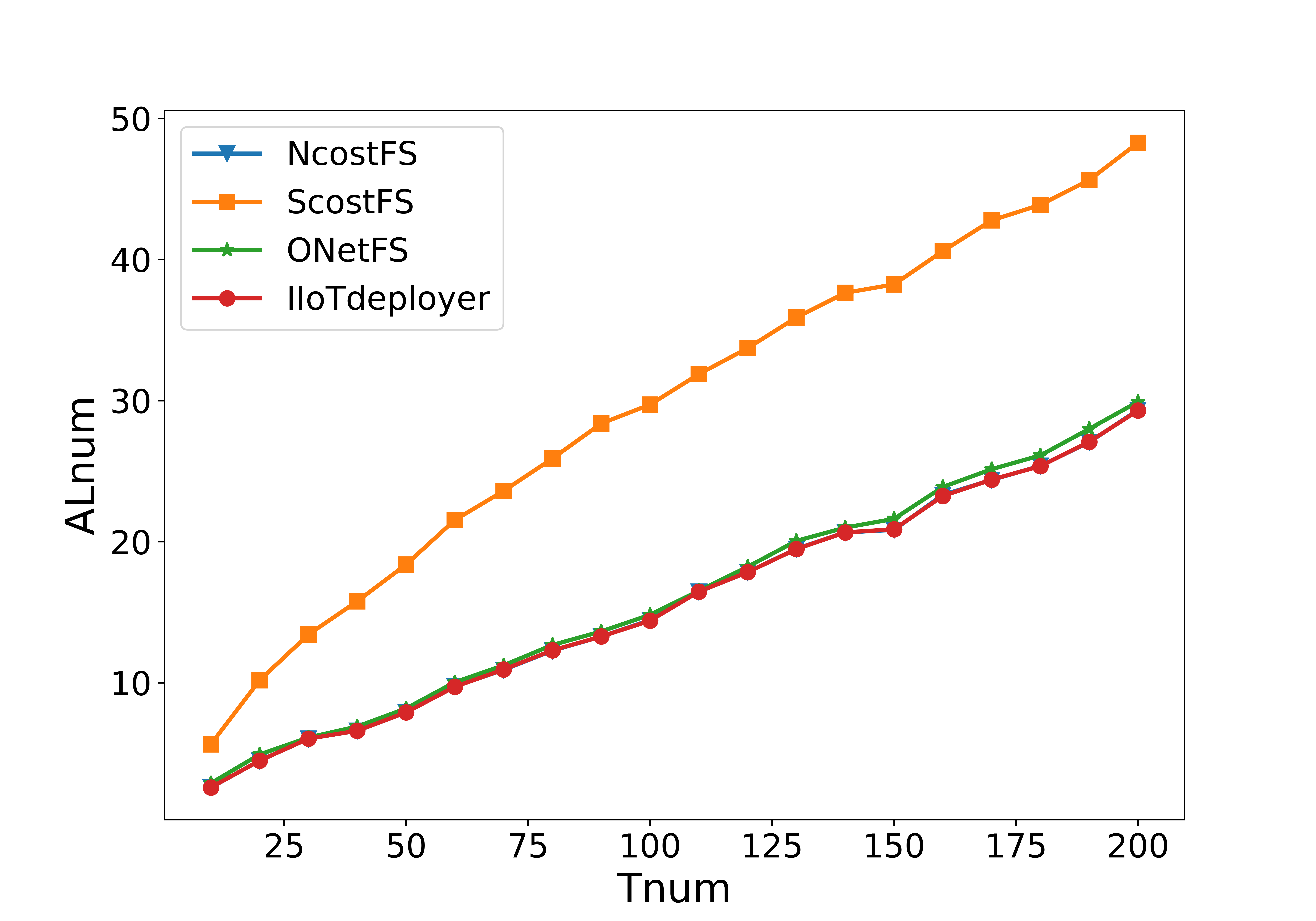}
		\end{minipage}
	}
	\caption{The cost variations of upgrading IIoT networks with the increases of IIoT periodic time-critical computing tasks under system settings of $c_s=100 \& c_l=1$ and $c_s=100 \& c_l=1000$. }
	\label{fig:deployer-varytnum-Costs}
\end{figure*}

\begin{figure}[ht!]
    \centering
    \setcounter{subfigure}{0}
    \subfigure[]{
		\begin{minipage}{0.45\columnwidth}
			\centering
			\includegraphics[scale=.18]{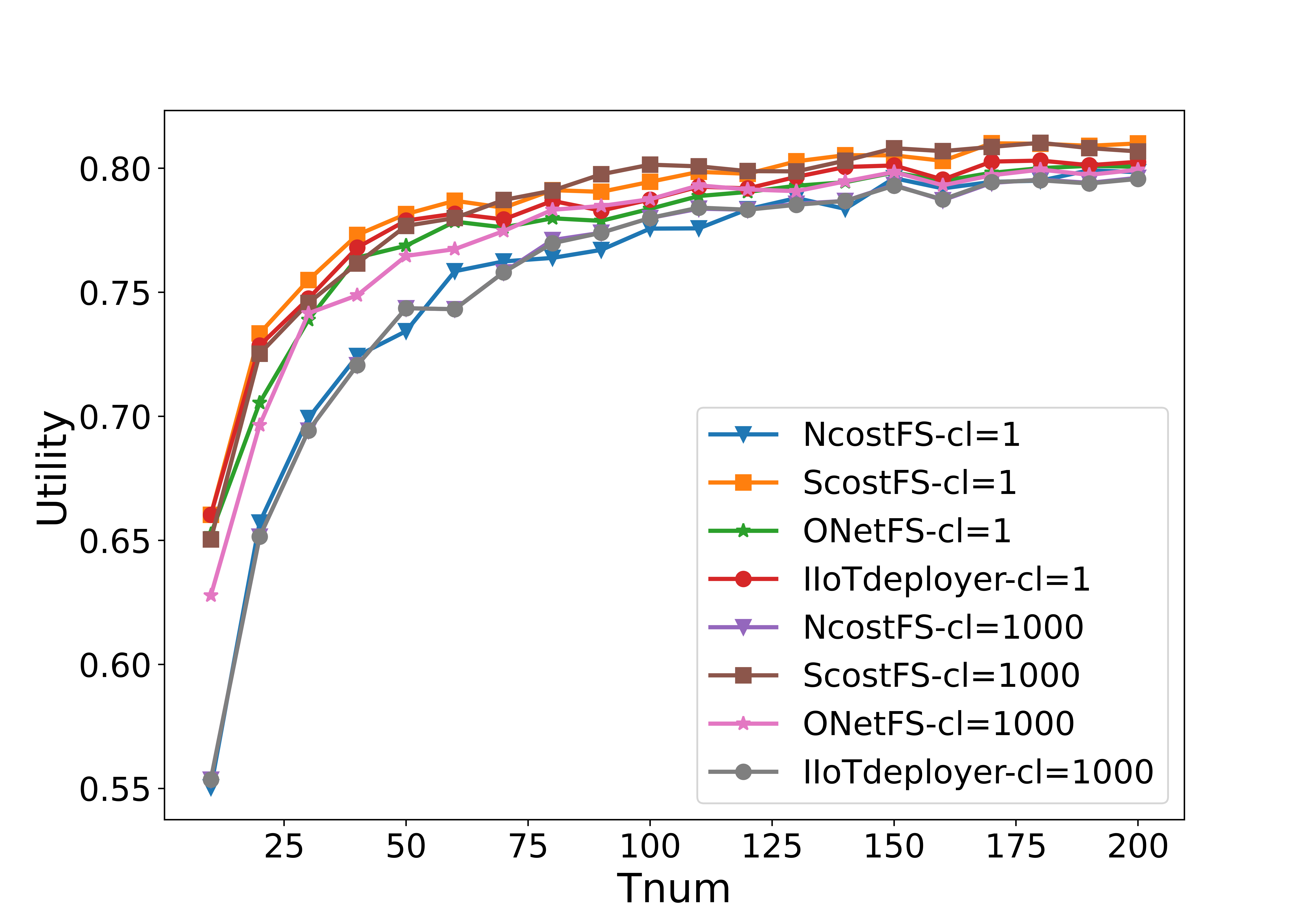}
		\end{minipage}
	}
        \subfigure[]{
		\begin{minipage}{0.45\columnwidth}
			\centering
			\includegraphics[scale=.18]{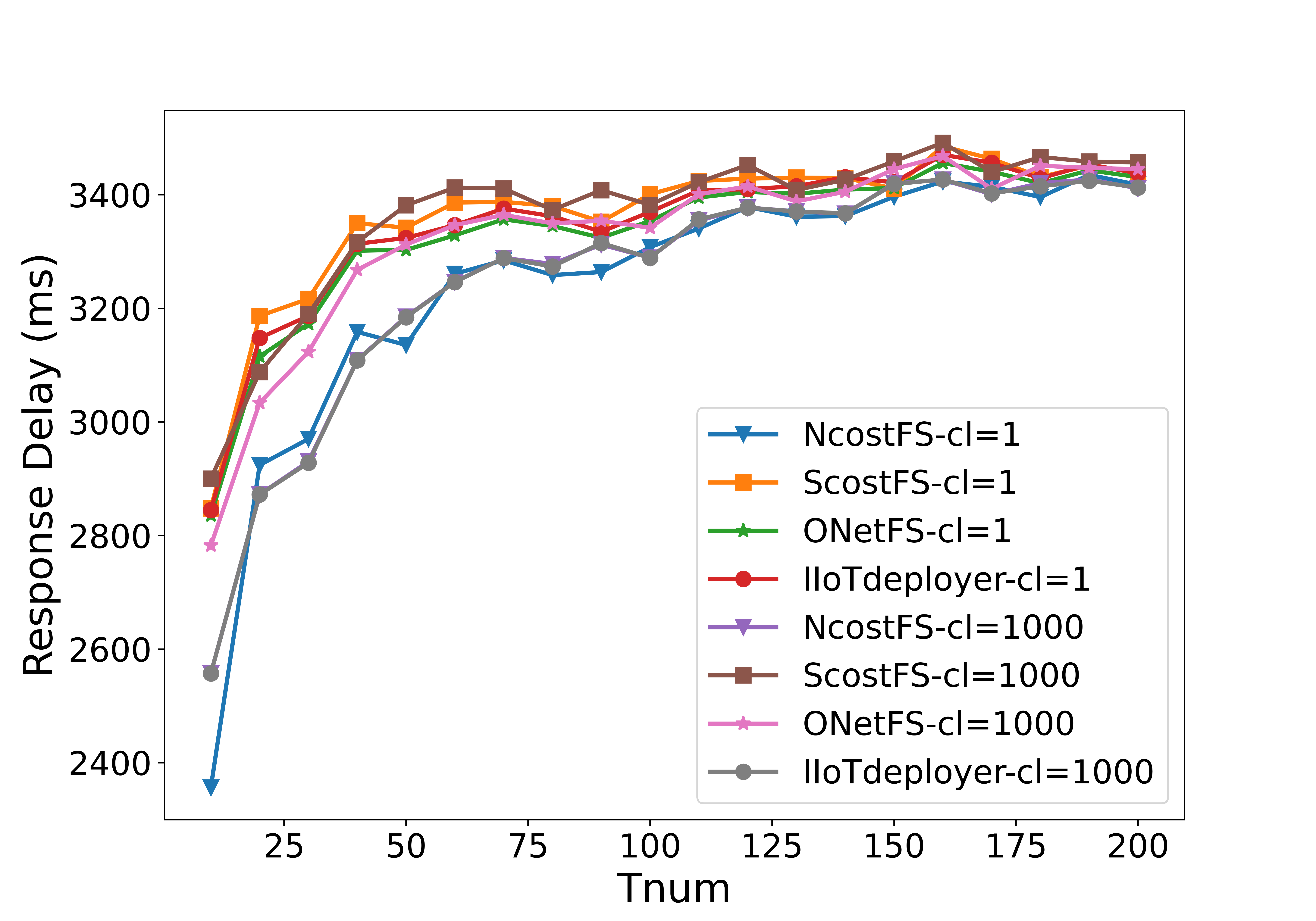}
		\end{minipage}
	}
    \caption{The variations of utility and response delay with the increases of IIoT periodic time-critical computing tasks under system settings of $c_s=100 \& c_l=1$ and $c_s=100 \& c_l=1000$. }
    \label{fig:deployer-varytnum-utilization-delay}
\end{figure}

IIoTBroker does not impose any restrictions on the feeding order of IIoT tasks during planning. Theoretically, IIoTBroker can be compatible with any feeding order of the IIoT tasks. We study the impacts of scheduling decision of IIoTBroker with different sorting criteria, including \textit{Period}, \textit{Base}, \textit{Random}, \textit{CTimeA} and \textit{CTimeB}. The results are shown in Fig.~\ref{fig:IIoTBroker-varysort}, where two cases are provided, namely $Tnum=50$ and $Tnum=100$. In Fig.~\ref{fig:IIoTBroker-varysort}(a), the scheduling performances under the five sorting criteria are very close, and the results of period ascending sorting are slightly better. The utilities shown in Fig.~\ref{fig:IIoTBroker-varysort}(b) and response delays shown in Fig.~\ref{fig:IIoTBroker-varysort}(c) are also similar. Scheduling with \textit{Period} sorting realizes a slightly higher utilization rate compared to other criteria, and planning with \textit{CTimeA} sorting can gain lower response delay. As shown in Fig.~\ref{fig:IIoTBroker-varysort}(d), scheduling with randomly sorting consumes shorter time. Therefore, based on the simulation results, the period ascending sorting method is suggested if users care most about computational cost reduction. If users want to reduce computational costs while maintaining low response delay, \textit{CTimeA} is recommended. If users want to make an effective scheduling decision quickly, randomly sorting is recommended.

\begin{figure*}[h!]
	\centering
        \setcounter{subfigure}{0}
	\subfigure[]{
		\begin{minipage}{0.6\columnwidth}
			\centering
			\includegraphics[scale=.23]{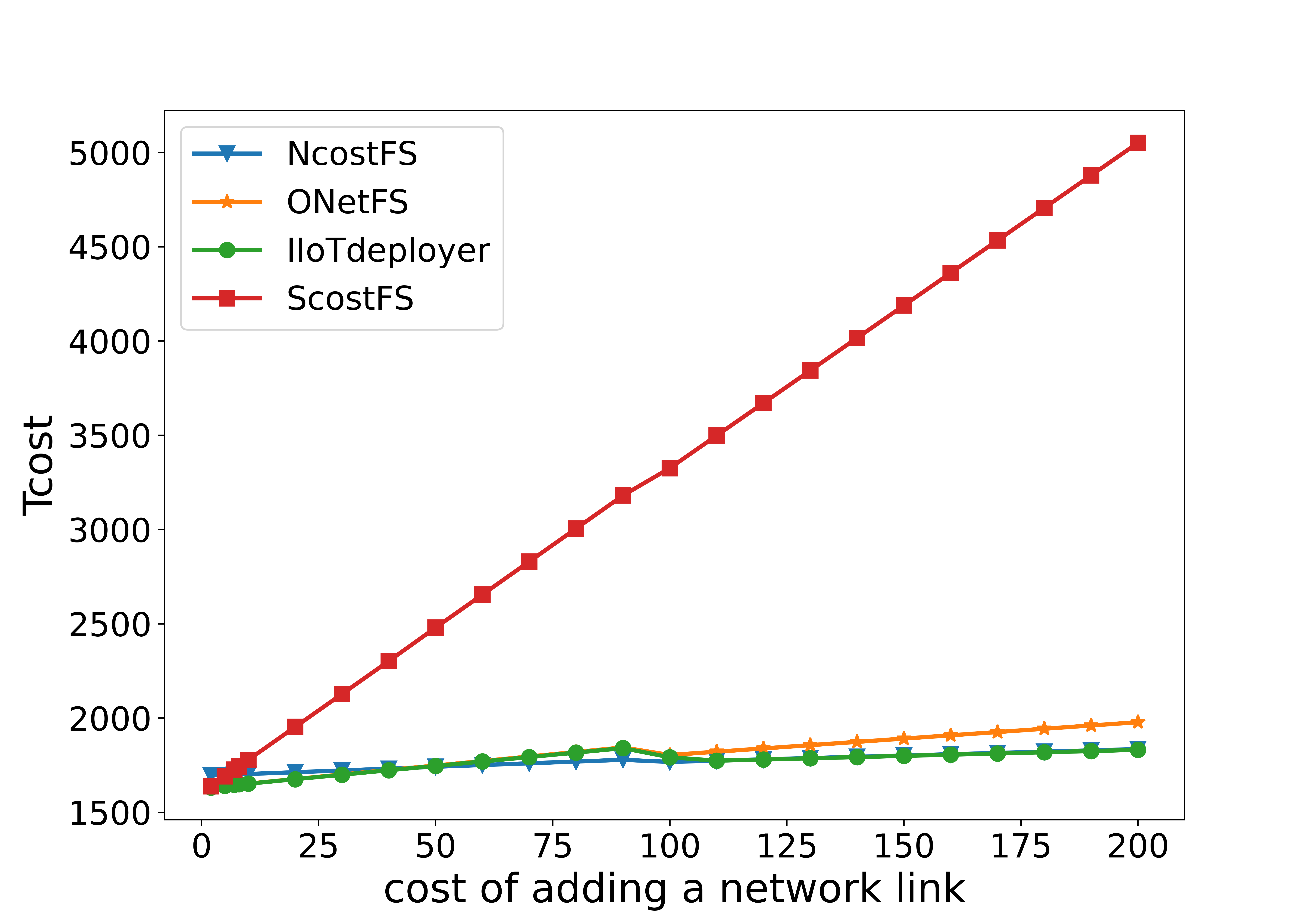}
		\end{minipage}
	}
        \subfigure[]{
		\begin{minipage}{0.6\columnwidth}
			\centering
			\includegraphics[scale=.23]{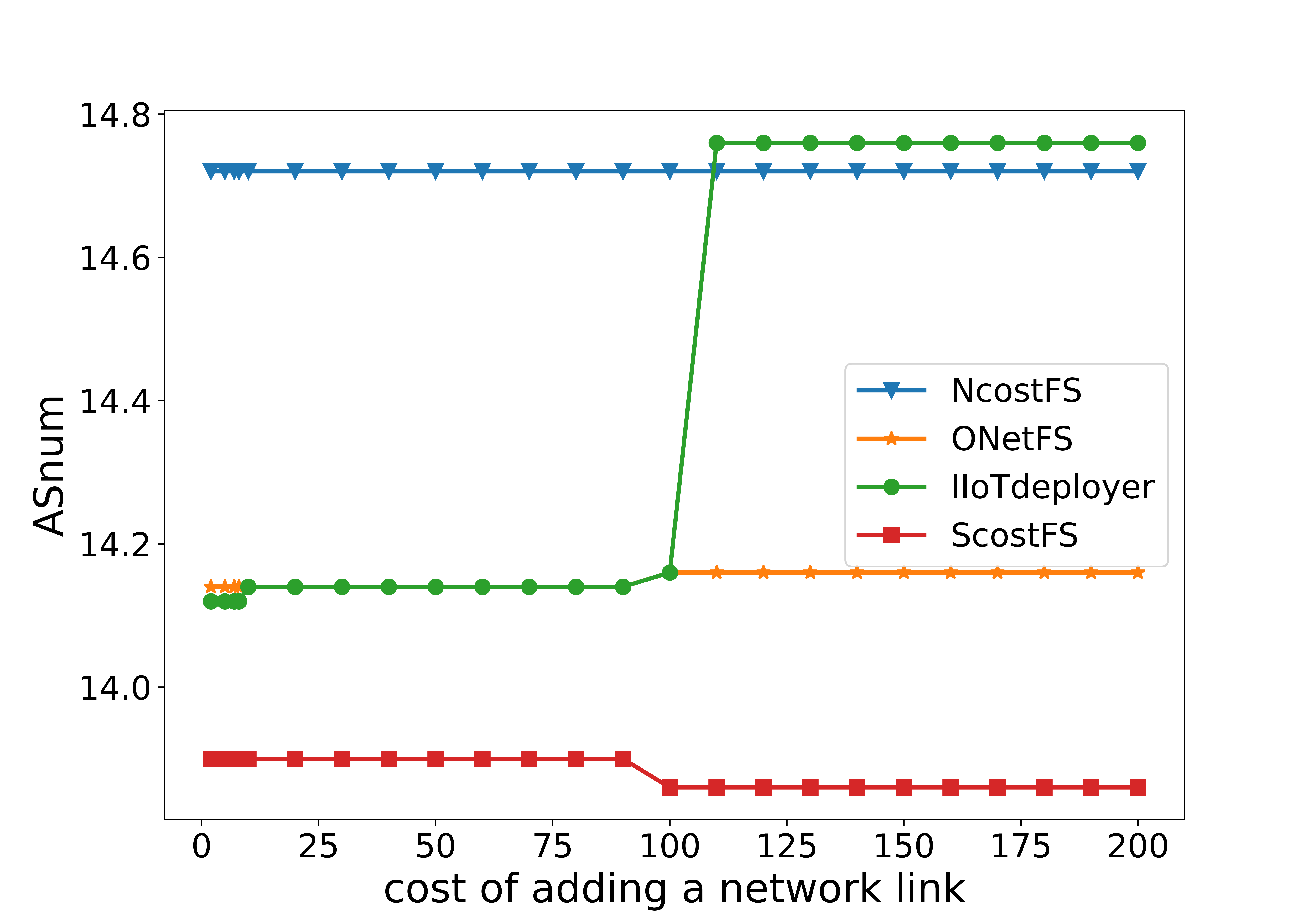}
		\end{minipage}
	}
        \subfigure[]{
		\begin{minipage}{0.6\columnwidth}
			\centering
			\includegraphics[scale=.23]{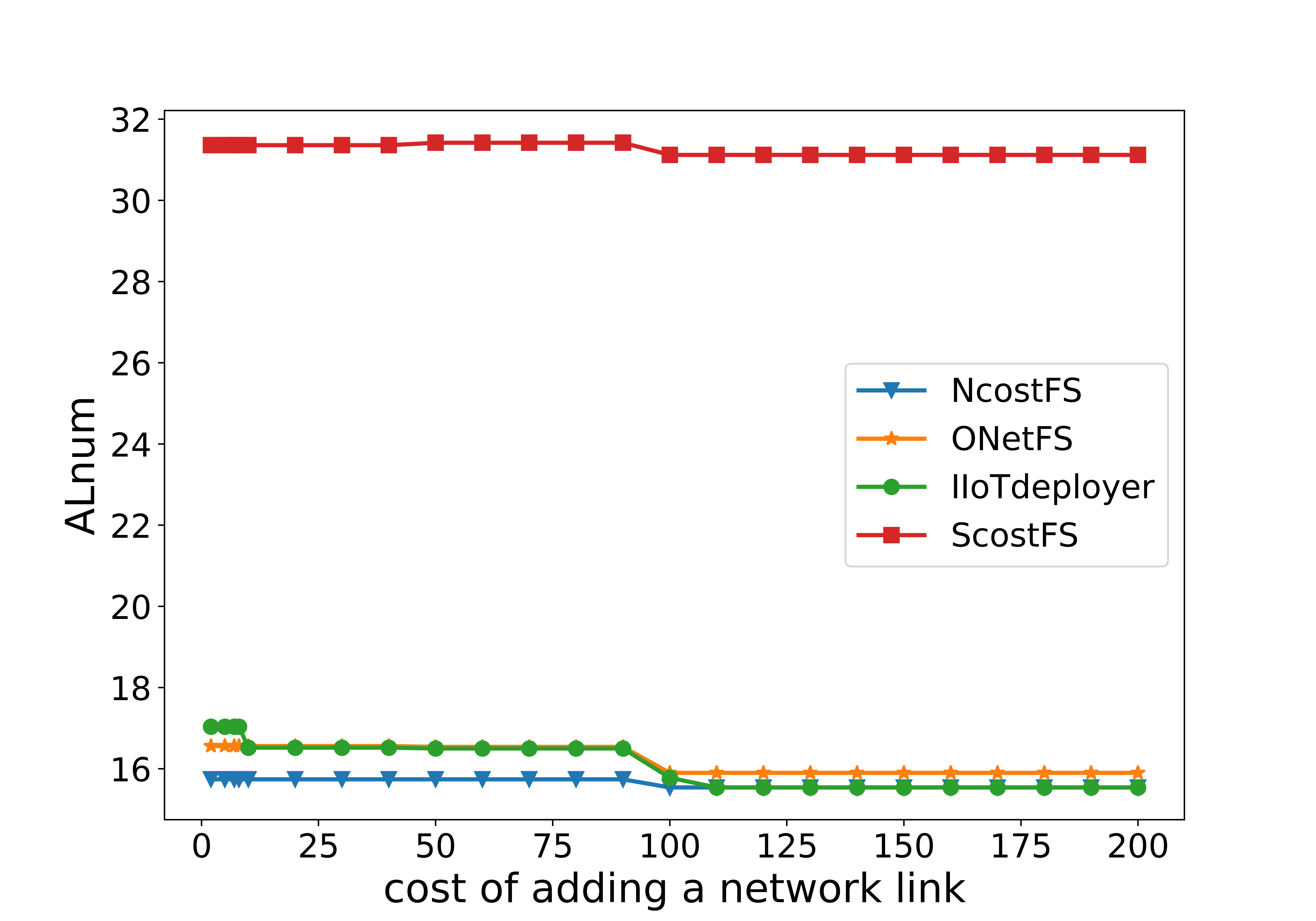}
		\end{minipage}
	}
        \subfigure[]{
		\begin{minipage}{0.6\columnwidth}
			\centering
			\includegraphics[scale=.23]{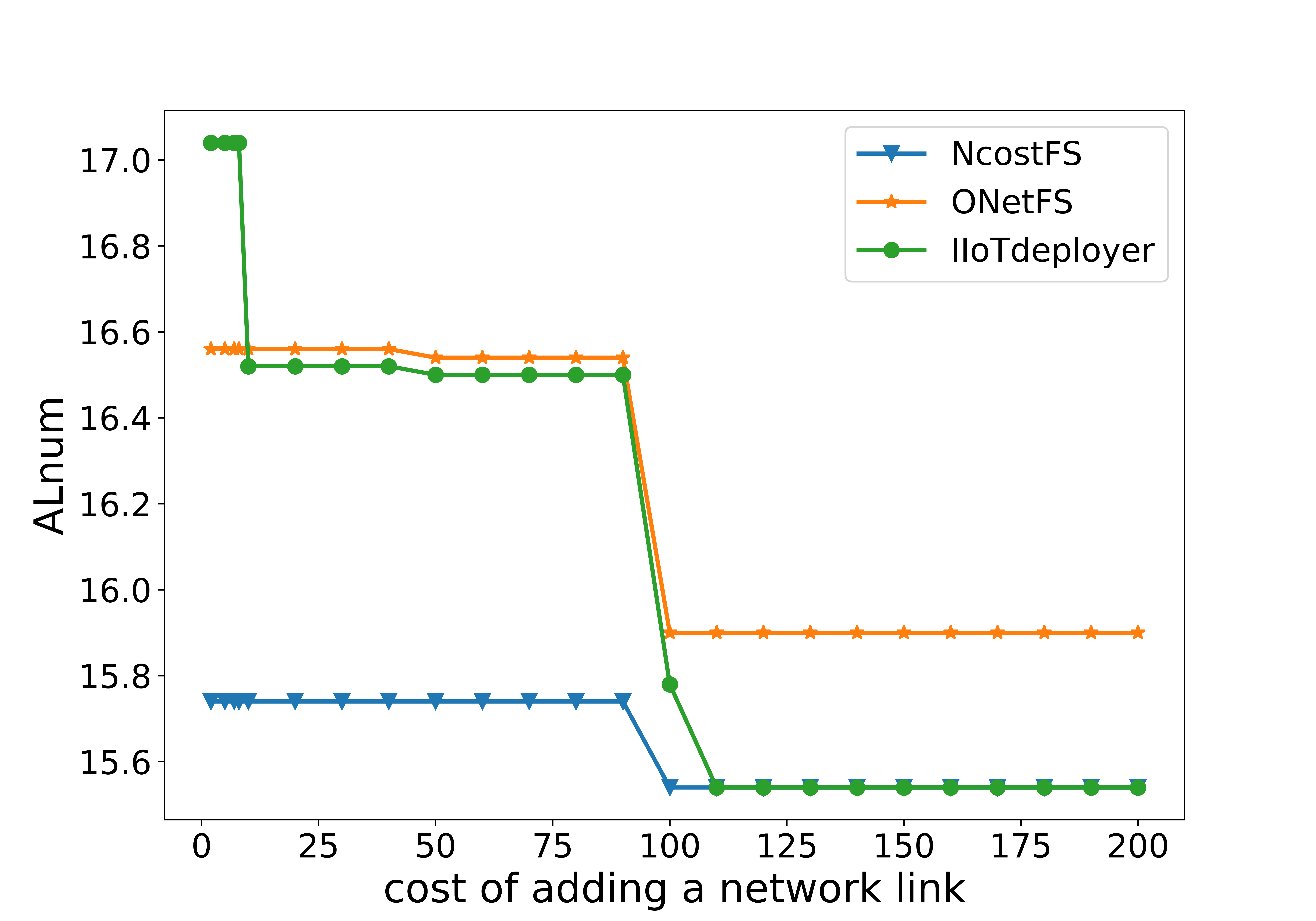}
		\end{minipage}
	}
        \subfigure[]{
		\begin{minipage}{0.6\columnwidth}
			\centering
			\includegraphics[scale=.23]{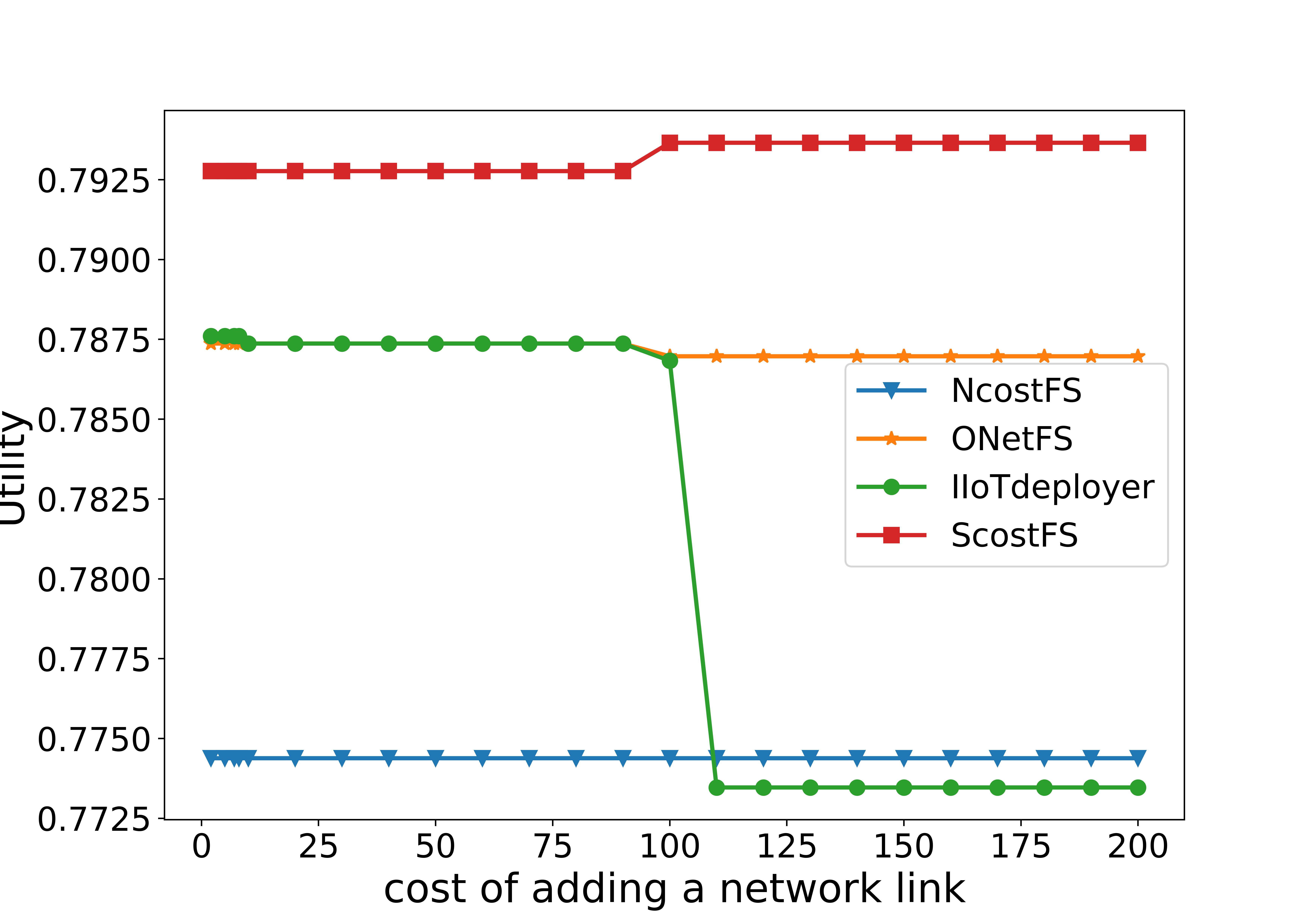}
		\end{minipage}
	}
        \subfigure[]{
		\begin{minipage}{0.6\columnwidth}
			\centering
			\includegraphics[scale=.23]{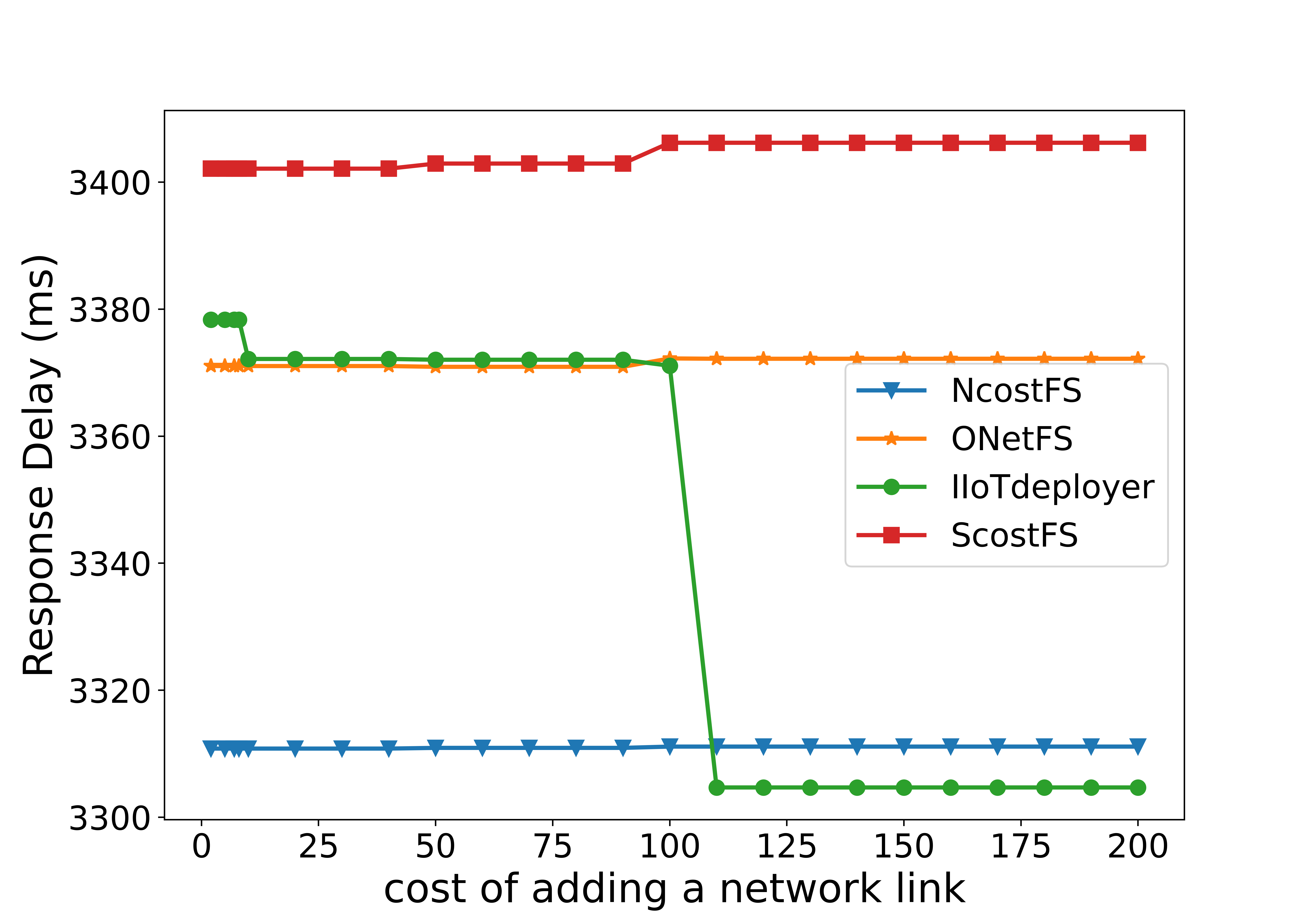}
		\end{minipage}
	}
	\caption{The performance variations of algorithms with $c_s = 100$ and $c_l$ varies in $\{2,5,7,8,10, 20,30,...,200\}$ when $Tnum=100$. }
	\label{fig:deployer-vary-cl-tum=100}
\end{figure*}

\subsection{IIoTDeployer-related Result Analysis}
We compare the performance of IIoTDeployer with other three baselines on the IIoT network structure optimization problems with scales $Tnum \in \{10, 50, 100\}$. The cost of servers and links can vary as well. In the simulations, we first study on two different combinations of $c_s$ and $c_l$, i.e. $c_s=100 \& c_l=1$ and $c_s=100 \& c_l=1000$. 
The purpose of validating performance on such two settings is to observe the differences between scheduling decisions when $c_l \ll c_s$ and $c_l \gg c_s$. In the situation of $c_l \ll c_s$, intuitively, it is advisable to reuse IIoT servers as much as possible by increasing network links. In contrast to the situation, when $c_l \gg c_s$, it is intuitively recommended to add some servers instead of deploying network links when existing computation and network resources are not sufficient to support IIoT tasks.  

According to the results in Table~\ref{table:IIoTDeployer-cl=1}, in general, IIoTDeployer can get high-quality upgrading schemes with lower costs. The total costs of IIoTBroker and ScostFS over three scales are very close, as well as the number of added IIoT servers. However, ScostFS introduces more network links in the upgrading schemes than other algorithms. NcostFS can always provide schemes with least links upgrades, but the number of servers increases much in the schemes. The phenomenons are in line with our expectations. ScostFS cares more about the cost of adding servers and pay less attention to link consumption. NcostFS cares more about the cost of increasing network links instead of servers. Compared with them, IIoTBroker tries to balance costs of servers and links. ONetFS is relatively mediocre and cannot achieve better results under any metric. In terms of utility, ScostFS and IIoTDeployer that use less servers earn higher utilization rate, because the overall computation capacity requirements of IIoT tasks are fixed. As for delay, NcostFS and ONetFS have lower response delay since more servers are scheduled to complete IIoT tasks. From the perspective of runtime, IIoTBroker takes approximately twice as long as other algorithms, since it must run both LCFU and SCFU for each IIoT task, while NcostFS, ScostFS and ONetFS require running only one of them in most cases. 

The results in Table~\ref{table:IIoTDeployer-cl=1000} illustrate that IIoTDeployer can realize great performance when $c_l \gg c_s$. Different from Table~\ref{table:IIoTDeployer-cl=1} where the performance of IIoTDeployer is much closer to that of ScostFS, the results of IIoTBroker are similar with that of NcostFS in Table~\ref{table:IIoTDeployer-cl=1000}. Because the cost of adding network links is much higher than adding servers, IIoTBroker strives to avoid increasing network links in order to save costs. ScottFS can still provide the upgrading schemes with least server deployment, but the total upgrading cost is the highest due to expensive link deployment. NcostFS can generate good but not the best upgrade schemes, just like it does in the cases where $c_l \ll c_s$. Although IIoTDeployer and NcostFS realize lowest upgrading costs, introducing more IIoT servers into the IIoT network makes them difficult to achieve the highest server utilization. The benefit is that IIoTDeployer and NcostFS realize lower response delays. 

Then, we explore the performance changes when the problem scale varies. The results are summarized in Fig.~\ref{fig:deployer-varytnum-Costs} and Fig.~\ref{fig:deployer-varytnum-utilization-delay}. In general, all metrics are growing as the number of IIoT tasks increases. When $c_s=100 \& c_l=1$, Tcosts of NcostFS keeps slightly higher than other three methods in Fig.~\ref{fig:deployer-varytnum-Costs}(a), while the curves of ScostFS, ONetFS and IIoTBroker almost overlap. We further study the reasons that leads to such situation. 
According to the results in Fig.~\ref{fig:deployer-varytnum-Costs}(b) and Fig.~\ref{fig:deployer-varytnum-Costs}(c), NcostFS schedules more servers, which cost a lot. Although NcostFS deploys less links but the links are much cheaper in this case. ScostFS consumes less servers than others, which helps it save cost. However, ScostFS taking much more links weakens the cost benefits. 

When the system setting changes to $c_s=100 \& c_l=1000$, the cost keeps growing as the IIoT task number increases, but the performance is different from the system setting $c_l=1$. In Fig.~\ref{fig:deployer-varytnum-Costs}(d)-Fig.~\ref{fig:deployer-varytnum-Costs}(f), the curves of IIoTBroker and NcostFS are overlapped, since the too expensive link costs makes the scheduling decision of IIoTBroker be close to that of NcostFS. Tcosts of ScostFS are rapidly increasing as the number of IIoT tasks increases, since ScostFS introduces too many expensive network links to ensure computation reachability. 
The variations of utility and response delay with the increase of IIoT task quantity are depicted in Fig.~\ref{fig:deployer-varytnum-utilization-delay}. Overall, both utilization rate and response delay grow with the increase in the number of IIoT tasks, and tend to converge to similar levels.

In addition, we also explore the changes in the metrics of four algorithms as the cost of adding links increases in IIoT networks. We carry out simulations with the system setting: $Tnum=100$ and $c_s=100 \& c_l \in \{2, 5, 7, 8, 10, 20, 30, ..., 200\}$. From Fig.~\ref{fig:deployer-vary-cl-tum=100}(a), Tcosts of ScostFS significantly increase with $c_l$. The primary reason is that ALnum of ScostFS always keeps much more than others, shown in Fig.~\ref{fig:deployer-vary-cl-tum=100}(c). 
The turning point only occurs when $c_l=100$, and at this point, $c_s=c_l$, it indicates that from this point on, the cost of adding a server is equal to adding a link. 
According to  Fig.~\ref{fig:deployer-vary-cl-tum=100}(b), we can see that ASnum changes as well when $c_l=100$. However, the slight changes cannot help ScostFS reduce total costs a lot. 
Compared with ScostFS, the curves of NcostFs in Fig.~\ref{fig:deployer-vary-cl-tum=100} are very smooth.  It seems decisions of NcostFs are not affected by variations of $c_l$, because the primary intention behind NcostFs it to use less links. From Fig.~\ref{fig:deployer-vary-cl-tum=100}(b)-Fig.~\ref{fig:deployer-vary-cl-tum=100}(d), small turning points also occur at $c_l=100$ since servers start to be relatively cheaper. 

Tcosts of IIoTBroker keep lowest in most cases as shown in Fig.~\ref{fig:deployer-vary-cl-tum=100}(a). IIoTBroker changes ASnum from near ScostFS to approaching NcostFS in Fig.~\ref{fig:deployer-vary-cl-tum=100}(b), meanwhile, it maintains that a few links are required to be added in Fig.~\ref{fig:deployer-vary-cl-tum=100}(c)-Fig.~\ref{fig:deployer-vary-cl-tum=100}(d). IIoTBroker has twice turns at $c_l=100$ and $c_l=110$. When $c_l=100$, the cost of adding a server is equivalent to adding a link, and IIoTBroker starts to attempt to reduce added link number. When $c_l=110$, adding a link becomes more expensive than adding a server, thus IIoTBroker further reduce the number of required links. 
ONetFS can adjust decisons based on variations of $c_l$, but its performance is worse than IIoTBroker's. The changes of utility in Fig.~\ref{fig:deployer-vary-cl-tum=100}(e) and response delay in Fig.~\ref{fig:deployer-vary-cl-tum=100}(f) have negative correlation with the change of ASnum in Fig.~\ref{fig:deployer-vary-cl-tum=100}(b). When ASnum increases, both utility and response delay reduce. 

The results in Fig.~\ref{fig:deployer-varytnum-Costs} and Fig.~\ref{fig:deployer-vary-cl-tum=100} illustrate that regardless of the scales of the optimization problem and the cost settings of upgrading, IIoTDeployer can always obtain high-quality IIoT network structure upgrade scheme. 

\section{Conclusion} \label{section:conclusion}
Driven by the requirements of IIoT intelligent production and management, we identify two challenges that have never been discussed before, i.e. providing deterministic response delay services for IIoT periodic time-critical computing tasks, and upgrading network structure to ensure sufficient network and computing resources for the tasks. We theoretically analyze the conditions of resource sharing conflicts and derive some theorems that help us design IIoTBroker and IIoTDeployer to address the two challenges. 
Abundant simulations are conducted to verify the efficiency of the proposed algorithm under different parameter settings. The results demonstrate that IIoTBroker can provide higher-quality scheduling decisions to ensure deterministic response for IIoT tasks than baselines. The computational efficiency of IIoTBroker is higher than others as well, since IIoTBroker consumes shorter runtime. IIoTDeployer can provide cost-friendly IIoT network structure optimization solutions, without considering the number of IIoT tasks and the cost settings for adding servers and links.

In the future, we will study deterministic scheduling of IIoT periodic time-critical computing tasks on IIoT networks with heterogeneous servers. Furthermore, we will explore sharing mechanisms to support deterministic scheduling of both TSN traffics and IIoT tasks on the same IIoT networks. The optimization of IIoT network structure will continue to be researched, so that IIoT networks can be compatible with more heterogeneous devices and servers.

\bibliographystyle{IEEEtranbst}
\bibliography{huyujiao}

\begin{IEEEbiography}[{\includegraphics[width=1in,height=1.25in,clip,keepaspectratio]{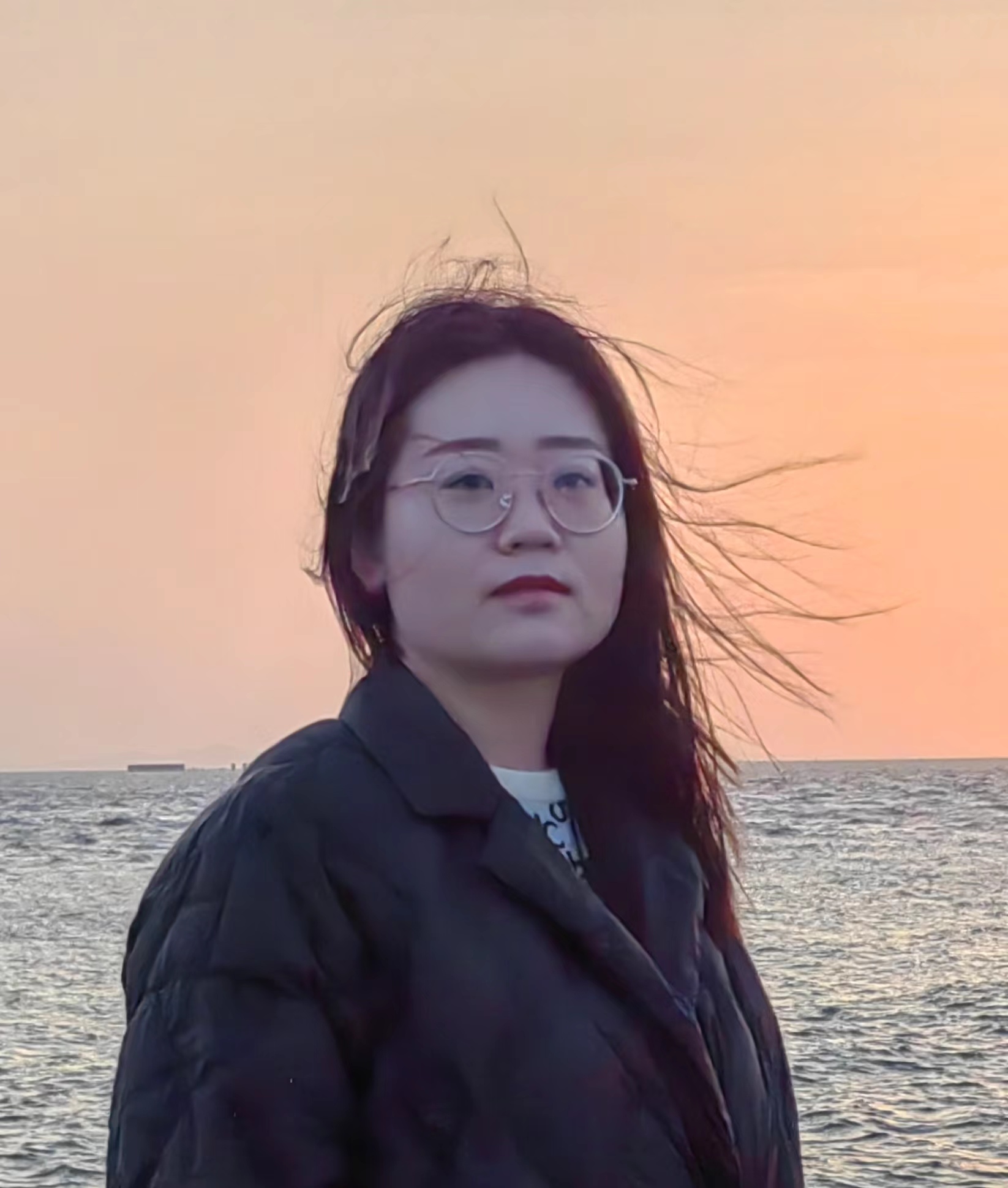}}]{Yujiao Hu}
    (Memeber, IEEE) received her Bachelor and PhD degrees from the Department of Computer Science of Northwestern Polytechnical University, Xi'an, China, in 2016 and 2021 respectively. From Nov. 2018 to March 2020, she was a visiting PhD student in National University of Singapore. Currently, she is a faculty member in Purple Mountain Laboratories. She focuses on deep learning, edge computing, multi-agent cooperation problems and time sensitive networks.    
\end{IEEEbiography}

\begin{IEEEbiography}[{\includegraphics[width=1in,height=1.25in,clip,keepaspectratio]{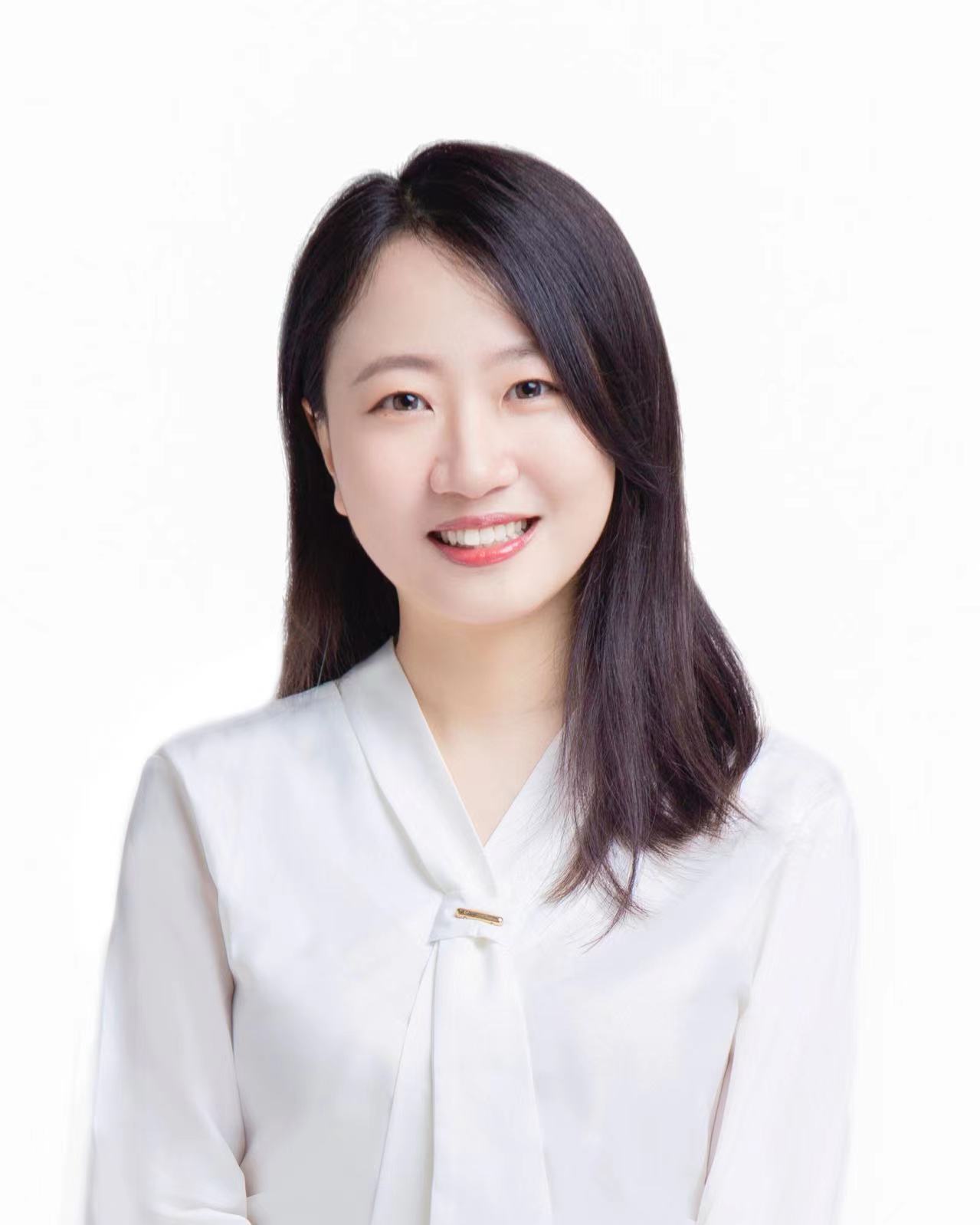}}]{Yining Zhu} 
received her B.E. degree from Zhejiang University, China, in 2015, and her Ph.D. degree from Northwestern University, the U.S., in 2019. She is currently an Associate Professor in the School of Computer Science, Northwestern Polytechnical University in China. Her research interests include network resource market, embodied AI and edge computing.
\end{IEEEbiography}

\begin{IEEEbiography}[{\includegraphics[width=1in,height=1.25in,clip,keepaspectratio]{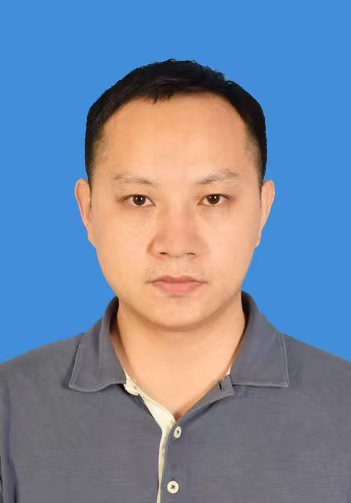}}]{Huayu Zhang} (Member, IEEE) received the B.E. degree from the Huazhong University of Science and Technology, China, in 2010, and the Ph.D. degree from Peking University, China, in 2017. He is currently a Researcher with Purple Mountain Laboratories, Nanjing, China. His research interests include distributed systems, graph theory and time sensitive networks. 
\end{IEEEbiography}

\begin{IEEEbiography}[{\includegraphics[width=1in,height=1.25in,clip,keepaspectratio]{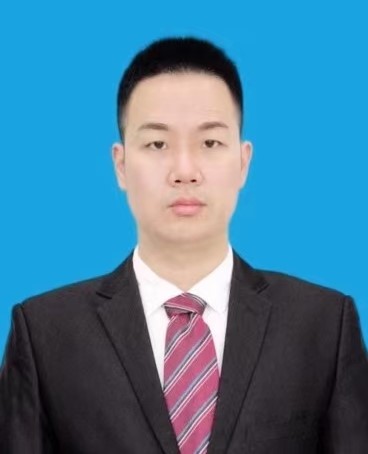}}]{Yan Pan} 
	 is currently a lecture at the Science and Technology on Information Systems Engineering Laboratory, National University of Defense Technology, Changsha, China, since Dec. 2020. Before that, he respectively received the B.S. degree in 2013 and the PhD. degree in 2021 from Northwestern Polytechnical University, Xi'an, China. He was a visiting student to University of Maryland, United States, during Jan. 2017 and Nov. 2018. His research interests include Industrial Internet of Things, industrial robots, and edge computing.
\end{IEEEbiography}

\begin{IEEEbiography}[{\includegraphics[width=1in,height=1.25in,clip,keepaspectratio]{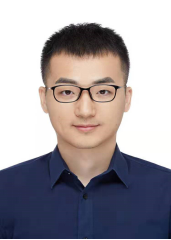}}]{Qingmin Jia}
    is currently a Researcher in Future Network Research Center of Purple Mountain Laboratories. He received the B.S. degree from Qingdao University of Technology in 2014, and received the Ph.D. degree from Beijing University of Posts and Telecommunications (BUPT) in 2019. His current research interests include edge computing, edge intelligence, IIoT and future network architecture. He has served as a Technical Program Committee Member of IEEE GLOBECOM 2021, HotICN 2021. 
\end{IEEEbiography}

\begin{IEEEbiography}[{\includegraphics[width=1in,height=1.25in,clip,keepaspectratio]{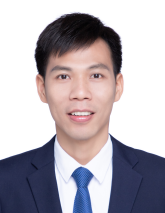}}]{Renchao Xie} 
    (Senior Member, IEEE) received the Ph.D. degree from the School of Information and Communication Engineering, Beijing University of Posts and Telecommunications (BUPT), Beijing, China, in 2012. He is a Professor with BUPT and Purple Mountain Laboratories. From November 2010 to November 2011, he visited Carleton University, Ottawa, ON, Canada, as a Visiting Scholar. His current research interests include edge computing, time-sensitive networks, and future network architecture. Dr. Xie has served as a Technical Program Committee Member of numerous conferences, including IEEE Globecom, IEEE ICC, EAI Chinacom, and IEEE VTC-Spring. 
\end{IEEEbiography}

\begin{IEEEbiography}[{\includegraphics[width=1in,height=1.25in,clip,keepaspectratio]{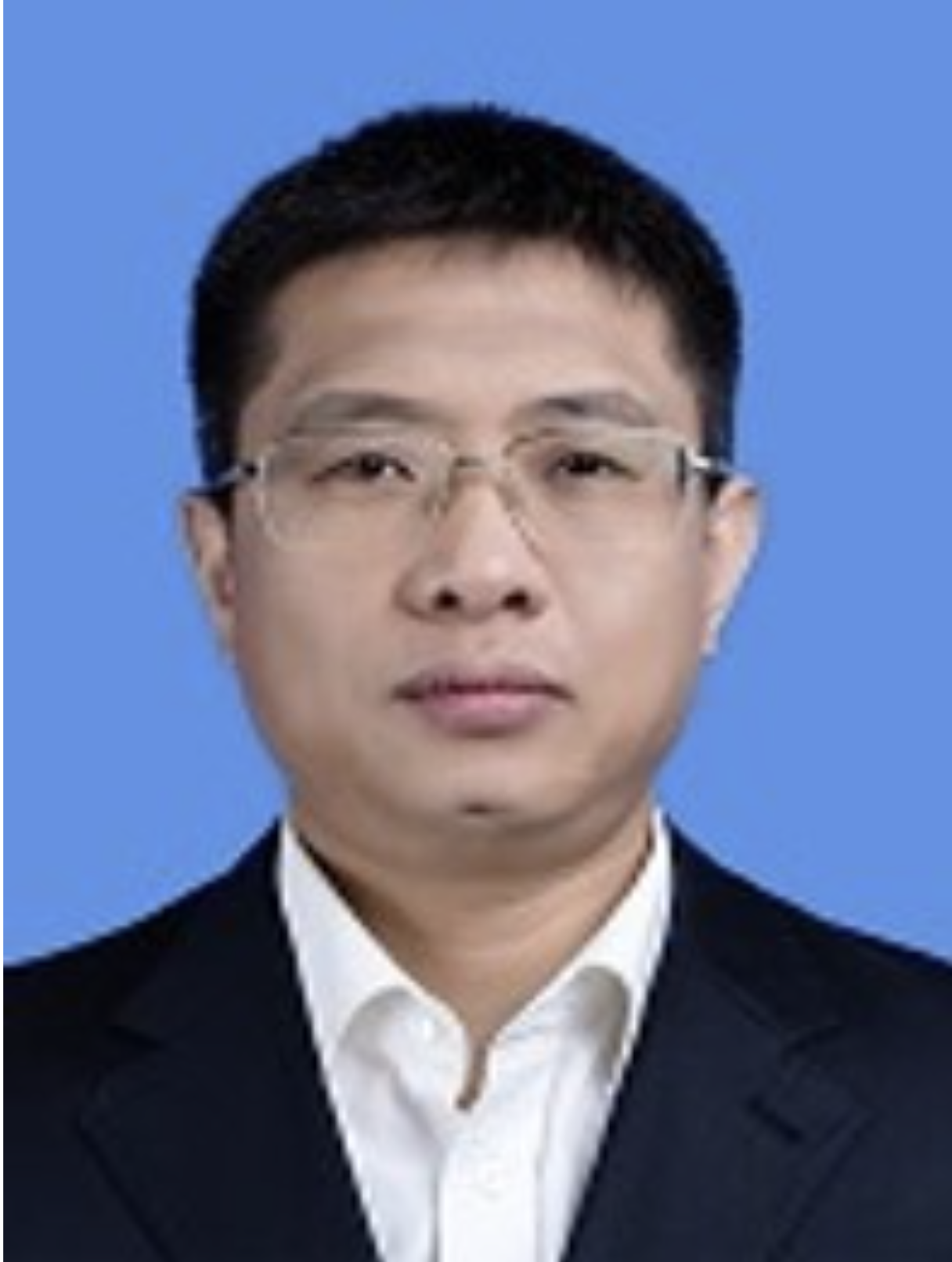}}]{Gang Yang} 
    (Member, IEEE) received the BE degree from the School of Automation from Second Artillery Engineering College of PLA, Xi’an, China, in 1998 and the MS and PhD degrees in computer science from Northwestern Polytechnical University, Xi’an, China, in 2002 and 2006, respectively. He is currently a professor with the School of Computer Science, Northwestern Polytechnical University, Xi’an, China. His research interests include distributed computing systems, cyber-physical systems and industrial Internet of Things.
\end{IEEEbiography}

\begin{IEEEbiography}[{\includegraphics[width=1in,height=1.25in,clip,keepaspectratio]{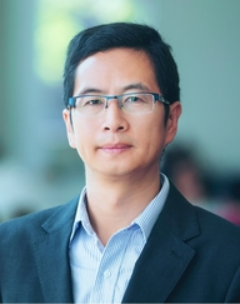}}]{F. Richard Yu } 
	(Fellow, IEEE) received the Ph.D. degree in electrical engineering from the University of British Columbia, Vancouver, BC, Canada, in 2003. From 2002 to 2006, he was with Ericsson, Lund, Sweden, and a start-up in California, USA. He joined Carleton University, Ottawa, ON, Canada, in 2007, where he is currently a Professor. His research interests include cyber–physical systems, connected/autonomous vehicles, distributed ledger technology, Industrial Internet of Things and deep learning. He is a Distinguished Lecturer, the Vice President (Membership), and an Elected Member of the Board of Governors of the IEEE Vehicular Technology Society. He is a Fellow of the IEEE, Canadian Academy of Engineering (CAE), Engineering Institute of Canada (EIC), and Institution of Engineering and Technology (IET).
\end{IEEEbiography}
\end{document}